\documentclass[fleqn,usenatbib]{mnras}

\usepackage{newtxtext,newtxmath}
\usepackage{float}
\usepackage{graphicx}

\usepackage{threeparttable}
\usepackage{lineno}

\usepackage[T1]{fontenc}

\usepackage[utf8]{inputenc}

\DeclareRobustCommand{\VAN}[3]{#2}
\let\VANthebibliography\thebibliography
\def\thebibliography{\DeclareRobustCommand{\VAN}[3]{##3}\VANthebibliography}

\usepackage{graphicx}	
\usepackage{amsmath}	
\usepackage{threeparttable}
\usepackage{outlines}
\usepackage{float}
\usepackage{multirow}
\usepackage{orcidlink}

\title[AT2024lhc and AT2024kmq as featureless TDEs]{AT2024lhc and AT2024kmq in the landscape of featureless tidal disruption events}

\author[Y. Yao et al.]{Yuhan Yao$^{1,2,3}$\thanks{E-mail: yuhanyao@berkeley.edu}\orcidlink{0000-0001-6747-8509},
Ryan Chornock$^{2,3}$\orcidlink{0000-0002-7706-5668},
Andrew Mummery$^{4}$,
Raffaella Margutti$^{2,3,5}$\orcidlink{0000-0003-4768-7586},
Marat Gilfanov$^{6,7}$,
\newauthor
Muryel Guolo$^{8}$\orcidlink{0000-0002-5063-0751},
Eric R. Coughlin$^{9}$\orcidlink{0000-0003-3765-6401},
Wenbin Lu$^{5,10,3}$\orcidlink{0000-0002-1568-7461},
Joheen Chakraborty$^{11}$\orcidlink{0000-0002-0568-6000},
Dheeraj R. Pasham$^{12,13}$\orcidlink{0000-0003-1386-7861},
\newauthor 
Kate D. Alexander$^{14}$\orcidlink{0000-0002-8297-2473},
Olivia Aspegren$^{2}$\orcidlink{0000-0001-5674-8403},
Charlotte R. Angus$^{15}$\orcidlink{0000-0002-4269-7999},
Xinze Guo$^{2,3,16}$\orcidlink{0009-0002-9727-8326},
Xander J. Hall$^{17}$\orcidlink{0000-0002-9364-5419},
\newauthor
Erica Hammerstein$^{2,3}$\orcidlink{0000-0002-5698-8703}, 
K.-Ryan Hinds$^{18}$\orcidlink{0000-0002-0129-806X},
Anna Y. Q. Ho$^{19}$\orcidlink{0000-0002-9017-3567},
Xiaoshan Huang$^{20}$\orcidlink{0000-0003-2868-489X},
Elias Kammoun$^{21}$\orcidlink{0000-0002-0273-218X},
\newauthor
Natalie LeBaron$^{2,3}$\orcidlink{0000-0002-2249-0595},
Matteo Lucchini$^{22}$\orcidlink{0000-0002-2235-3347},
Zoë McGrath$^{18}$\orcidlink{0009-0006-0726-1328},
Matt Nicholl$^{15}$\orcidlink{0000-0002-2555-3192},
Daniel A. Perley$^{18}$\orcidlink{0000-0001-8472-1996},
\newauthor
R. Michael Rich$^{23}$\orcidlink{0000-0003-0427-8387},
Genevieve Schroeder$^{19}$\orcidlink{0000-0001-9915-8147},
Xinyue Sheng$^{15}$\orcidlink{0000-0002-6527-1368},
Jesper Sollerman$^{24}$\orcidlink{0000-0003-1546-6615},
\newauthor
Jean Somalwar$^{2,3,25}$\orcidlink{0000-0001-8426-5732},
Jacob R. Wise$^{18}$\orcidlink{0000-0003-0733-2916},
Michael W. Coughlin$^{26}$\orcidlink{0000-0002-8262-2924},
Andrew Drake$^{27}$\orcidlink{0000-0003-0228-6594},
\newauthor
Matthew J. Graham$^{27}$\orcidlink{0000-0002-3168-0139},
George Helou$^{28}$\orcidlink{0000-0003-3367-3415},
Joahan C. Jaimes$^{21}$\orcidlink{0000-0002-0987-3372},
Mansi M. Kasliwal$^{27}$\orcidlink{0000-0002-5619-4938},
\newauthor
Ashish A. Mahabal$^{21,30}$\orcidlink{0000-0003-2242-0244},
Pavel Medvedev$^{6}$\orcidlink{0000-0002-9380-8708},
Josiah Purdum$^{31}$\orcidlink{0000-0003-1227-3738},
Ben Rusholme$^{28}$\orcidlink{0000-0001-7648-4142},
Rashid Sunyaev$^{6,7}$
\\
\\
Aﬃliations are listed at the end of the manuscript.
}

\date{Accepted XXX. Received YYY; in original form ZZZ}

\pubyear{\the\year{}}

\begin{document}

\label{firstpage}
\pagerange{\pageref{firstpage}--\pageref{lastpage}}
\maketitle

\begin{abstract}
We study AT2024kmq and AT2024lhc, two tidal disruption events (TDEs) with blue featureless spectra associated with high-mass black holes ($M_{\rm BH}\sim 10^8\,M_\odot$). Both events show optical precursors consistent with shock dissipation from stream self-intersection. Their X-ray emission is luminous ($L_{\rm X}\sim 10^{44}\,{\rm erg\,s^{-1}}$), highly variable (with minimum observed variability timescales of 1.3\,hr and 4.8\,hr for factor of $\sim3$ flux changes), long-lasting ($>1\,\rm yr$), emerging no later than the optical peak, and well characterized by power-laws with $1.7<\Gamma<3$ (where $f_\nu \propto \nu^{1-\Gamma}$). The X-ray properties and radio non-detections support a compact corona ($\lesssim 10 r_{\rm g}$) producing Comptonized X-ray emission. Using all published featureless TDEs, we find statistically significant bimodality in the distribution of their peak UV/optical blackbody luminosities and radii. We assemble a comparison TDE sample with early-time X-ray observations with eROSITA, in which we find different $M_{\rm BH}$ distributions in TDEs with different X-ray spectral evolution properties: low-mass black holes ($M_{\rm BH} \sim 10^6 M_\odot$) remain soft ($\Gamma>4$) within $t\lesssim 2$\,yr, intermediate masses ($\sim 10^7 M_\odot$) transition from soft to hard at $\sim$1 yr, while high masses ($\sim 10^8 M_\odot$) are hard ($1.5<\Gamma\lesssim 3$) from the outset. We interpret this result as evidence that the soft-to-hard state transition in TDEs occurs at the critical threshold of $\dot{M}_{\rm acc} \sim 0.03 \dot M_{\rm Edd}$ (similar to X-ray binaries), using the fact that the transition timescale predicted by simple disk theory scales with black hole mass as $t_{\rm tr}\propto M_{\rm BH}^{-3/4}$. 
\end{abstract}




\section{Introduction} \label{sec:intro}

Tidal disruption events (TDEs) occur when a star passes sufficiently close to a massive black hole (MBH) to be torn apart by tidal forces. 
The canonical tidal disruption radius 
\begin{align}
    r_{\rm T} = R_\star (M_{\rm BH}/M_\star)^{1/3}=3.2\times 10^{13}M_{\rm BH, 8}^{1/3}(\rho_\star / \rho_\odot)^{-1/3}\,{\rm cm} \label{eq:rt},
\end{align}
where $M_{\rm BH, 8} \equiv M_{\rm BH} / (10^8\,M_\odot)$; $R_\star$, $M_\star$, and $\rho_\star$ are radius, mass, and mean density of the star.
First theoretically predicted in the 1970s \citep{Hills1975, lacy82}, early observational candidates were identified in the soft X-ray \citep[e.g.,][]{Komossa1999}, where emission originates from the inner accretion disk. In recent years, optical time-domain surveys have come to dominate TDE discoveries.

In TDEs, the mass fall-back rate is expected to follow a power-law decline $\dot{M}_{\rm fb} \propto t^{-5/3}$ \citep{Rees1988, Phinney1989}. Mass accretion onto the black hole is delayed since it takes time for material to redistribute its angular momentum. Nevertheless, the mass accretion rate still varies by orders of magnitude over timescales of years (see Appendix A of \citealt{Mummery2025_calorimetry}). This rapid evolution makes TDEs ideal testbeds for whether MBHs undergo accretion state transitions analogous to those observed in stellar-mass X-ray binaries (XRBs). 

During the outbursts of transient BH XRBs, the systems transition between distinct spectral states as the accretion rate evolves. At high accretion rates, systems are in the high soft state (HSS), where X-ray emission is dominated by thermal disk radiation. The inner disk radius is located at the innermost stable circular orbit (ISCO). At low accretion rates, they occupy the low hard state (LHS), characterized by power-law spectra from Comptonization in a hot corona with photon index $\Gamma \sim 1.5$--2.0 \citep{Remillard2006, Done2007, gilfanov10}. Here, the inner disk is truncated at $r_{\rm tr}>10^2 r_{\rm g}$, where the gravitational radius is
\begin{align}
    r_{\rm g} = \frac{GM_{\rm BH}}{c^2} = 1.5\times 10^{13} M_{\rm BH, 8}\,{\rm cm}. \label{eq:rg}
\end{align}
Within $r_{\rm tr}$ there exists a hot accretion flow, and the exact value of $r_{\rm tr}$ likely depends on the mass accretion rate \citep{Yuan2004, Yuan2014}. At sufficiently low rates, the thin disk is absent. 

Between the HSS and the LHS lies the intermediate state (IMS), where the X-ray spectrum has comparable contributions from disk and power-law components and $\Gamma$ softens to 2.0--2.5. The IMS is further divided into the hard intermediate state (HIMS) and soft intermediate state (SIMS). The HIMS $\rightarrow$ SIMS transition is associated with the optically thick, geometrically thin accretion disk extending inward to replace the hot accretion flow \citep{Yuan2014}, and with crossing of the ``jet line'' on the hardness-intensity diagram, where compact jet radio emission peaks during the HIMS and collapses in the SIMS \citep{Fender2004, Belloni2010}.  

The transition between soft states (HSS and SIMS) and hard states (LHS and HIMS) is expected to occur around a critical dimensionless accretion rate \citep[][Equation 27]{Yuan2014}: 
\begin{align}
    \dot{m} \equiv {\dot M}_{\rm acc}/{\dot M}_{\rm Edd} \approx (0.06 \text{--}0.08)\alpha,
\end{align}
where $\dot M_{\rm acc}$ is the accretion rate, $\dot{M}_{\rm Edd}\equiv L_{\rm Edd}/(0.1c^2)$, $L_{\rm Edd}=1.26\times 10^{38}(M_{\rm BH}/M_\odot)\,{\rm erg\,s^{-1}}$ is the Eddington luminosity, and $\alpha$ is the ``viscous'' nuisance parameter of \citet{Shakura1973}. For typical values of $\alpha \sim 0.1$--0.4 \citep{King2007}, this yields $\dot{m} \sim 0.01$--0.03. Observations of BH XRBs are roughly consistent with this expectation: the soft-to-hard state transitions occur at $\sim$0.005--0.1$L_{\rm Edd}$ with a mean of $0.03 L_{\rm Edd}$ \citep{Maccarone2003, Dunn2010, Tetarenko2016, VahdatMotlagh2019}. 

Recent observations suggest that at least some TDEs exhibit state transitions \citep{Jonker2020, Wevers2021, Yao2022, Guolo2024, Hajela2025, Berger2026}, but whether this phenomenon is universal across the TDE population, and how it depends on $M_{\rm BH}$ and $\dot{m}$, remain open questions.

In the optical band, TDEs exhibit distinctive observational signatures that distinguish them from other nuclear transients. Their peak optical luminosities span a range similar to supernovae (SNe), with $-17\lesssim M_{g, \rm peak}\lesssim -23$ \citep{Yao2023}. However, unlike SNe, TDEs maintain persistently hot temperatures: blackbody fits to UV/optical spectral energy distributions (SEDs) yield $T_{\rm bb}\sim {\rm few}\times 10^4$\,K throughout their evolution \citep{Hammerstein2023}, with minimal color evolution. At late times (months to years post-peak), TDEs exhibit a characteristic UV/optical plateau phase that is consistent with a viscously spreading outer accretion disk \citep{vanvelzen19_late_time_UV, Mummery2020_14li, Mummery2024_fundamental_scaling, Alush2025_plateau, Guolo2026}.

The optical spectra of many TDEs are characterized by broad emission lines of hydrogen, helium, and Bowen fluorescence features (see a thorough analysis of spectral evolution by \citealt{Charalampopoulos2022}). 
Using 17 TDEs selected from the first 1.5\,yr of the Zwicky Transient Facility (ZTF; \citealt{Bellm2019b, Graham2019, Dekany2020}) optical sky survey, \citet{vanVelzen2021} proposed a spectroscopic classification scheme with three main classes: (i) TDE-H, exhibiting broad H$\alpha$ and H$\beta$ emission lines; (ii) TDE-H+He, showing broad Balmer lines plus a broad \ion{He}{ii} $\lambda4686$ complex, with most events also displaying Bowen fluorescence features (\ion{N}{iii}, \ion{O}{iii}); (iii) TDE-He, characterized by broad \ion{He}{ii} $\lambda4686$ emission only, with no broad Balmer lines.

\citet{Hammerstein2023} expanded the sample to 30 TDEs from the first 2.5 yr of ZTF and identified a fourth class: (iv) TDE-featureless, characterized by a blue continuum with no discernible emission or absorption features. Additionally, earlier studies have associated a fraction of coronal line emitters with TDEs (e.g., \citealt{Komossa2008, Wang2011, Wang2012, Onori2022}). This TDE-coronal spectroscopic class was also recovered through a retrospective search of TDEs in ZTF \citep{Yao2023}.

The four TDE-featureless events presented in \citet{Hammerstein2023} are all exceptionally luminous, with rest-frame $g$-band absolute magnitude $-23\lesssim M_{g, \rm peak}\lesssim -20.5$, where intrinsic TDE rates are low \citep{Yao2023}. Consequently, these events are typically discovered at high redshifts ($z>0.1$). \citet{Andreoni2022} suggested this population might harbor off-axis jets, as the UV/optical properties of luminous featureless TDEs and jetted TDEs are similar (see also spectral analysis in \citealt{Hammerstein2026}).
However, radio follow-up observations of a sample of eight overluminous TDEs (five of which are featureless) found no evidence of off-axis jets \citep{Yao2025_radio}. 
Recently, several subluminous featureless TDEs with $M_{g, \rm peak} \gtrsim -18$ have been identified \citep{Yao2022, Zhu2025}, raising the question of whether the featureless class exhibits a continuous luminosity distribution or is intrinsically bimodal. 

Understanding the population of luminous, spectroscopically featureless TDEs is of vital importance for several reasons.
First, these events dominate the TDE population around high-mass black holes  ($M_{\rm BH}\sim 10^8\,M_\odot$), where $r_{\rm T}$ (Eq.~\ref{eq:rt}) becomes comparable to $r_{\rm g}$ (Eq.~\ref{eq:rg}).
In this regime, general relativistic (GR) effects fundamentally shape the debris dynamics and energy dissipation. Luminous featureless TDEs therefore provide unique laboratories for testing shock and accretion physics in the strong-field limit.
Second, at the high-mass end, measuring TDE rates as a function of black hole mass constrains the event horizon suppression effect, which depends on both black hole spin and stellar age \citep{beloborodov92, Kesden12_maximum_Mbh, dorazio19, coughlin22, Huang2024_tde_spin}. The observed rate decline at $M_{\rm BH}>10^7\,M_\odot$ thus probes the average spin of MBHs in quiescent galaxies \citep{Du2022} --- a quantity difficult to measure through other means. 
Finally, luminous TDEs will dominate samples at high redshift, where only the most energetic events remain detectable. Next-generation surveys, particularly the Vera C. Rubin Observatory's Legacy Survey of Space and Time (LSST; \citealt{Ivezic2019}) and the Nancy Grace Roman Space Telescope, will discover hundreds to thousands of such events at $z>1$. However, the current sample remains small ($\sim 10$ events), with limited multi-wavelength characterization. Establishing a comprehensive understanding of their properties and rates is therefore essential to enable their use as probes of MBHs and stars at high redshifts.

In this work, we present comprehensive multi-wavelength observations of two luminous featureless TDEs discovered by ZTF: AT2024kmq (ZTF24aapvieu) and AT2024lhc (ZTF24aaoxmyb). These events are particularly notable because they exhibit luminous hard X-ray emission contemporaneous with their optical peaks --- a characteristic previously unobserved in optically selected TDEs. We describe our discovery, classification, and basic properties in Section~\ref{sec:basic}, followed by multi-wavelength observations and analysis in Section~\ref{sec:obs}. In Section~\ref{sec:discuss}, we discuss the likely origin of the X-ray emission and highlight challenges in explaining the UV/optical peak emission using currently available models. We conclude in Section~\ref{sec:conclusion}. This work includes follow-up observations obtained through January 2026.

We adopt a standard flat $\Lambda$CDM cosmology with matter density $\Omega_{\rm M} = 0.3$ and the Hubble constant $H_0=70\,{\rm km\,s^{-1}\,Mpc^{-1}}$. We use the notation $\mathcal{Q}_n \equiv \mathcal{Q}/10^n$ for quantities in CGS units, with the exception that $M_{{\rm BH},n} \equiv M_{\rm BH}/(10^n M_\odot)$ for black hole masses. All logarithms are base 10 unless otherwise stated. Uncertainties are reported at 68\% confidence (1$\sigma$), and upper limits at 3$\sigma$. We correct for Galactic extinction using the \citet{Cardelli1989} extinction law with $R_V = 3.1$ and the \citet{Schlafly2011} dust map.

\section{Discovery, Classification, and Basic Information} \label{sec:basic}

The discovery and identification of AT2024kmq as a TDE have been published by \citet{Ho2025}, which presented follow-up observations obtained prior to October 2024. AT2024lhc was classified as a TDE by the ZTF group \citep{Yao2024_24lhc_astronote, Chornock2024_24lhc_CR}. 

\begin{table}
    \caption{Basic information for the two TDEs.\label{tab:basic_info}}
\begin{threeparttable}
    \begin{tabular}{ccc} 
    \hline
    \hline
	Parameter & AT2024kmq & AT2024lhc\\
    \hline
    \hline
    ZTF name & ZTF24aapvieu & ZTF24aaoxmyb\\
       $t_0$ (MJD)     & 60460.24495 & 60424.40988 \\ 
       $z$        & $0.192$ &  $0.2045$\\
       $E_{B-V, \rm MW}$ (mag)  & 0.0169 & 0.0233\\
       $N_{\rm H, MW}$ ($\rm cm^{-2}$) & $1.77\times 10^{20}$ & $2.34\times 10^{20}$\\
       log($M_{\rm gal}/M_\odot)$ & $11.22_{-0.04}^{+0.03}$ & $11.39_{-0.06}^{+0.05}$ \\
       $E_{\rm B-V, \rm host}$ (mag) & $0.016_{-0.011}^{+0.022}$ & $0.027_{-0.016}^{+0.020}$   \\
       $\sigma_\ast$ (km\,s$^{-1}$) & $181\pm 21$ & $230.5\pm 19.8$\\
       \hline
       log($M_{\rm BH}/M_\odot)^{a}$ & $8.30 \pm 0.37 $ & $8.76 \pm 0.34$ \\
       log($M_{\rm BH}/M_\odot)^{b}$ & $8.54\pm0.37$ & $8.81\pm0.43$\\
       log($M_{\rm BH}/M_\odot)^{c}$ & $7.66^{+0.24}_{-0.20}$ & $7.93^{+0.26}_{-0.24}$\\
       \hline
       0.2--2.3\,keV $L_{\rm X}$ (erg\,s$^{-1}$) & $<1.6\times 10^{42}$ & $<8.5\times 10^{41}$ \\
       $L_{[\ion{O}{III}]}$ (erg\,s$^{-1}$) & $<3.0\times 10^{39}$ & $<5.2\times 10^{39}$ \\
       $W1 - W2$ (mag)  &   $0.308\pm0.020$   &    $0.307\pm0.017$  \\
    \hline
    \hline
\end{tabular}
\begin{tablenotes}
    \item $^{a}$ MBH mass estimated using the \citet{Kormendy2013} $M_{\rm BH}$--$\sigma_\ast$ relation.
    \item $^{b}$ MBH mass estimated using the $M_{\rm gal}$--$M_{\rm BH}$ scaling relation presented in \citet{Yao2023}.
    \item $^{c}$ MBH mass estimated using the \texttt{tidalspin} package \citep{Mummery2024_spin}, treating the $\sigma_\ast$-based estimates as priors. See text in \S\ref{subsec:Mbh}.
\end{tablenotes}
\end{threeparttable}
\end{table}

We use $t_{\rm obs}$ ($t_{\rm rest}$) to denote the observer-frame (rest-frame) time relative to $t_0$, which is time of the most recent ZTF upper limit prior to the first optical detection. 
The basic information for the two objects is presented in Table~\ref{tab:basic_info}, including the optical reference epoch $t_0$, redshift $z$, Milky-Way extinction $E_{B-V, \rm MW}$, the Galactic hydrogen-equivalent column density $N_{\rm H, MW}$ \citep{Willingale2013}, the host galaxy total stellar mass $M_{\rm gal}$ and extinction $E_{B-V, \rm host}$ (see \S\ref{subsec:prospector}), the stellar velocity dispersion $\sigma_\ast$ (see \S\ref{subsec:sigma}), the black hole masses estimated using host-galaxy scaling relations, the upper limits of the host-galaxy X-ray and [\ion{O}{III}] $\lambda5007$ narrow line luminosities (see \S\ref{subsec:not_agn}), and the $W1-W2$ color in the unWISE catalog (Vega system, \citealt{Schlafly2019}).

\subsection{Host Stellar Population Synthesis} \label{subsec:prospector}

To estimate global host galaxy properties, we model the pre-flare archival UV--IR SED of the host galaxy using population synthesis via the \texttt{prospector} software \citep{Johnson2021} built on \texttt{FSPS} \citep{Conroy2009}, following the same data collecting and fitting procedures as adopted by \citet[][Section 3]{vanVelzen2021}. The galaxy stellar mass $M_{\rm gal}$ is derived from the surviving fraction of stars formed. The host galaxy extinction $E_{B-V, \rm host}$ is computed as $0.268\times \texttt{dust2}$, where \texttt{dust2} is the the optical depth at 5500\,\AA\ of the \citet{Calzetti2000} attenuation curve. 

\subsection{Stellar Velocity Dispersion} \label{subsec:sigma}

For AT2024kmq, the central velocity dispersion was initially reported as $175 \pm 27\,{\rm km\,s^{-1}}$ from fitting the 4200--6800\,\AA\ Gemini Multi-object Spectrograph (GMOS) spectrum taken at $t_{\rm rest}=+8$\,d\footnote{At this epoch, the transient flux is $<10\mu$Jy ($>21.4$\,mag, see \S\ref{subsec:ztf_atlas}), contributing negligible light to the spectrum.} \citep{Ho2025}. However, we later identified flux calibration issues at the blue end. We therefore reextracted and refit the spectrum using the 5030--5600\,\AA\ region, obtaining $\sigma_\ast = 181 \pm 21\,{\rm km\,s^{-1}}$, which we adopt in this work.

For AT2024lhc, the host galaxy has been observed by the Sloan Digital Sky Survey (SDSS, \citealt{Gunn2006}) on 2003 June 1 ($t_{\rm rest}=-6337$\,d) and by the Dark Energy Spectroscopic Instrument (DESI; \citealt{DESI2025}) on 2021 May 1 ($t_{\rm rest}=-904$\,d). \citet{Simard2011} measured that the galaxy half-light radius is $r_{1/2}=16.2$\,kpc (4.8\arcsec), and the effective semi-major radius of the galaxy bulge is $r_{\rm e} = 20.7$\,kpc (i.e., 6.2\arcsec).
The SDSS (3\arcsec\ fiber size) pipeline measures $\sigma_\ast = 230.5\pm19.8\,{\rm km\,s^{-1}}$, and running the \texttt{fastspecfit} software \citep{Moustakas2023} on the DESI spectrum  (1.5\arcsec\ fiber size; \citealt{DESI2022_instrument}) gives $\sigma_\ast = 270\pm 26\,{\rm km\,s^{-1}}$. We adopt the SDSS value since its fiber size better matches $r_{1/2}$ and $r_{\rm e}$ where $\sigma_\ast$ (in the $M_{\rm BH}$--$\sigma_\ast$ relationship) is defined.

\subsection{Black Hole Masses} \label{subsec:Mbh}

The $M_{\rm BH}$–$\sigma_\ast$ relation gives ${\rm log}(M_{\rm BH}/M_\odot)$ values of $8.30\pm0.37$ for AT2024kmq and $8.76\pm0.34 $ for AT2024lhc. These estimates must be reconciled with the Hills mass constraint --- the maximum black hole mass capable of tidally disrupting a star before it is swallowed whole \citep{Hills1975}.

For an observable TDE, the pericenter distance must exceed the innermost bound circular orbit radius, $r_{\rm IBCO}$, which depends on the angle between the star's orbit with respect to the black hole’s spin axis $\phi$, and the dimensionless spin parameter $a$ \citep{Bardeen1972}. For non-spinning BHs (i.e., $a=0$) $r_{\rm IBCO} = 4r_{\rm g}$, and for prograde equatorial orbits with $a=1$ we have $r_{\rm IBCO}=r_{\rm g}$. An analytic solution for the Hills mass accounting for spin effects is provided by \citet{Mummery2024_spin}.

For solar-type stars, tidal disruption remains possible around $M_{\rm BH}\sim {\rm few}\times 10^8\,M_\odot$ black holes if they are rapidly spinning. However, disruption by a $M_{\rm BH}\sim 10^9\,M_\odot$ black hole, as suggested by the $\sigma_\ast$-based estimate for AT2024lhc, would typically require a more evolved, lower-density star. Treating the $\sigma_\ast$-based estimates as priors and incorporating the TDE identification as evidence, we use the \texttt{tidalpin}\footnote{We note that this package assumes a stellar mass function in the form of the Kroupa inital mass function \citep{Kroupa2001}, which might not be accurate in the nuclei of TDE host galaxies.} Bayesian framework \citep{Mummery2024_spin} to infer revised black hole masses. This analysis yields ${\rm log}(M_{\rm BH}/M_\odot) = 7.66^{+0.24}_{-0.20}$ and $|a_{\rm BH}| = 0.63_{-0.20}^{+0.38}$ for AT2024kmq, and ${\rm log}(M_{\rm BH}/M_\odot) = 7.93^{+0.26}_{-0.24}$ and $|a_{\rm BH}| = 0.84_{-0.15}^{+0.16}$ for AT2024lhc.

In summary, while precise black hole mass constraints remain challenging, both events are hosted by high-mass black holes with $M_{\rm BH}\sim 10^8\,M_\odot$. 

\subsection{Lack of X-ray/Optical/IR AGN Signatures} \label{subsec:not_agn}

The host-galaxy X-ray luminosity upper limits (see Table~\ref{tab:basic_info}) are constrained using archival observations from the eROSITA telescope \citep{Predehl2021} onboard the \textit{Spektrum-Roentgen-Gamma} (SRG) satellite \citep{Sunyaev2021}, obtained during 2020--2022. 

In the GMOS spectrum of AT2024kmq and the SDSS/DESI spectra of AT2024lhc, we do not identify narrow emission lines typically observed in active galactic nuclei (AGN). We place upper limits on their [\ion{O}{III}] $\lambda5007$ line luminosities (see Table~\ref{tab:basic_info}) by modeling the spectra using the \texttt{pPXF} software \citep{Cappellari2023}, following the procedures adopted by \citet{Zhang2026_eRO}.

AGN can be broadly divided into radiative mode ($L/L_{\rm Edd}\gtrsim0.01$) and jet mode ($L/L_{\rm Edd}\lesssim0.01$). The general picture is that radiative-mode AGN possess a geometrically thin, optically thick accretion disk with the inner disk radius at ISCO, whereas jet-mode AGN are surrounded by an inner hot accretion flow and an outer truncated disk --- similar to the soft and hard states in XRBs \citep{Heckman2014}. 

For radiative-mode AGN, the [\ion{O}{III}] $\lambda5007$ line luminosity correlates with the bolometric luminosity \citep{Heckman2014}. Using the bolometric correction of \citet{Heckman2004}, we estimate the AGN bolometric luminosities to be $< 1.0\times 10^{43}\,{\rm erg\,s^{-1}}$ and $<1.8\times 10^{43}\,{\rm erg\,s^{-1}}$ for AT2024kmq and AT2024lhc, respectively. These upper limits rule out the existence of radiative-mode AGN in their host galaxies, consistent with the fact that their MIR $W1-W2$ colors are not AGN-like \citep{Stern2012}.

While the host galaxies may harbor jet-mode AGN (discussed in \S\ref{subsec:radio}), the absence of radiative-mode AGN indicates there is no pre-existing disk extending to the ISCO. Therefore, the luminous optical transients cannot be caused by sudden changes in the mass accretion rate of a pre-existing AGN disk.

\begin{figure*}
\centering
    \includegraphics[width=\textwidth]{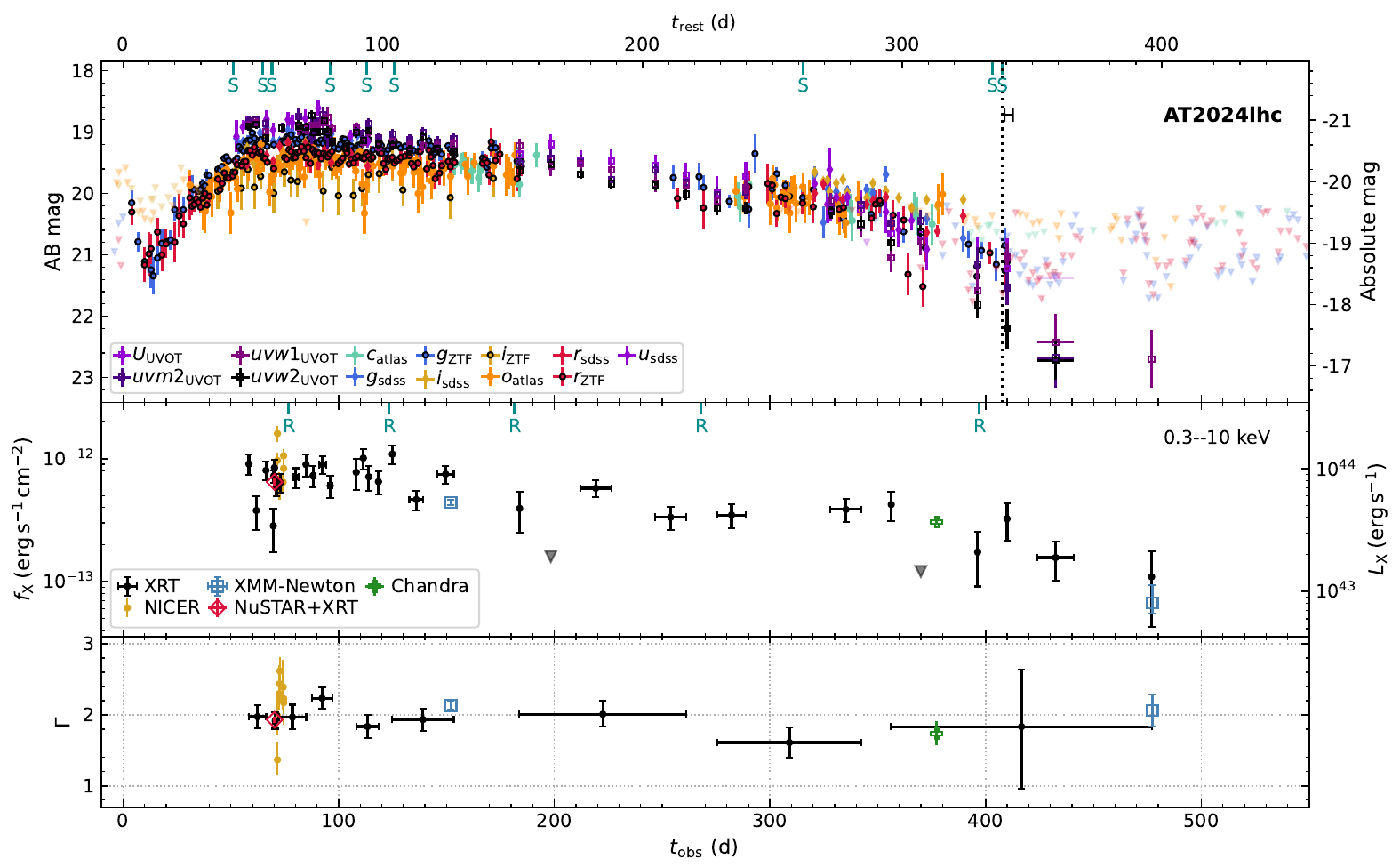}
    \caption{UV, optical and X-ray evolution of AT2024lhc. 
    UV and optical light curves are shown in the upper panel, with epochs of optical spectroscopy marked with letter ``S'', and the epoch of HST spectroscopy marked with a vertical dotted line. 
    Solid points represent detections above 3$\sigma$ in the optical and above 2$\sigma$ in the UV; semitransparent downward triangles indicate 3$\sigma$ upper limits.
    The observed 0.3--10\,keV X-ray light curves are shown in the middle panel, with epochs of radio continuum observations marked with letter ``R''.
    The evolution of the X-ray power-law photon index $\Gamma$ is shown in the lower panel.
    \label{fig:lc_overview_24lhc}}
\end{figure*}

\begin{figure*}
\centering
    \includegraphics[width=\textwidth]{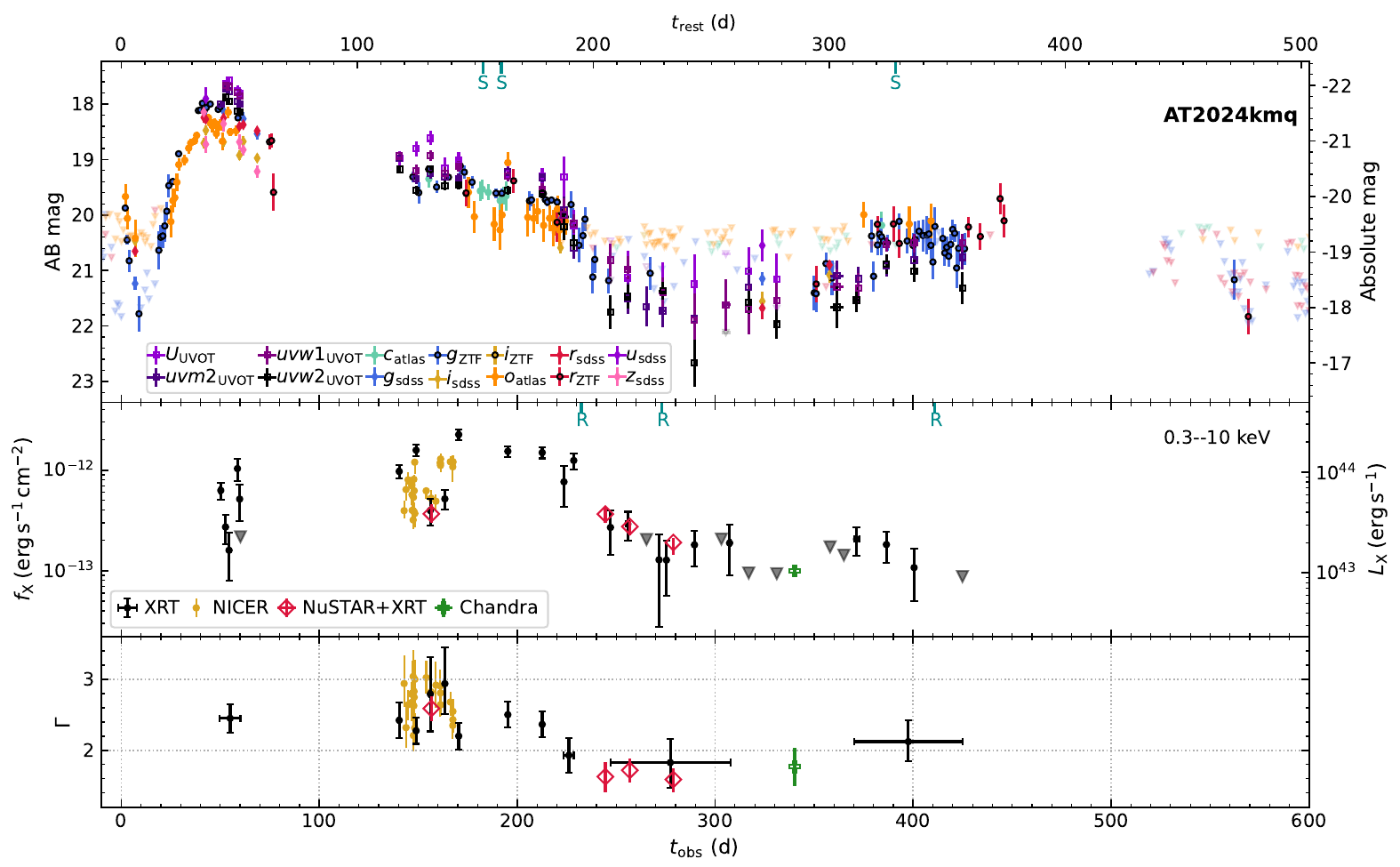}
    \caption{UV, optical and X-ray evolution of AT2024kmq. We only indicate epochs of optical spectroscopy and radio observations first presented in this work, see \citet{Ho2025} for earlier observations.
    \label{fig:lc_overview_24kmq}}
\end{figure*}

\section{Observations and Analysis} \label{sec:obs}

\subsection{Optical Photometry}
\label{subsec:opt_phot}

\subsubsection{ZTF and ATLAS} \label{subsec:ztf_atlas}
We obtained ZTF \citep{Masci2019, Masci2023} and ATLAS \citep{Tonry2018, Smith2020, Shingles2021} forced photometry. 
For ATLAS data, we cleaned and corrected the photometry using the \texttt{ATClean} method \citep{Rest2025}. 
Baseline correction was performed using the methods outlined in \citet{Yao2019}. 

\begin{figure}
    \centering
    \includegraphics[width=\columnwidth]{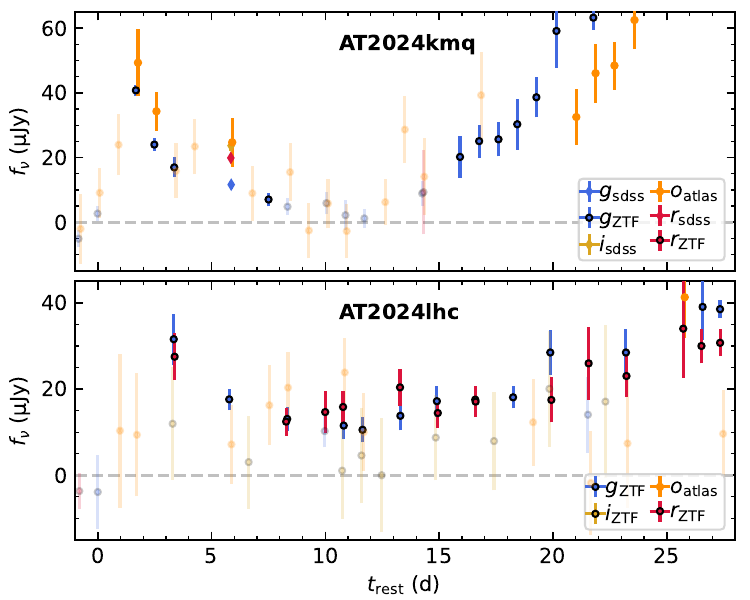}
    \caption{Early-phase optical light curves of AT2024kmq and AT2024lhc in flux space, showing the precursor emission detected in both events. Solid and semitransparent points represent $>3\sigma$ detections and other observations, respectively. The dashed horizontal line at zero is the average value pre-transient. \label{fig:precursor}}
\end{figure}

The Galactic extinction-corrected optical light curves of AT2024lhc and AT2024kmq are shown in the upper panels of Figure~\ref{fig:lc_overview_24lhc} and Figure~\ref{fig:lc_overview_24kmq}, respectively. They exhibit significant optical precursors at $t_{\rm rest}<10$\,d. The precursor in AT2024kmq is particularly prominent: it was initially identified as an independent fast-evolving transient before the main TDE flare emerged. Figure~\ref{fig:precursor} highlights these precursor light curves in flux space. We discuss their physical interpretation in \S\ref{subsec:precursor}.

\subsubsection{LT}

We obtained follow-up photometry of both events using the IO:O Optical Imager on the Liverpool Telescope (LT; \citealt{Steele2004}). Image subtraction was performed using reference images from Pan-STARRS1 (PS1; \citealt{Chambers2016}) for the $g$, $r$, $i$, and $z$ bands, and from SDSS \citep{SDSS2017} for the $u$ band.

For AT2024lhc, we performed aperture photometry on the LT data using the Photometry Sans Frustration ({\texttt{PSF}}; \citealt{Nicholl2023}) routine, which uses an optimal aperture that captures 90\% of point spread function flux after template subtraction. For AT2024kmq, photometry was obtained using a PSF-fitting methodology based on the technique described in \citet{Fremling2016}.

In addition, for AT2024kmq we include LT and Lowell Discovery Telescope (LDT) photometry published by \citet{Ho2025}. The LT and LDT photometry are shown in Figures~\ref{fig:lc_overview_24lhc}--\ref{fig:lc_overview_24kmq}. Both instruments employ SDSS filters.

The LT $z$ band data was not included for further analysis, as we found that photometric uncertainties are underestimated in this band.

\subsection{Optical Spectroscopy}
\label{subsec:opt_spec}

We obtained low-resolution optical spectra using the Kast spectrograph on the Shane 3-m telescope at Lick Observatory \citep{Miller1993}, the Low Resolution Imaging Spectrograph (LRIS; \citealt{Oke1995}) on the Keck-I telescope, and the Alhambra Faint Object Spectrograph and Camera (ALFOSC\footnote{\href{http://www.not.iac.es/instruments/alfosc}{{http://www.not.iac.es/instruments/alfosc}}}) at the 2.56\,m Nordic Optical Telescope (NOT) located at the Roque de los Muchachos Observatory on La Palma (Spain).
These observations were coordinated using the \textit{fritz.science} instance of \texttt{SkyPortal} \citep{vanderWalt2019, Coughlin2023}. 
The observing log is provided in Table~\ref{tab:spec_lowres}. Epochs of spectroscopic observations are marked with `S' in Figures~\ref{fig:lc_overview_24lhc}--\ref{fig:lc_overview_24kmq}.

Kast observations were obtained with the D57 dichroic, 600/4310 grism (blue), and 300/7500 grating (red), unless otherwise noted in Table~\ref{tab:spec_lowres}. Reductions followed \citet{Silverman2012}. 
LRIS observations used the 600/4000 grism (blue), D56 dichroic, and 400/8500 grating (red). 
ALFOSC observations employed grism 4 and were reduced using a custom \texttt{PypeIt} fork \citep{pypeit:zenodo, pypeit:joss_arXiv, pypeit:joss_pub}.

\subsubsection{AT2024kmq}
\begin{figure}
    \includegraphics[width=\columnwidth]{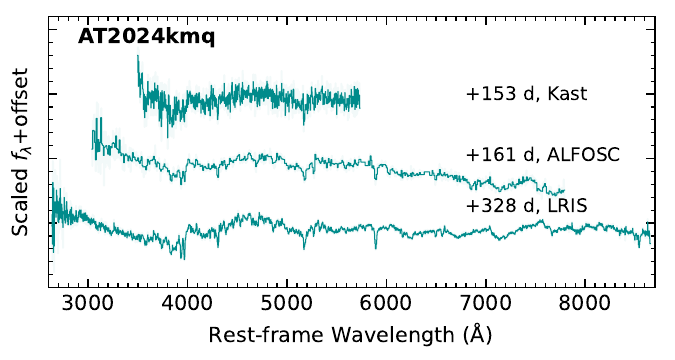}
    \caption{Optical spectra of AT2024kmq obtained after October 2024 (see \citealt{Ho2025} for earlier epochs). Strong telluric features in the ALFOSC spectrum are masked. No prominent broad emission lines characteristic of TDEs are detected.\label{fig:opt_spec_24kmq}}
\end{figure}

Figure~\ref{fig:opt_spec_24kmq} shows optical spectra of AT2024kmq obtained after October 2024; earlier spectra are presented in \citet{Ho2025}. No prominent broad emission lines characteristic of TDEs are detected at any epoch, confirming the spectroscopically featureless nature reported by \citet{Ho2025} at earlier times. We do not perform host–transient decomposition due to the lack of a pre- or post- flare host spectrum template.

\subsubsection{AT2024lhc}

\begin{figure}
    \includegraphics[width=\columnwidth]{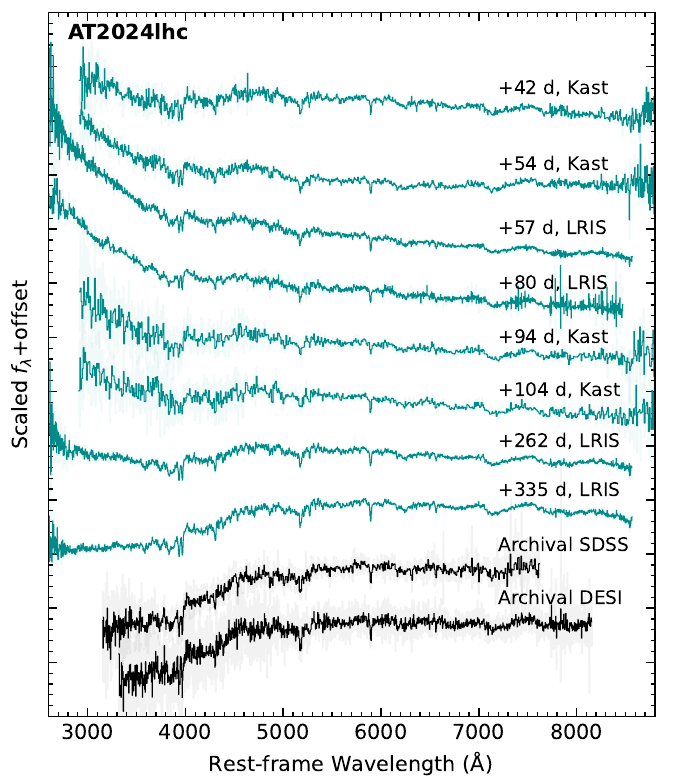}
    \caption{Optical spectroscopic observations of AT2024lhc. All epochs remain spectroscopically featureless (see Figure~\ref{fig:24lhc_opt_spec_sub}).\label{fig:opt_spec_24lhc}}
\end{figure}

\begin{figure}
    \centering
    \includegraphics[width=\columnwidth]{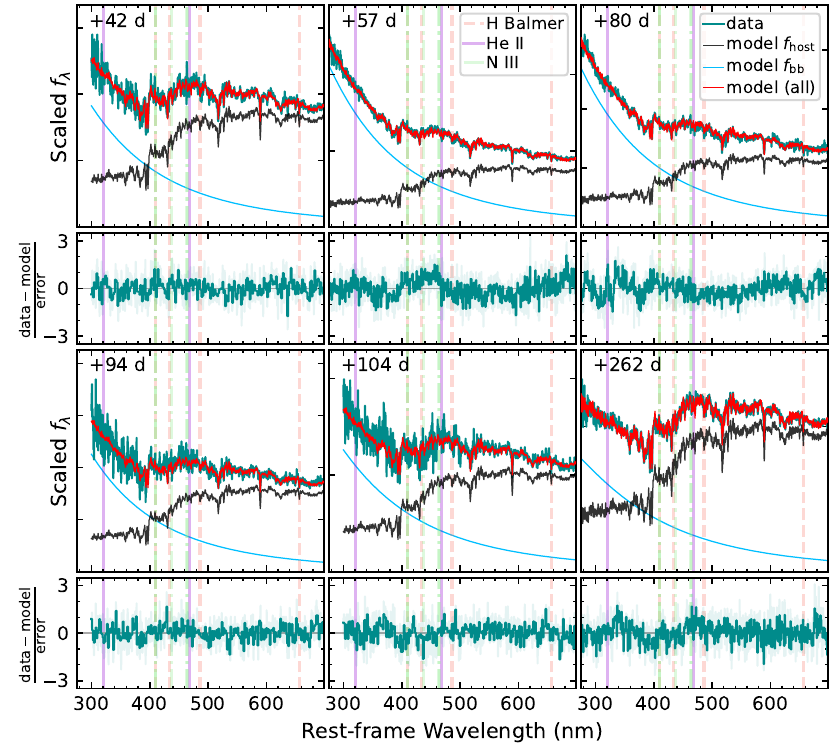}
    \caption{Host–transient decomposition for the optical spectra of AT2024lhc. Galactic extinction-corrected spectra (black) are modeled as the sum of host galaxy emission (gray) and blackbody continuum (blue), with the total model shown in red. Residuals show no significant broad features, confirming AT2024lhc as spectroscopically featureless from $t_{\rm rest}=42$\,d to 262\,d. \label{fig:24lhc_opt_spec_sub}}
\end{figure}

Figure~\ref{fig:opt_spec_24lhc} shows optical spectra of AT2024lhc. 
To search for transient spectral features, we model each spectrum as the sum of host galaxy and blackbody continuum contributions. The +335\,d LRIS spectrum serves as the host template: at this epoch, the transient had faded to $g= 21.2$\,mag and contributed negligibly to the observed flux.

Figure~\ref{fig:24lhc_opt_spec_sub} shows the host--blackbody decomposition. At all epochs, the combination of host and blackbody reproduces the observed spectra with no significant residual features. We therefore classify AT2024lhc as spectroscopically featureless throughout the observed evolution from $t_{\rm rest}= +42$\,d to $+262$\,d, with no detection of the broad H, He, or Bowen emission lines characteristic of other optically selected TDEs.

\subsection{Swift/UVOT}\label{subsec:uvot}

Both AT2024lhc and AT2024kmq were observed by the Ultra-Violet/Optical Telescope (UVOT; \citealt{Roming2005}) on board Swift. We obtain the host-subtracted UVOT fluxes using the same procedures as outlined in \citet{Ho2025}. Due to the lower S/N at late times, some of the adjacent observations are stacked using the same procedures as in \citet{Yao2024_22cmc}.

\subsection{X-ray Observations} \label{subsec:xray}

We obtained X-ray observations with the X-Ray Telescope (XRT; \citealt{Burrows2005}) on board Swift, the Neutron Star Interior Composition Explorer (NICER; \citealt{Gendreau2016}), the Nuclear Spectroscopic Telescope ARray (NuSTAR; \citealt{Harrison2013}), the Chandra X-ray Observatory (CXO), and the XMM-Newton telescope. 

The Swift/XRT, NICER, and NuSTAR data presented here are processed using \texttt{HEASoft} version 6.34. 
The XMM-Newton data is reduced with the \textit{XMM-Newton} Science Analysis System \citep[SAS;][]{Gabriel_04}.
The Chandra data is reduced using the Chandra Interactive Analysis of Observations (\texttt{CIAO}; \citealt{Fruscione2006}) software package (v4.17).

\subsubsection{Swift/XRT} \label{subsubsec:xrt}

We processed Swift/XRT data following standard procedures outlined in \citet{Ho2025}. To improve the SNR at late times, we stacked consecutive observations. XRT spectra were binned using the optimal scheme of \citet{Kaastra2016} (minimum one count per bin via \texttt{ftgrouppha}) and modeled with $W$-statistics \citep{Cash1979} using an absorbed power law (\texttt{tbabs*zashift*powerlaw} in \texttt{xspec}), where $N_{\rm H}$ is fixed at the Galactic value given in Table~\ref{tab:basic_info}. This simple model provides adequate fits to all spectra. From each best-fit model, we derived the 0.3--10 keV count-rate-to-flux conversion factor. Figures~\ref{fig:lc_overview_24lhc}--\ref{fig:lc_overview_24kmq} (middle and bottom panels) shows the evolution of the observed-frame 0.3--10 keV flux ($f_{\rm X}$) and power-law photon index $\Gamma$.

\subsubsection{NICER} \label{subsubsec:nicer}
AT2024lhc was observed by the NICER X-ray Timing Instrument in July 2024 (obsID 7204630101--7204630108) for a total of 13.4~ks (PI: Yao).
AT2024kmq was observed by NICER in October--November 2024 (obsID 7204930101--7641010107) and February--March 2025 (obsID 7641010107--7641010119, 7716010101--7716010129; PIs: Guolo, Pasham).
For AT2024kmq, an offset of $\sim 0.7\arcmin$ was applied to NICER pointing to minimize contamination from a nearby X-ray source.

As both sources are faint and hard, we followed the time-resolved spectroscopy approach for reliable estimation of faint-source light curves outlined in Section 2.1 of \cite{Chakraborty2024}, which we summarize here. We divided the data into continuous good-time intervals (GTIs) determined by the \texttt{nimaketime} routine. We then fit the spectrum of each GTI with the \texttt{SCORPEON}\footnote{\href{https://heasarc.gsfc.nasa.gov/docs/nicer/analysis_threads/scorpeon-overview/}{https://heasarc.gsfc.nasa.gov/docs/nicer/\\analysis\_threads/scorpeon-overview}} model over a broadband energy range (0.25--10 keV) for data taken in orbit night, and a slightly restricted range (0.38--10 keV) during orbit day. \texttt{SCORPEON} is a semi-empirical, physically motivated background model which explicitly includes components for the cosmic X-ray background, fluorescence emission from the solar wind charge exchange, and non-X-ray noise events (e.g., precipitating electrons and cosmic rays), allowing these to be fit alongside the source for joint uncertainty estimation. 
We grouped our spectra with the optimal binning scheme of \cite{Kaastra2016}, i.e., \texttt{grouptype=optmin} with \texttt{groupscale=10} in the \texttt{ftgrouppha} command, and performed all spectral fitting with the Cash statistic. 
We used a $\Delta C$-stat threshold of 25 to claim a source detection, and retained only GTIs with a best-fitting $C$-stat/d.o.f $\leq 1.5$. The evolution of $f_{\rm X}$ and $\Gamma$ are shown in Figures~\ref{fig:lc_overview_24lhc}--\ref{fig:lc_overview_24kmq}.

\subsubsection{NuSTAR}

We obtained NuSTAR observations for both events under pre-approved Target of Opportunity (ToO) programs, summarized in Table~\ref{tab:nustar}. 

To generate the source NuSTAR spectra for the two photon counting detector modules (FPMA and FPMB), source photons were extracted from a circular region with a radius of $r_{\rm src}=35\arcsec$ centered on the optical position of the source. The background was extracted from a $r_{\rm bkg}=80\arcsec$ region located on the same detector. 
We selected energy ranges where the net count rate is above the background for spectral fitting. 
The best-fit absorbed power-law models, shown in Figure~\ref{fig:nustar}, provide adequate description to the data. 
The best-fit values of $\Gamma$ and $f_{\rm X}$ are given in Table~\ref{tab:nustar}, and plotted in Figures~\ref{fig:lc_overview_24lhc}--\ref{fig:lc_overview_24kmq}.

\subsubsection{XMM-Newton}

AT2024lhc was observed with \textit{XMM-Newton} at two epochs. We processed the observation data files (ODFs) following the procedures outlined in \citet{Guolo2024}. Our analysis focuses on EPIC-pn data, which offers a higher SNR than the MOS detectors.

We group the spectrum using the \citet{Kaastra2016} scheme and simultaneously ensure at least 10 counts per bin. Using $W$-statistics, we fit in the energy range where the source dominates, which is 0.3--2.9\,keV for the first epoch and 0.3--2.5\,keV for the second epoch. The fitting result is shown in Table~\ref{tab:xmm_cxo} and Figure~\ref{fig:lc_overview_24lhc}.

\subsubsection{Chandra} \label{subsubsec:cxo}

For both events, we obtain one epoch of observation with the CXO. We used the Advanced CCD Imaging Spectrometer (ACIS; \citealt{Garmire2003}), with the aim point on the back illuminated S3 chip. Both objects are clearly detected. 
We extract the source spectrum using a source region of $r_{\rm src}=2.0^{\prime\prime}$ centered on the apparent X-ray position of each object. 
A total of 65 and 193 (0.5--8\,keV) counts are detected within the source regions of AT2024kmq and AT2024lhc, respectively. 
The background spectra are extracted using nearby source-free regions. 
We group the Chandra spectra to at least one count per bin, and modeled the 0.3--8\,keV data using $W$-statistics. Using \texttt{tbabs*zashift*powerlaw}, we obtained good fits with best-fit parameters shown in Table~\ref{tab:xmm_cxo} and Figures~\ref{fig:lc_overview_24lhc}--\ref{fig:lc_overview_24kmq}.


\subsubsection{X-ray Variability Timescale} \label{subsubsec:t_X_var}

\begin{table*}
\begin{center}
    \caption{Rest-frame X-ray variability timescales. \label{tab:t_x_var}}
    \begin{tabular}{ccccccc} 
    \hline
	Name & Instrument & Phase $t_{\rm rest}$ (d) & $t_{\rm X, var}$ (d) & Significance ($\sigma$) & Flux ratio\\
    \hline
      & & 124.0 & 0.108 & 3.7 & 2.1\\
     AT2024kmq & NICER & 124.4 & 0.053 & 4.0 & 3.2 \\
     & & 134.2 & 1.840 & 4.8 & 2.4\\
     \hline
     AT2024lhc &XRT & 57.4 & 0.199 & 3.1 & 2.6\\
    \hline
    \end{tabular}
\end{center}
\end{table*}

Both AT2024kmq and AT2024lhc exhibit significant X-ray variability on short timescales. To quantify this, we systematically test all observation pairs separated by less than two days (observer frame) for statistically significant flux changes. For a pair to be classified as variable, we require both (1) a significance level exceeding $3\sigma$ and (2) a flux ratio exceeding a factor of two.

For NICER observations (flux-calibrated measurements from \S\ref{subsubsec:nicer}), we compute the significance of flux variations as $\sigma = |F_2 - F_1| / \sqrt{\sigma_{F_1}^2 + \sigma_{F_2}^2}$, where $F_1$ and $F_2$ are the fluxes in two consecutive observations with uncertainties $\sigma_{F_1}$ and $\sigma_{F_2}$. For Swift XRT observations (net count rate measurements from \S\ref{subsubsec:xrt}), we similarly compute the significance of net count rate variations. The flux ratio is defined as ${\rm max}(F_2/F_1, F1 / F2)$ (or equivalently for count rates).

NICER detected rapid variability in AT2024kmq, with three significant events at rest-frame phases $t_{\rm rest} \sim 124$--134\,d (Table~\ref{tab:t_x_var}). The minimum detected variability timescale is $\Delta t_{\rm rest} = 0.053\,{\rm d}\approx 1.3$ hr, with a flux ratio of 3.2 and significance of $4.0\sigma$. 
Swift/XRT detected one significant variation in AT2024lhc at $t_{\rm rest} = 57.4$ d, with $\Delta t_{\rm rest} = 0.199\,{\rm d}\approx 4.8$\,hr (flux ratio 2.6, $3.1\sigma$).

\subsection{Radio: VLA} \label{subsec:radio}

\begin{figure}
    \centering 
    \includegraphics[width=0.95\columnwidth]
    {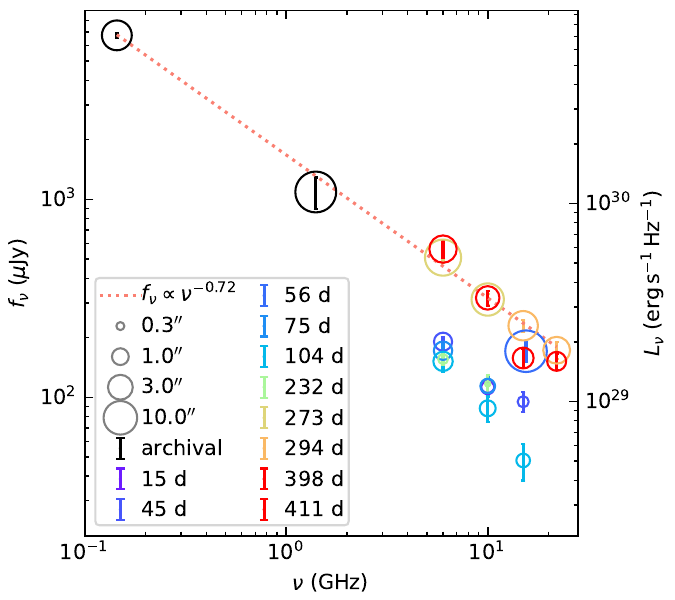}
    \caption{The SED of a radio source at the nucleus of AT2024kmq's host galaxy, color-coded by $t_{\rm obs}$ (shown in the legend). Data at $t_{\rm obs}>200$\,d are presented in this paper (the integrated flux in Table~\ref{tab:vla_24kmc}), and is compared to earlier data presented in \citet{Ho2025}. Symbol size reflects the synthesized beam size (full width at half-power) at each frequency. Observations with larger beams resolve extended emission from the pre-existing radio source, yielding higher integrated flux densities than measurements at higher resolution. }
    \label{fig:24kmq_radio_flux}
\end{figure}

\begin{figure}
    \centering
    \includegraphics[width=0.95\columnwidth]{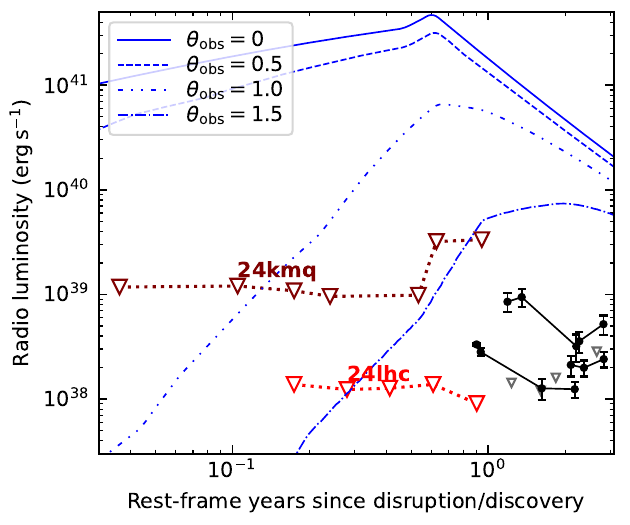}
    \caption{Radio upper limits of AT2024kmq, AT2024lhc, compared with the sample of optically overluminous TDEs presented in \citet{Yao2025_radio}. The blue lines show model radio light curves for a jet with the same intrinsic properties as the best-fit model of the jetted TDE Swift J1644+57 \citep{Beniamini2023}, viewed at different angles. Assuming all other parameters are identical, even the most off-axis jet models are ruled out for both AT2024kmq and AT2024lhc. \label{fig:radio_lc} }
\end{figure}

We obtained radio observations using the Very Large Array (VLA; \citealt{Perley2011}) under Programs 24A-487, 24A-494, and 25A-152 (PI: Y.~Yao). The data were analyzed following the standard radio continuum image analysis procedures in the Common Astronomy Software Applications (\texttt{CASA}; \citealt{CASATeam2022}). 
We used \texttt{tclean} to produce radio images. 

For AT2024lhc, we measured the flux density as the maximum pixel value within a region with the size of the synthesized beam, centered on the optical coordinate of the TDE. The uncertainty was estimated as the root-mean-square (rms) of the pixel values in a nearby source-free region of the image. 
AT2024lhc was not detected throughout the observations, and 3$\sigma$ upper limits are reported in Table~\ref{tab:vla_24lhc}. 

For AT2024kmq, \citet{Ho2025} reported that the host is a radio-bright AGN with extended radio emission. Therefore, we used the interactive 2D fit tool in CASA viewer, which fits Gaussians to two dimensional intensity distributions. The measured peak and integrated fluxes are reported in Table~\ref{tab:vla_24kmc}. The integrated flux can be directly compared with earlier measurements reported in \citet{Ho2025}. Figure~\ref{fig:24kmq_radio_flux} shows the radio flux measurements from both \citet{Ho2025} and this work. Fitting observations obtained using large beam sizes ($>3\arcsec$), we obtained $f_\nu \propto \nu^{-0.72}$, which is typical for radio AGN \citep{Condon2002}. Due to the presence of the AGN, we are not able to set strong constraints on the radio emission of the TDE. Following \citet{Ho2025}, we assume that the transient flux is lower than the measured flux densities.

\subsubsection{Constraints on Transients: No J1644-like Jets} 
Figure~\ref{fig:radio_lc} presents the radio upper limits of AT2024kmq and AT2024lhc with other optically overluminous TDEs selected by ZTF \citep{Yao2025_radio}. These limits rule out off-axis jets with properties that are the same as those observed in the jetted TDE Swift\,J1644+57 (see figure caption for details).

\subsubsection{Constraints on Host Galaxies} \label{subsubsec:jet_agn}

As discussed in \citet{Ho2025}, the pre-TDE radio spectral index and luminosity of AT2024kmq are consistent with jet-mode AGN. Its 1.4\,GHz luminosity ($L_{\rm 1.4\,GHz}= 4.6\times 10^{23}\,{\rm W\,Hz^{-1}}$) indicates a jet mechanical power of $P_{\rm mech} = 3.5\times 10^{36}\,{\rm W}$, using the $P_{\rm mech}$--$L_{\rm 1.4\,GHz}$ relation by \citet[][Equation 2]{Heckman2014}. The Eddington-scaled accretion rate is therefore $\dot m \approx L_{\rm mech}/L_{\rm Edd}\sim 0.003$. At such an accretion rate, the inner disk radius is likely $r_{\rm tr}\sim {\rm few}\times 10 r_{\rm g}\sim 10^{15}\,{\rm cm}$.

\begin{figure*}
\includegraphics[width=0.45\textwidth]{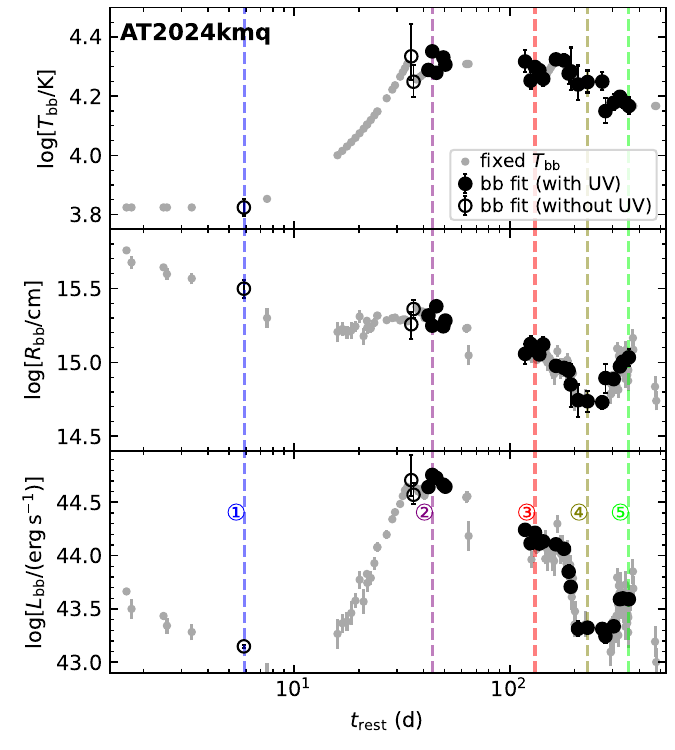}
\includegraphics[width=0.45\textwidth]{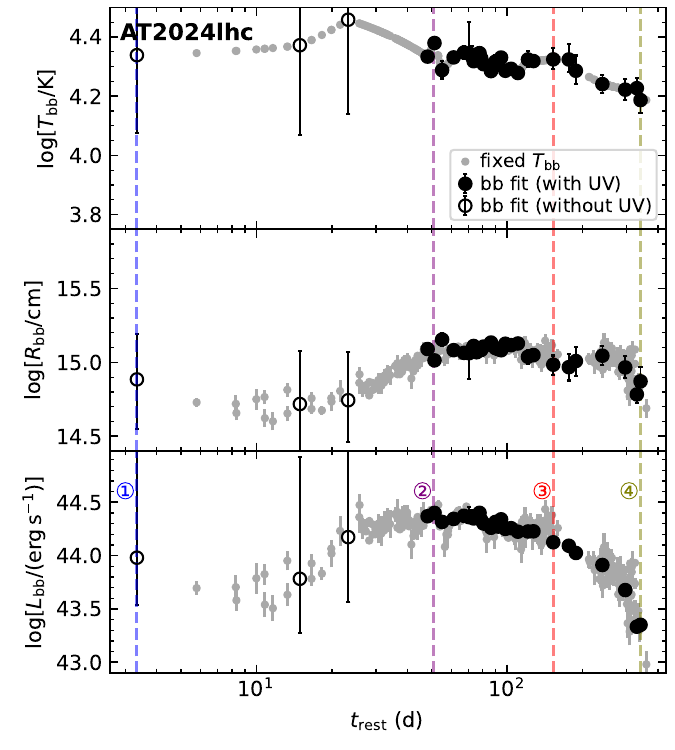}
\caption{Evolution of the UV/optical blackbody properties of AT2024lhc and AT2024kmq. Epochs where both $T_{\rm bb}$ and $R_{\rm bb}$ are fitted are shown in black, where solid and hollow markers indicate epochs with and without UVOT observations, respectively. Epochs scaled assuming fixed $T_{\rm bb}$ are shown in gray. The vertical dashed lines mark epochs where detailed broadband SEDs are presented in Figure~\ref{fig:broad_sed}. \label{fig:bb_pars}}
\end{figure*}

\subsection{Blackbody Modeling} \label{subsec:bbfit}

Following previous TDE studies, we model the UV and optical photometry using a Planck function, where the SED is characterized by an effective temperature $T_{\rm bb}$ and a photospheric radius $R_{\rm bb}$. 
We assume negligible host extinction based on two facts: (1) host galaxy population synthesis modeling yields very small reddening values (see Table~\ref{tab:basic_info}), and (2) the X-ray spectra can be well fit by a simple power-law absorbed by Galactic ISM, with no need to add absorption from extragalactic ISM (see \S\ref{subsec:xray}). 

At epochs where both UV and optical photometry are available, we perform the fit using Monte Carlo Markov Chain (MCMC) simulations with \texttt{emcee} \citep{Foreman-Mackey2013}. Both detections and non-detections are incorporated in the fitting following the procedures outlined in \citet{Laskar2014} and \citet{Eftekhari2024}. 
We allow both $T_{\rm bb}$ and $R_{\rm bb}$ to vary freely. To account for systematic uncertainties in the photometry and potential deviations of the true SED from a pure blackbody, we use a Gaussian likelihood function that includes an excess variance term $\sigma_0^2$ that is constant across all bands.
We set the priors to be uniformly distributed in the following ranges: log($T_{\rm bb}$/K) in [3.2, 5.1], log($R_{\rm bb}$/cm) in [12, 16.5], and log$\sigma_0$ in [$-6$, $-0.5$].
Within the ensemble, we use 100 walkers, each of which is run until convergence, where the convergence test follows the steps outlined in \citet{Yao2019}. 

At epochs with only optical data, we fix $T_{\rm bb}$ to the linearly interpolated value derived from epochs with UV coverage. We estimate $R_{\rm bb}$ directly from the radio of $\nu L_\nu$ and the color correction for blackbody spectrum. 

The results are shown in Figure~\ref{fig:bb_pars}. AT2024kmq and AT2024lhc exhibit several notable similarities:
\begin{enumerate}
    \item Both events show precursor emission lasting $\sim 8$\,d above survey sensitivity and peaking at $10^{43.5}$--$10^{44}\,{\rm erg\,s^{-1}}$, though with different thermal properties: AT2024lhc has $T_{\rm bb}\sim 2\times 10^4\,{\rm K}$ and $R_{\rm bb}\sim 10^{15}\,{\rm cm}$, while AT2024kmq exhibits a lower temperature ($T_{\rm bb}\sim 6600\,{\rm K}$) but larger radius ($R_{\rm bb}\sim 3\times 10^{15}\,{\rm cm}$). 
    \item During the main peak phase, which lasts 180\,d (AT2024kmq) and 290\,d (AT2024lhc) in the rest frame, both events maintain relatively constant temperatures of $T_{\rm bb}\sim 2\times 10^4\,{\rm K}$. The maximum radius is $R_{\rm bb}\sim 10^{15} \, {\rm cm}\sim 70r_{\rm g} M_{\rm BH, 8}^{-1}$ (see Eq.~\ref{eq:rg}). Their peak luminosities ($L_{\rm bb}\sim 10^{44.5}\,{\rm erg\,s^{-1}}$) are at the high-end of the TDE population \citep{Yao2023}. Each event shows a sharp drop at the end of this phase driven by the decreasing $R_{\rm bb}$. 
    \item Additionally, beginning at $t_{\rm rest}\approx 310$\,d, AT2024kmq exhibits a rebrightening phase, which appears to have ended by $t_{\rm rest}\approx 480$\,d. 
\end{enumerate}

We discuss the possible power sources of the UV/optical emission during and after the precursor in \S\ref{subsec:precursor} and \S\ref{subsec:opt_peak}, respectively. 

\subsection{HST UV Spectroscopy} \label{subsec:hst}

\begin{figure}
    \centering
    \includegraphics[width=\columnwidth]{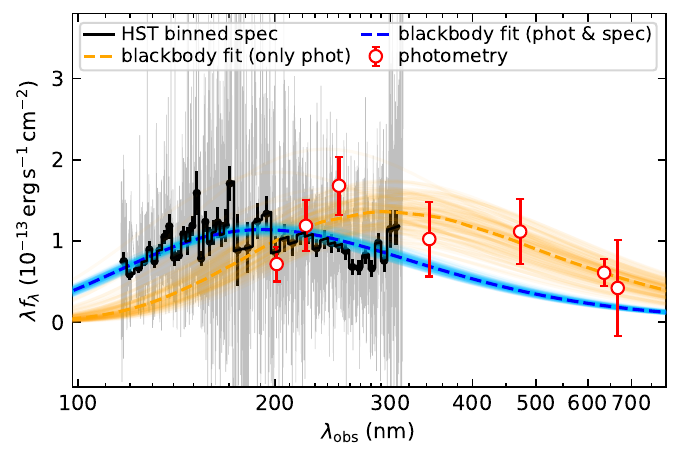}
    \caption{HST UV spectrum of AT2024lhc. The unbinned and binned data are shown in thick black and thin gray lines, respectively. Regions aﬀected by strong geocoronal absorption or emission have been masked. 
    The blue lines show the blackbody model fitted to both spectroscopy and photometry, while the orange lines show the model fitted to photometry only.
    \label{fig:24lhc_hst_spec} }
\end{figure}

We obtained 8 orbits of ultraviolet spectroscopy of AT2024lhc with the Space Telescope Imaging Spectrograph (STIS; \citealt{Woodgate1998}) MAMA detectors onboard the Hubble Space Telescope (GO-17767; PI: Chornock) on 2025 June 05 and 06 ($t_{\rm rest}\approx338$\,d). The original observation plan included 4 orbits of far-UV (FUV) spectroscopy with the G140L grism and 4 orbits of near-UV (NUV) spectroscopy with the G230L grism spread over three visits. However, two of the visits suffered guide star failures, and usable data were obtained for 3 orbits of G140L observations and 1 orbit of G230L, for total exposure times of 7199~s and 1919~s in the FUV and NUV, respectively. In addition, between triggering and acquisition of the observations, AT2024lhc faded significantly 
(Figure~\ref{fig:lc_overview_24lhc}). The resultant low signal-to-noise ratio (SNR) of the observations caused the default pipeline reduction to misidentify the object trace in two exposures. We manually re-ran the \texttt{x1d} pipeline extraction with the \texttt{A2CENTER} parameter set to be near the actual location of the source. The individual object exposures were combined using the Hubble Advanced Spectral Products (HASP; \citealt{Debes2024}) script using default parameters. 

The final HST spectrum is shown in Figure~\ref{fig:24lhc_hst_spec}. 
The gray line shows the raw data, and the black line shows the binned spectrum. 
We binned the spectra by a factor of 40 pixels, where each binned flux value is the mean of 40 consecutive pixels, and the flux uncertainty is estimated as the standard error of the mean within each bin. 
The red data shows the UV and optical photometry obtained at $t_{\rm rest}=340.3$\,d, which is very close in time to the HST observation. We see that in the NUV band where both photometry and spectrum are available, the data are consistent with each other at the $<2\sigma$ level.

To assess whether the blackbody function employed in \S\ref{subsec:bbfit} also provides a good description of the FUV data, in Figure~\ref{fig:24lhc_hst_spec}, we show the best-fit blackbody model obtained in \S\ref{subsec:bbfit} using the dashed thick orange line, and 100 random draws from the MCMC posterior using the solid thin orange lines (see \S\ref{subsec:bbfit}). 
As can be seen, when fitting only to photometry, the blackbody model, with ${\rm log}(T_{\rm bb}/{\rm K}) = 4.18_{-0.04}^{+0.05}$ and ${\rm log}(R_{\rm bb}/{\rm cm}) = 14.99_{-0.10}^{+0.09}$, systematically underestimates the FUV flux at $\lambda_{\rm obs} < 1700$\,\AA.
When we instead fit the blackbody model to both the HST spectroscopy and photometry simultaneously, we obtain $\log(T_{\rm bb}/{\rm K}) = 4.36\pm0.01$ and $\log(R_{\rm bb}/{\rm cm}) = 14.48_{-0.02}^{+0.03}$, shown as blue lines in Figure~\ref{fig:24lhc_hst_spec}. We note that because the photometric uncertainties are large, this joint fit is effectively dominated by the HST spectrum. This fit slightly underestimates the optical photometry, though the discrepancy remains within the photometric uncertainties.

Our HST observations therefore suggest that the blackbody parameters derived in \S\ref{subsec:bbfit} should be taken with caution, and that a single blackbody might not provide an adequate description of the underlying intrinsic FUV--optical SED shape.

\begin{figure*}
    \includegraphics[width=0.45\textwidth]{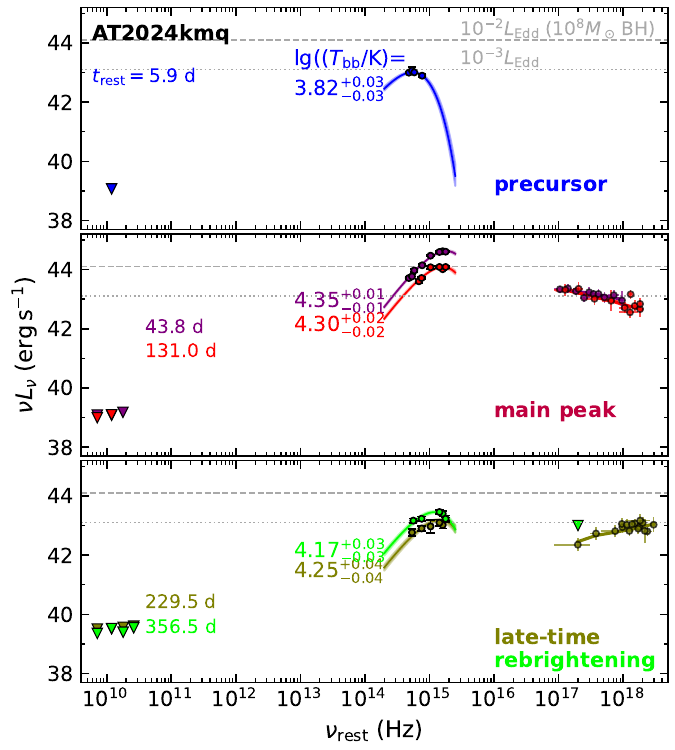}
    \includegraphics[width=0.45\textwidth]{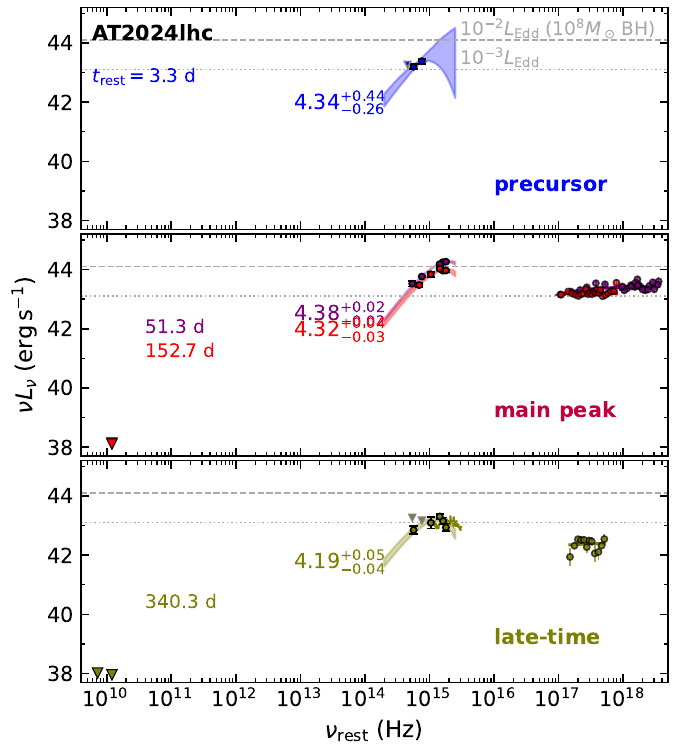}
    \caption{Broadband SEDs of AT2024kmq and AT2024lhc at representative epochs. 
    In each panel, the horizontal dashed and dotted lines show $10^{-2}L_{\rm Edd}$ and $10^{-3}L_{\rm Edd}$ for $10^8\,M_\odot$ BHs, where $L_{\rm Edd} = 1.26\times 10^{38}(M_{\rm BH}/M_\odot)\,{\rm erg\,s^{-1}}$. \label{fig:broad_sed}}
\end{figure*}

\subsection{Broadband SED} \label{subsec:sed}

We construct broadband SEDs from the radio to hard X-ray bands at representative epochs in the light curve evolution (marked by vertical dashed lines in Figure~\ref{fig:bb_pars}). 
Figure~\ref{fig:broad_sed} shows the SEDs at each epoch, comprising: available radio upper limits, UV/optical photometry with the 68\% confidence interval of the blackbody fits, and X-ray spectra or upper limits. 

We compute the bolometric luminosity following \citet{Yao2022}. The rest-frame spectrum is divided into three frequency ranges: (1) Between $\nu_{\rm rest}=10^{14}$\,Hz and $10^{15.5}$\,Hz, we integrate beneath the blackbody model fitted to UV/optical photometry (\S\ref{subsec:bbfit}). (2) Between $\nu_{\rm rest}=10^{17}$\,Hz and $10^{18.5}$\,Hz, we integrate  beneath the best-fit X-ray spectral model (\S\ref{subsec:xray}). (3) Between $\nu_{\rm rest}=10^{15.5}$\,Hz and $10^{17}$\,Hz (the EUV gap), we assume the spectrum is continuous and approximate it with a power law connecting the UV and X-ray components. When X-ray observations are unavailable or yield only upper limits, we exclude the X-ray and EUV components, treating the UV/optical integral as a lower limit on $L_{\rm bol}$. 

Table~\ref{tab:Lbol} presents the computed bolometric luminosities and the corresponding Eddington ratios $\lambda_{\rm Edd}\equiv L_{\rm bol} / L_{\rm Edd}$. 

\begin{table}
    \caption{Bolometric luminosity at representative phases. \label{tab:Lbol}}
\begin{threeparttable}
    \begin{tabular}{ccccc} 
    \hline
    \hline
	Name & $t_{\rm rest}$ (d) & $L_{\rm bol}\,({\rm erg\,s^{-1}})$&  $\lambda_{\rm Edd}^{a}$ & $\lambda_{\rm Edd}^{b}$\\
    \hline
    \hline
    \multirow{5}{*}{2024kmq} &5.9 & $>1.5\times 10^{43}$ & $>6.0\times 10^{-4}$ & $>2.6\times 10^{-3}$\\
                            & 43.8 & $3.3\times 10^{45}$ & $0.13$ & $0.57$\\
                            & 131.0 & $3.7\times 10^{44}$ &  $0.015$ & $0.064$ \\
                            & 229.5 & $5.2\times 10^{43}$ &  $2.1\times 10^{-3}$ & $9.0\times 10^{-3}$ \\
                            & 356.5 & $>3.9\times 10^{43}$  & $>1.6\times 10^{-3}$ &  $>6.8\times 10^{-3}$ \\
    \hline
    \multirow{4}{*}{2024lhc} & 3.3 & $>3.3\times 10^{43}$ & $>4.6\times 10^{-4}$ & $>3.1\times 10^{-3}$  \\
                            & 51.3 & $5.2\times 10^{44}$ & $7.2\times 10^{-3}$   & $0.048$ \\
                            & 152.7 & $3.5\times 10^{44}$ & $4.8\times 10^{-3}$  & $0.033$ \\
                            & 340.3 & $4.8\times 10^{43}$ & $6.6\times 10^{-4}$  & $4.4\times 10^{-3}$ \\
    \hline
    \hline
    \end{tabular}
\begin{tablenotes}
    \item $^{a}$ Eddington ratio adopting $M_{\rm BH}$ estimated from $\sigma_\ast$, see Table~\ref{tab:basic_info}.
    \item $^{b}$ Eddington ratio adopting $M_{\rm BH}$ estimated from \texttt{tidalspin}, see Table~\ref{tab:basic_info}. 
\end{tablenotes}
\end{threeparttable}
\end{table}

\section{Discussion}
\label{sec:discuss}

\begin{figure*}
    \centering
    \includegraphics[width=0.9\textwidth]{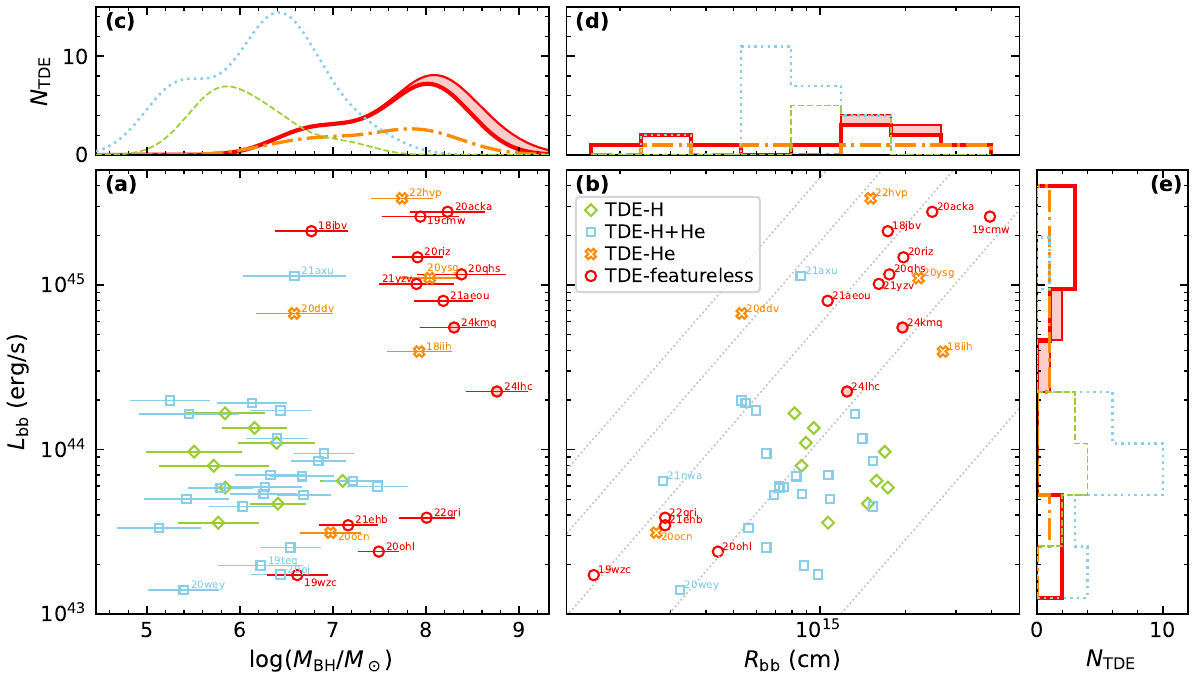}
    \caption{AT2024kmq and AT2024lhc compared with other TDEs, color-coded by spectral subclasses.
    Panel (a): TDEs in the phase space of peak $L_{\rm bb}$ and black hole mass, where $M_{\rm BH}$ is directly estimated using host-galaxy scaling relations. 
    Panel (b): TDEs in the phase space of peak $L_{\rm bb}$ and $R_{\rm bb}$ at peak luminosity. 
    Dotted lines show constant temperatures of $T_{\rm bb}=[1,2,3,4,5]\times 10^4$\,K. 
    Panel (c): Distribution of $M_{\rm BH}$ for TDEs of four spectral subclasses, estimated using kernel density estimation with bandwidth equal to the uncertainty in $M_{\rm BH}$. 
    Panels (d) and (e): Histograms of $L_{\rm bb}$ and $R_{\rm bb}$ at maximum light for the four spectral subclasses. 
    For the TDE-featureless class (panels c--e), thin red lines show the results including AT2024kmq and AT2024lhc, while thick red lines show the distribution of the comparison sample only. The red shaded regions indicate the contribution of these two events. 
    \label{fig:uvo_sptype_summary1} }
\end{figure*}

Here we discuss AT2024kmq and AT2024lhc in the context of spectroscopically featureless TDEs. In \S\ref{subsec:fbnt_class}, we examine the optical properties and black hole masses of featureless TDEs in comparison to broad-line TDEs. We then discuss the origin of their X-ray emission in \S\ref{subsec:origin_Xrays}, the optical precursor emission in \S\ref{subsec:precursor}, and the post-precursor UV/optical emission in \S\ref{subsec:opt_peak}.

\subsection{Featureless blue TDEs: two sub-classes?} \label{subsec:fbnt_class}

To place AT2024kmq and AT2024lhc in the context of known spectroscopically featureless TDEs, we compile a sample of ZTF TDEs. We start with 46 TDEs with known spectroscopic subtypes from ZTF sample papers \citep{Hammerstein2023, Yao2023, Yao2025_radio}, excluding AT2021qth/ZTF21abhrchb (the only TDE-coronal source) and AT2021gje/ZTF21aapvvtb (an event with peculiar spectral features). We also include two events classified as TDE-featureless (AT2022gri/ZTF20aahmtso, AT2019wzc/ZTF19acnskyy/Ansky) by \citet{Zhu2025}, and AT2020ohl/ASASSN-20hx/ZTF20abjpxei, previously classified as an ambiguous nuclear transient (ANT) by \citet{Hinkle2022}. We reclassify AT2020ohl as TDE-featureless because, unlike other ANTs (see examples in \citealt{Frederick2021, Wiseman2025}), it lacks broad Balmer lines and exhibits a featureless blue continuum throughout its evolution, similar to AT2022gri and AT2019wzc. It is also the only source among 19 ANTs examined by \citet{Hinkle2024} that lacks a dust reprocessing echo.

\begin{table}
    \caption{Bootstrap bimodality test using the comparison sample.  \label{tab:bimodality_test} }
\begin{threeparttable}
    \begin{tabular}{cccc}
    \hline
        Spectral Subtype & $N_{\rm TDE}^{a}$ &log[$L_{\rm bb}/({\rm erg\,s^{-1}})$]$^{b}$ & log[$R_{\rm bb}/\rm cm$]$^{b}$\\
    \hline
        TDE-H+He & 24 & 0.0\% & 0.2\% \\
        TDE-H & 9 & 2.4\% & 52.6\% \\
        TDE-He & 5 & ---$^{c}$ & 43.0\% \\
        TDE-featureless & 11&  100.0\% & 86.8\% \\
    \hline
    \end{tabular}
\begin{tablenotes}
    \item $^{a}$ Number of TDEs in each spectral type.
    \item $^{b}$ The fraction of 1000 bootstrap realizations that satisfy both conservative bimodality criteria (silhouette score $>0.7$ and Cohen's $d>3.0$) for each spectral subtype and observable.
    \item $^{c}$ Cohen's $d$ cannot be reliably computed in the majority of bootstrap realizations because one of the two clusters contains fewer than two events.
\end{tablenotes}
\end{threeparttable}
\end{table}

This gives us a comparison sample of 49 TDEs. The number of events in each spectral subtype is given in Table~\ref{tab:bimodality_test}. We characterize their UV and optical light curves following the same procedures as outlined in \citet{Yao2023}. The light curve properties we measure are UV/optical blackbody luminosity and radius ($L_{\rm bb}$ and $R_{\rm bb}$) at light curve maximum. Of the 49 TDEs in our sample, 29 have published velocity dispersion measurements \citep{Ahumada2020, Hammerstein2023_KCWI, Yao2023, Yao2025_radio}, to which we add AT2022gri with $\sigma_\ast = 155\pm4\,{\rm km\,s^{-1}}$ derived from our analysis of its DESI spectrum. For these 30 objects, we apply the \citet{Kormendy2013} $M_{\rm BH}$--$\sigma_\ast$ relation. For the remaining 19 TDEs without velocity dispersion measurements, we use the $M_{\rm BH}$--$M_{\rm gal}$ scaling relation (Eq.~5 of \citealt{Yao2023}).

Figure~\ref{fig:uvo_sptype_summary1} presents the comparison sample across $M_{\rm BH}$, $L_{\rm bb}$, and $R_{\rm bb}$, as well as the distributions. Because individual black hole mass estimates for TDE hosts carry large uncertainties, we represent the $M_{\rm BH}$ distributions using kernel density estimation with bandwidths set by the measurement uncertainties. In contrast, the uncertainties on $L_{\rm bb}$ and $R_{\rm bb}$ at peak are relatively small, so simple histograms provide an adequate representation of their distributions. Similar to what was reported in \citet{Hammerstein2023}, we see that the TDE-featureless and TDE-He subtypes are generally hosted by higher-mass black holes, and they also exhibit higher peak blackbody temperatures compared to the TDE-H and TDE-H+He subtypes. 

Before discussing the distributions of the blackbody parameters in detail, we first caution that a single blackbody may not provide an adequate description of the TDE UV/optical continuum SED shape. As shown in \S\ref{subsec:hst} and in multiple previous studies \citep{Cenko2016, vanVelzen2020, Hung2021, Zhu2025}, a single-temperature blackbody fit to NUV and optical photometry often underestimates the FUV continuum. Nevertheless, the blackbody function has the advantage of simplicity and provides estimates of the characteristic size and temperature of the UV/optical emitting region, albeit with systematic uncertainties.

With this caveat in mind, we note that the TDE-featureless class appears bimodal in peak $L_{\rm bb}$ (panel e): one population has  $2\times 10^{44}\lesssim[L_{\rm bb}/({\rm erg\,s^{-1}})]\lesssim 3\times 10^{45}$, and another has $10^{43}\lesssim[L_{\rm bb}/({\rm erg\,s^{-1}})]\lesssim 4\times 10^{43}$. This bimodality is also evident in ${\rm log}[R_{\rm bb}/({\rm cm})]$ (panel d).

To assess bimodality robustness, we perform a conservative bootstrap analysis using $K$-means clustering. For each spectral subtype, we perform 1000 bootstrap iterations where each data point is perturbed by Gaussian noise ($\sigma=0.05$\,dex for log$T_{\rm bb}$ and $\sigma=0.02$\,dex for log$R_{\rm bb}$)\footnote{These are the medians of measurement uncertainties (see, e.g., Tab.~6 of \citealt{Hammerstein2023}). We have verified that choosing slightly different values of $\sigma$ does not significantly change the results of our analysis.} to account for measurement uncertainties. 
In each iteration, we fit a two-component $K$-means model and compute: 
(1) the silhouette score\footnote{\url{https://en.wikipedia.org/wiki/Silhouette_(clustering)}}, which quantifies cluster quality by measuring how well-separated points are between clusters relative to their dispersion within clusters, and 
(2) the separation ratio (Cohen's $d$)\footnote{\url{https://en.wikipedia.org/wiki/Effect_size\#Cohen's_d}}, defined as the distance between cluster centers divided by the pooled within-cluster standard deviation. 
We impose stringent dual criteria for claiming bimodality: silhouette score $>0.7$ (indicating excellent cluster separation) and separation ratio $>3.0$ (indicating clusters separated by more than $3\sigma$). A distribution is classified as robustly bimodal only if $>80$\% of bootstrap samples simultaneously satisfy both criteria.

Table~\ref{tab:bimodality_test} presents the results using only the comparison sample of 49 TDEs. The TDE-featureless class exhibits robust bimodality in both log$L_{\rm bb}$ (100.0\% of bootstrap samples) and log$R_{\rm bb}$ (86.8\%), while all other spectral subtypes are consistent with unimodal distributions. Including AT2024kmq and AT2024lhc in the test leads to slightly different fractions: log$L_{\rm bb}$ (99.9\% of bootstrap samples) and log$R_{\rm bb}$ (93.5\%), and evidence for bimodality remains strong. We caution that given the small sample size and potential selection biases, this bimodal structure should be validated with larger, systematically selected samples in future work.

The bimodality seen among the TDE-featureless class may be closely connected to the physical conditions required to suppress line formation in luminous transients. In particular, the radiative transfer models of \citet{Aspegren2026} show that featureless blue spectra can arise in multiple regimes that lead to high ionization. The ionization state is set by the Saha equation, which depends on the temperature --- the hotter the medium, the more ionized it is --- so sustaining high temperatures is key to suppressing line formation. 
Within this framework, the two subclasses we identify may represent two ways that inhibit line formation: a high source luminosity regime (where AT2024kmq and AT2024lhc reside) and a compact photosphere regime (containing AT2019wzc, AT2020ohl, AT2021ehb, and AT2022gri). Consistent with this interpretation, panel (b) of Figure~\ref{fig:uvo_sptype_summary1} shows that the TDE-featureless and TDE-He subtypes generally exhibit higher peak blackbody temperatures compared to the TDE-H+He subtype, with the TDE-H subtype exhibiting the lowest temperatures.


Higher-mass black holes may also favor featureless spectra since stronger GR effects could boost outflow velocities, which help suppress intrinsically strong UV metal lines \citep{Aspegren2026}. Indeed, \citet{Zhu2025} presented HST UV spectra of three subluminous featureless TDEs (AT2019wzc, AT2021ehb, AT2022gri), demonstrating their lack of spectral features even in the UV. This suggests that high outflow velocities may also play a role in establishing featureless spectra. The HST UV spectrum of AT2024lhc presented here is not of high enough SNR to search for spectral lines --- future UV spectroscopy studies of luminous featureless TDEs will be useful to test this hypothesis.

\subsection{Origin of the TDE hard X-ray emission} \label{subsec:origin_Xrays}

A common behavior between AT2024kmq and AT2024lhc is their luminous ($L_{\rm X, peak}\sim 10^{44}\,{\rm erg\,s^{-1}}$) X-ray emission appearing at early times. Their peak X-ray luminosities exceed the pre-flare host galaxy levels (Table~\ref{tab:basic_info}) by more than two orders of magnitude, ruling out contamination from pre-existing AGN. The X-ray spectral hardness is inconsistent with thermal emission from an accretion disk. In both events, NuSTAR detects high-energy photons extending to $\sim$20 keV (Figure~\ref{fig:nustar}).

The minimum detected X-ray variability timescales are 1.3\,hr and 4.8\,hr for both AT2024kmq and AT2024lhc, respectively (see Table~\ref{tab:t_x_var}), providing upper limits on the intrinsic variability timescales. For a $10^8\,M_\odot$ black hole, this corresponds to approximately ten light-crossing times of the gravitational radius ($t_{\rm g}=r_{\rm g}/c\approx500$\,s). The rapid X-ray variability therefore constrains the X-ray emitting region to $\lesssim 10 r_{\rm g}$. This strongly suggests a compact emitting region. The absence of radio detections of the transients out to $t_{\rm rest}\sim1$ yr (\S\ref{subsec:radio}) disfavors a jet, though a highly off-axis configuration \citep{Sfaradi2024} cannot be excluded.

Power-law X-ray continua can arise from two primary mechanisms: thermal Comptonization, in which soft disk photons are inverse Compton scattered by thermal electrons in a hot, optically thin, compact corona \citep{st80}, or bulk Comptonization, in which photons are upscattered by electrons with non-uniform transrelativistic bulk motions \citep{Blandford1981}. In AGN and XRBs, thermal Comptonization dominates \citep{gm14, Kara2025}, though recent three-dimensional radiation magnetohydrodynamic simulations of AGN and TDE environments suggest bulk Comptonization may also contribute \citep{Jiang2025, Huang2026_disk_formation}.

The broad-band spectrum shape of both events during their main peaks are similar to those found in recent simulations due to bulk Comptonization \citep{Huang2026_disk_formation}. However, AT2024kmq exhibits a hard X-ray photon index at late times ($\Gamma \approx 1.6$; Figure~\ref{fig:lc_overview_24kmq}, Table~\ref{tab:nustar}), whereas the spectrum shape produced by bulk Comptonization is sensitive to local velocity field and rarely produce photon index of $\Gamma < 2$ \citep{Blandford1981}. We will later show that thermal Comptonization natually explains why hard X-rays are preferentially detected in high-$M_{\rm BH}$ TDEs (see below). Note that in the accretion disk plus corona geometry, one would expect reflection features (e.g., relativistically broadened iron line and Compton hump; \citealt{Fabian2016}) to be present. However, the faintness of these TDEs precludes detection of such features in our NuSTAR observations.

Previous X-ray studies of non-jetted TDEs have established that their emission is typically soft and consistent with thermal disk emission \citep{Auchettl2017, sazonov21, Guolo2024}. In some cases (e.g., AT2018fyk, AT2020ocn, and AT2021ehb), a spectral state transition of soft $\rightarrow$ hard (or soft $\rightarrow$ intermediate) have been observed at a few hundred days after the optical peak, which has been interpreted as coming from inverse Compton scattering of thermal disk photons in a region of hot corona \citep{Wevers2021, Yao2022, Guolo2024}. 
\citet{Yao2022} postulated that in those events, a few hundred days is needed for the magnetic field to build up via differential rotation, which then leads to a magnetically dominated hot corona.
In ASASSN-15oi, the soft $\rightarrow$ hard transition occurred around $\sim 10^3$\,d \citep{Hajela2025}.

In contrast, AT2024kmq and AT2024lhc exhibit hard X-ray emission at much earlier times ($t_{\rm rest}<50$\,d or no later than the optical peak), raising the question: what physical conditions enable rapid corona formation? 
To address this, we next contextualize our events within a control sample of ZTF TDEs with early-time X-ray constraints, and compare observations with accretion disk theories. 

\subsubsection{Evidence for Soft-to-hard State Transition at $\dot m =0.03$} \label{subsubsec:soft-to-hard}
We utilize X-ray data from the eROSITA-RU covering Galactic longitudes $0^{\circ}<l<180^{\circ}$. During its all-sky surveys (December 2019--February 2022, with 6-month cadence), eROSITA provides unbiased X-ray snapshots for ZTF TDEs. We selected all publicly classified TDEs passing the ZTF TDE selection filter \citep{vanVelzen2021} with at least one eROSITA observation between first optical light $t_{\rm fl}$ and 6 months after $t_{\rm fl}$. This yields a comparison sample of 30 TDEs constructed without prior knowledge of X-ray properties. Following the procedures adopted in \S\ref{subsec:fbnt_class}, we estimate the black hole masses of this comparison sample\footnote{We note that \citet{Brightman2021} reported $\sigma_\ast =213\pm12\,{\rm km\,s^{-1}}$ for ZTF19acymzwg (SDSS\,J1430+43) using a Keck/LRIS spectrum, which does not have the resolution to measure $\sigma_\ast$. We therefore exclude this measurement and estimate the black hole mass from its $M_{\rm gal}$ instead. } using host-galaxy scaling relations: the $M_{\rm BH}$--$\sigma_\ast$ relation for galaxies with measured velocity dispersions, and the $M_{\rm BH}$--$M_{\rm gal}$ relation otherwise. 
We cross matched this sample with the eROSITA source catalog, resulting in 11 TDEs with matched sources. The eROSITA upper limits of the remaining 19 sources are all $<10^{43}\,{\rm erg\,s^{-1}}$.

\begin{figure*}
    \centering
    \includegraphics[width=\textwidth]{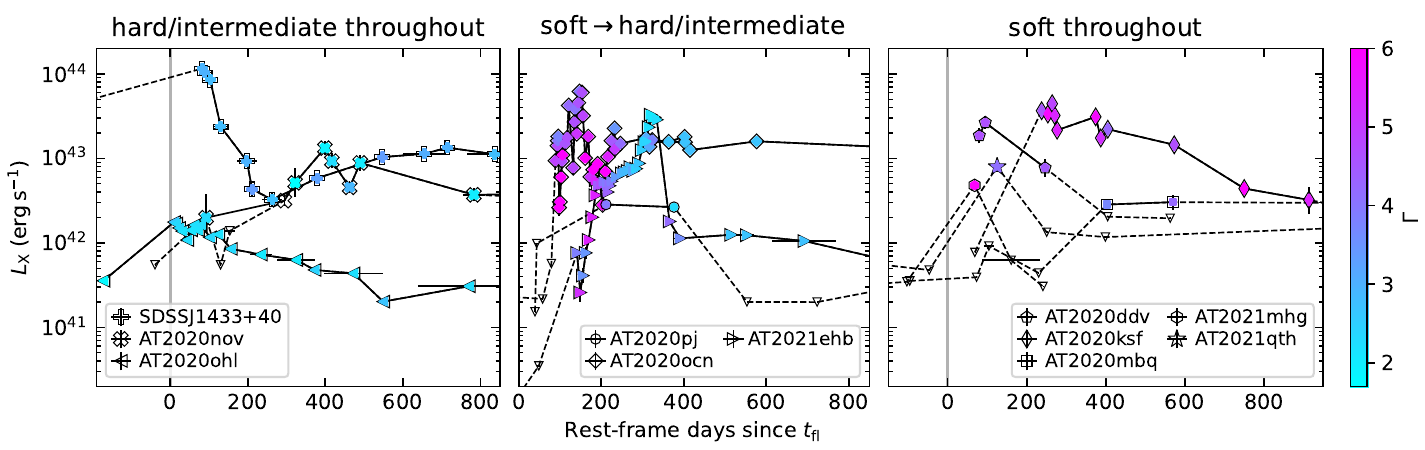}
    \caption{X-ray light curves of 11 ZTF TDEs with eROSITA detections within 6 months after first optical light. The data points are color-coded by the X-ray spectral power-law index $\Gamma$.  \label{fig:lc_Gamma}}
\end{figure*}

In Figure~\ref{fig:lc_Gamma}, we show the detailed X-ray (eROSITA+Swift/XRT) light curves of the 11 detected sources, color-coded by $\Gamma$. All X-ray spectra were modeled using \texttt{tbabs*zashift*powerlaw} following the same procedures outlined in \S\ref{subsubsec:xrt}. For ZTF19acymzwg (SDSS\,J1430+43), we further added a component of host absorption (\texttt{ztbabs}) with $N_{\rm H, host}$ fixed at $7\times10^{20}\,{\rm cm^{-2}}$, following \citet{Brightman2021}. Based on the values of $\Gamma$, we divide the X-ray states into soft ($\Gamma> 4$), intermediate ($\Gamma \sim 3$), and hard ($\Gamma \lesssim 2.5$). As indicated by Figure~\ref{fig:lc_Gamma}, the X-ray spectral evolution of the 11 TDEs with eROSITA detections can generally be divided into three categories: those with soft spectra throughout the first $\sim2$\,yr, those with hard/intermediate spectra throughout the first $\sim2$\,yr, and those with a spectral transition of ${\rm soft}\rightarrow {\rm hard/intermediate}$ around $\sim 1$\,yr. 

\begin{figure}
    \centering
    \includegraphics[width=\columnwidth]{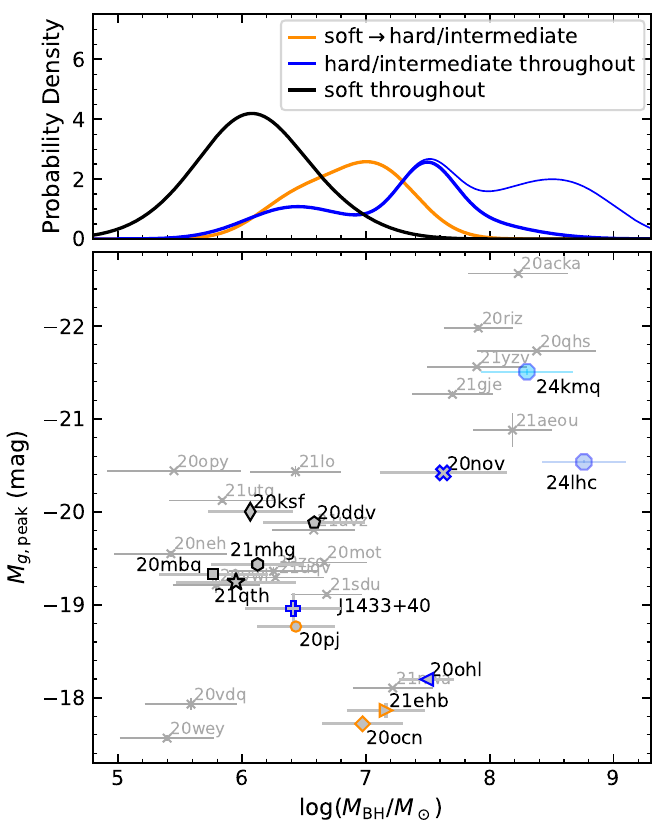}
    \caption{\textit{Upper}: Black hole mass distribution of X-ray detected TDEs obtained by kernel density estimation, using a kernel width equal to the uncertainty of $M_{\rm BH}$. For the ``hard/intermediate throughout'' class, the thin blue line shows the result including AT2024kmq and AT2024lhc, while the thick blue line shows the result with the comparison sample only. 
    \textit{Bottom}: Peak rest-frame $g$-band luminosity versus black hole mass for AT2024kmq, AT2024lhc, and the comparison sample, where $M_{\rm BH}$ is estimated using host-galaxy scaling relations. Markers follow the same scheme as in Figure~\ref{fig:lc_Gamma}. \label{fig:Mpeak_Mbh} }
\end{figure}

In Figure~\ref{fig:Mpeak_Mbh}, we show the distribution of the comparison sample in the $M_{g, \rm peak}$ vs. $M_{\rm BH}$ diagram, color-coded by the X-ray spectral class.  One striking result is the mass segregation of those TDEs that have X-ray spectra which are always soft (lowest masses), transition between states (middle masses) and are always hard (highest masses). This mass segregation of the X-ray spectral states of TDEs can be understood with classical time-dependent accretion disk theory, and the additional assumption that TDE disks transition to a harder accretion state at a fixed $\dot{m}$. Indeed, this exact mass-segregation result was predicted by \cite{MumBalb21b}. 

This result is simple to understand heuristically. First note that the viscous timescale in a TDE disk is (to leading order) independent of black hole mass 
\begin{align}
    t_{\rm visc} \approx \alpha^{-1} (h/r)^{-2} \sqrt{r_{\rm T}^3/GM_{\rm BH}} \approx \alpha^{-1}(h/r)^{-2}\sqrt{R_\star^3/GM_\star},
\end{align} 
where $h/r$ is the aspect ratio of the disk. We assume that a fraction $f_{\rm d}$ of the returning stellar material which forms a disk, such that $M_{\rm disk}=f_{\rm d} M_\star$ does not depend on $M_{\rm BH}$. Thus, the peak accretion rate (in unit of $\dot M_{\rm Edd}$) of a TDE disk is generically a decreasing function of black hole mass 
\begin{align}
    \dot m_{\rm peak} \equiv \frac{\dot M_{\rm acc, peak}}{\dot{M}_{\rm Edd}} \approx \frac{M_{\rm disk}}{t_{\rm visc}} \times \frac{1}{\dot{M}_{\rm Edd}} \propto \frac{1}{M_{\rm BH}}.
\end{align}
With a lower $\dot m_{\rm peak}$ and a similar evolutionary timescale, higher mass black hole TDEs will have $\dot m_{\rm peak} \sim 0.01 $ at earlier times, making them more likely to be represented in a sample of TDEs with coronae. Conversely, lower-mass black holes will stay above a given Eddington accretion ratio for longer. 

The above heuristic argument can be made quantitative. Indeed, Mummery in prep. calculated the time it takes for a TDE disk accretion rate to drop to a certain value of $\dot{m}$, which we denote $t_{\rm tr}$. For a full disruption, the transition time is
\begin{equation}
    t_{\rm tr} \approx 4000\, {\rm d} \, \alpha_{-1}^{-1/4}\theta_{-1}^{-1/2} \eta_{-1}^{3/4} f_{\rm d, -1}^{3/4} m_\star^{5/8} r_\star^{3/8} \dot{m}_{-2}^{-3/4} M_{\rm BH, 6}^{-3/4}.   \label{eq:t_tr}
\end{equation}
Here, $\eta$ is the radiative efficiency of the accretion disk, $m_\star \equiv M_\star/M_\odot$, $r_\star \equiv R_\star/R_\odot$, and $\theta \equiv (h/r)$. A subscript $-1 \,(-2)$ indicates that a value has been normalized by 0.1 (0.01) respectively.
The mass-segregation of the different spectral states can then be understood from the $t_{\rm tr}\propto M_{\rm BH}^{-3/4}$ scaling of the time to fall to a fixed value of $\dot{m}$. 

\begin{figure}
    \centering
    \includegraphics[width=0.99\linewidth]{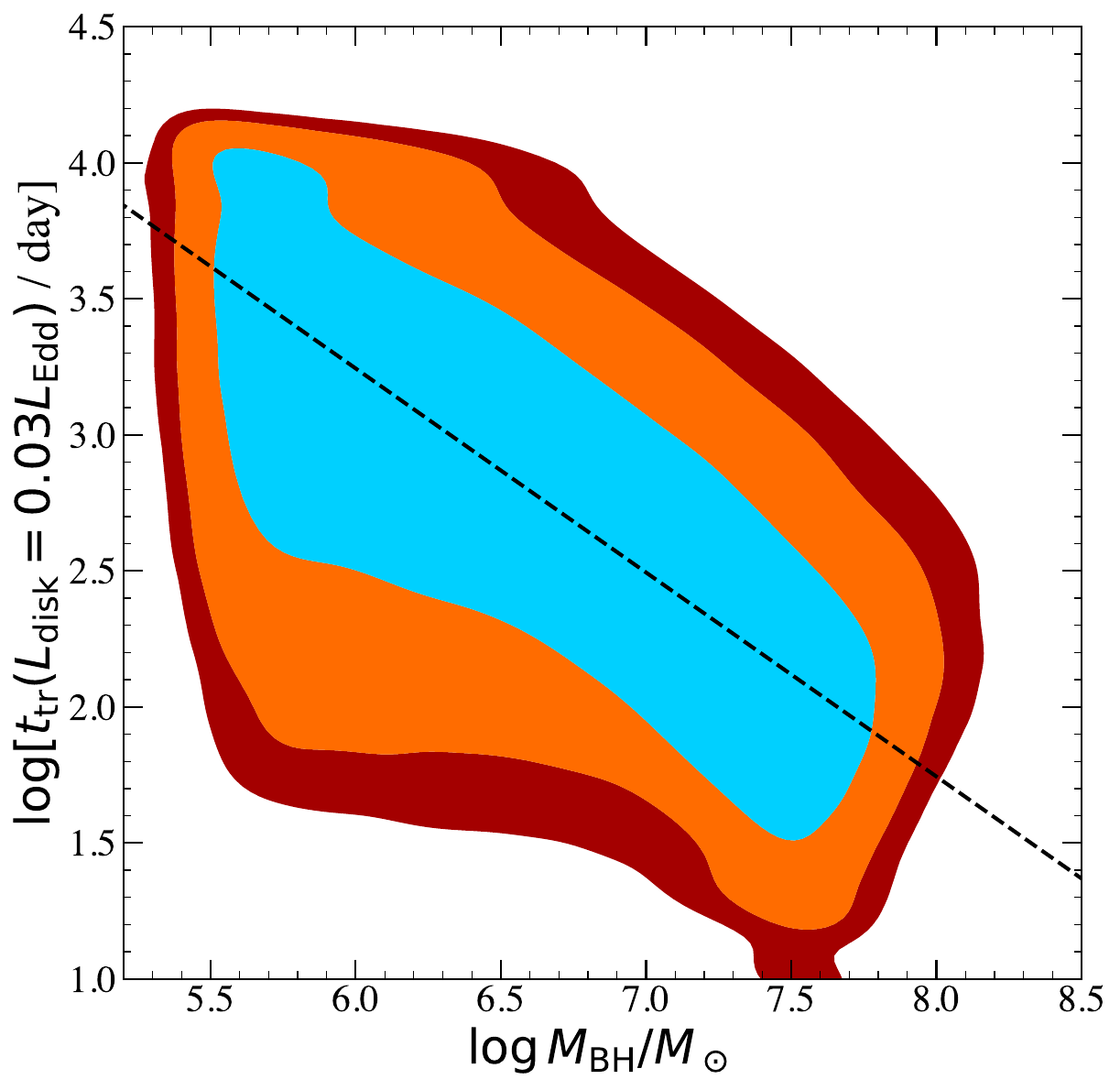}
    \caption{The time for a variety of different simulated TDE disk systems to fall to an Eddington ratio of 0.03 (see text for simulation details), as a function of black hole mass. The blue, orange and red contours show 1, 2 and 3 $\sigma$ confidence intervals respectively (i.e., they contain 68.3\%, 95.4\% and 99.7\% of the points). The black dashed curve shows the analytical scaling argument discussed in the text $t_{\rm tr} \propto M_{\rm BH}^{-3/4}$. We see that the analytical argument captures the main trend across the simulated population. The vast majority of sources with black holes mass $\log_{10} M_{\rm BH}/M_\odot \gtrsim 8$ did not ever exceed an Eddington ratio of 0.03, and are not displayed.  }
    \label{fig:fitted_simulation}
\end{figure}

To verify this quantitatively, we perform a numerical population synthesis calculation of the time required to reach $\dot{m} = 0.03$. This threshold is motivated by observations of black hole XRBs (see \S\ref{sec:intro}). We sample black hole masses from a log-uniform distribution spanning $10^{5.5}$--$10^{8.5} M_\odot$ and, for each black hole mass, draw stellar\footnote{Note that for modelling convenience, our simulated sample does not include partial disruptions with $\beta<1$, where $\beta$ is defined as the ratio between the pericenter distance and the tidal radius. } and disk properties following \citet{Mummery2024_fundamental_scaling}. 
For every sampled system, we solve the time-dependent disk equations using the \texttt{FitTeD} code \citep{Mummery2025_FitTeD} to compute the transition time $t_{\rm tr}$ at which the disk bolometric luminosity reaches $0.03 L_{\rm Edd}$. Figure~\ref{fig:fitted_simulation} shows the resulting distribution of $M_{\rm BH}$ versus $t_{\rm tr}$ from $N = 10^5$ simulated systems. The black dashed curve shows the analytical prediction from Equation~\ref{eq:t_tr}, which accurately captures the mean trend and validates the $t_{\rm tr} \propto M_{\rm BH}^{-3/4}$ scaling.

\begin{figure}
    \centering
    \includegraphics[width=0.95\linewidth]{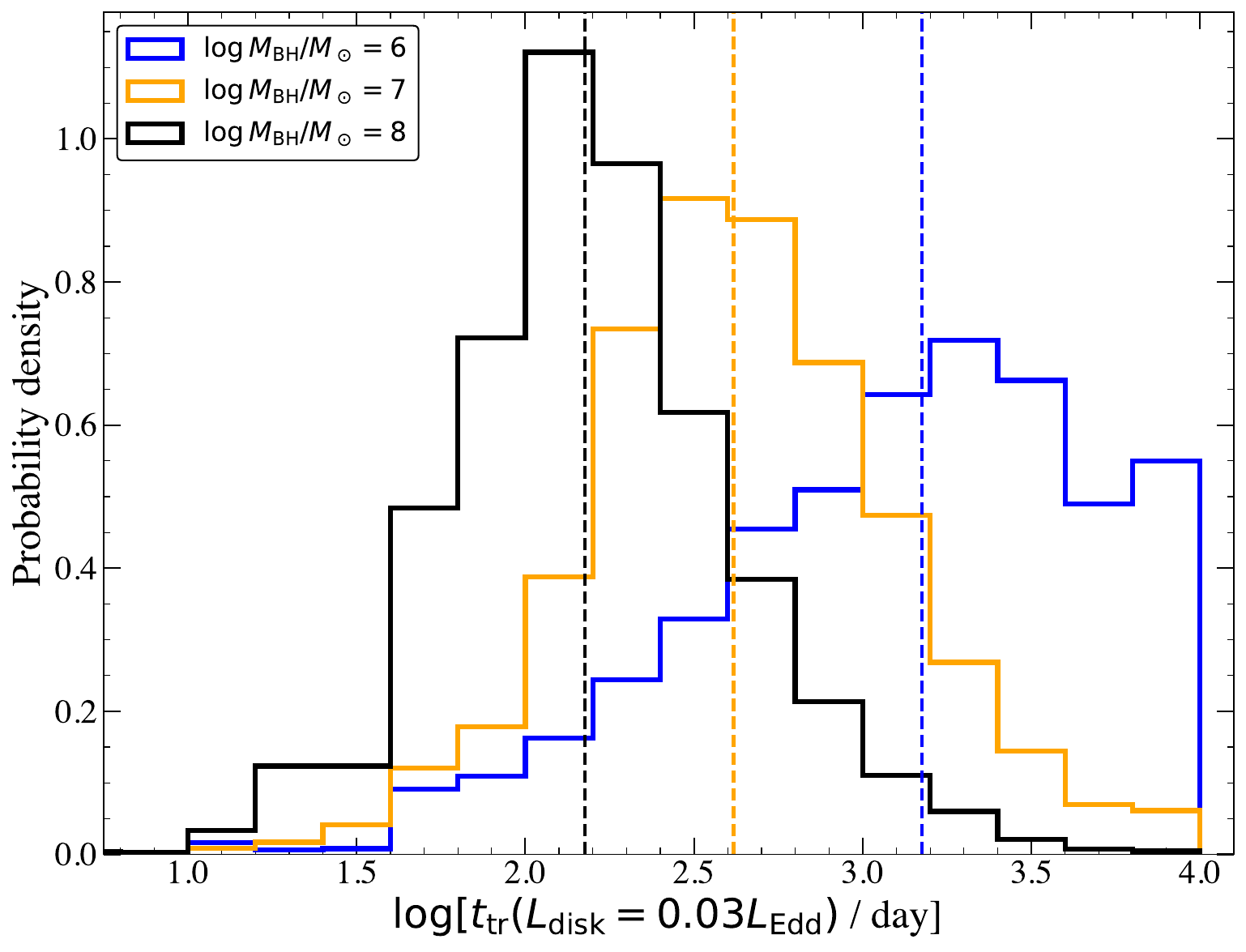}
    \includegraphics[width=0.95\linewidth]{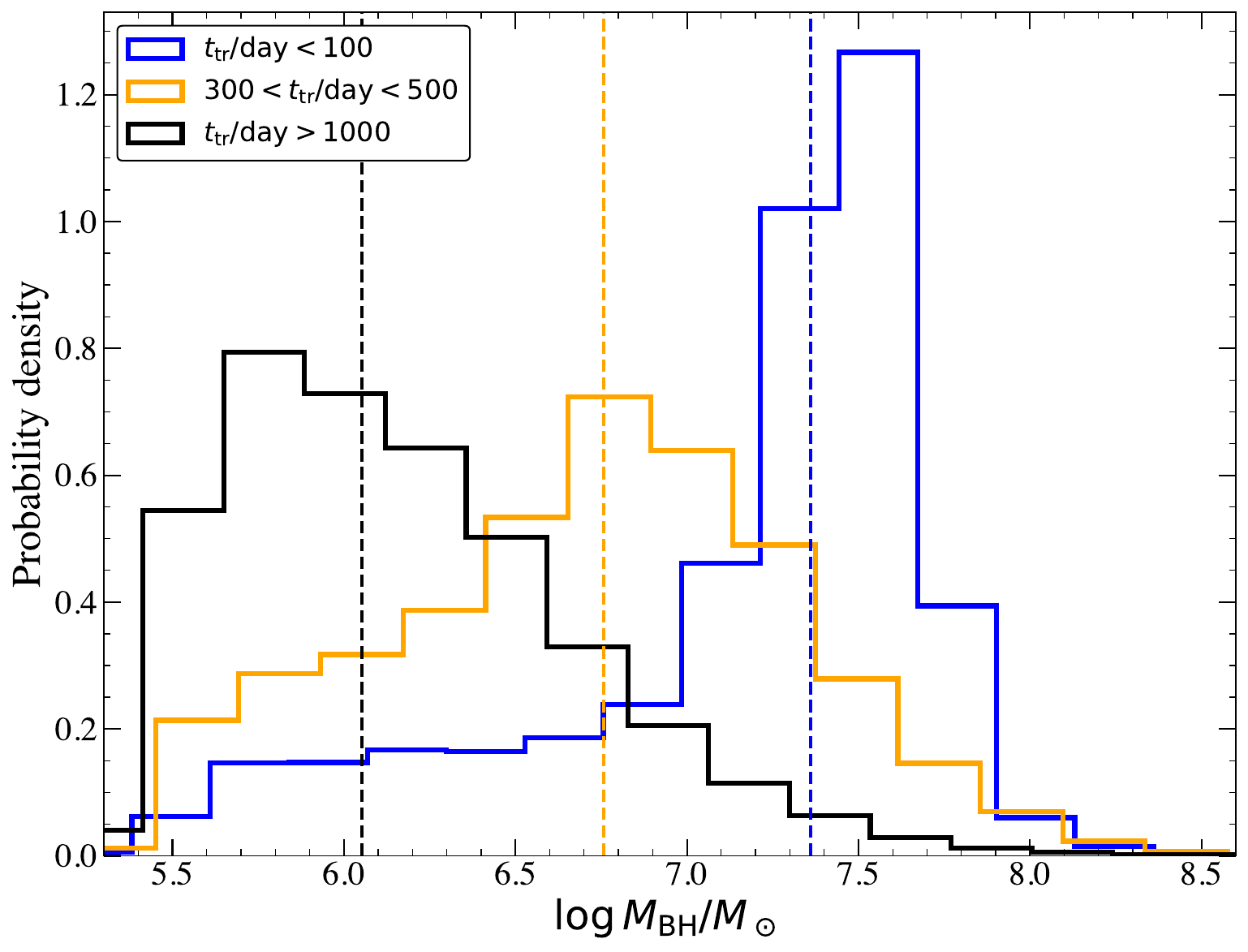}
    \caption{\textit{Upper}: one dimensional probability distributions of the transition time for disk systems within a factor two of the black hole masses denoted on the plot. We see that $M_{\rm BH} \sim 10^6M_\odot$ TDE systems typically take $\sim 10^3$\,d (multiple years), $M_{\rm BH} \sim 10^7M_\odot$ systems typically transition after $\sim 1$\,yr, while $M_{\rm BH} \sim 10^8M_\odot$ systems transition rapidly at $t\lesssim 100$\,d, and will likely only ever be observed in a harder state. \textit{Lower}: one dimensional projection of the black hole masses for systems which transition through $\dot{m}= 0.03$ in different temporal bins: rapidly $t_{\rm tr} < 100$\,d (blue posterior), on intermediate timescales $t_{\rm tr} \in (300, 500)$\,d (orange posterior), and on long timescales $t_{\rm tr} > 10^3$\,d. These posteriors are in good agreement with that seen in the data (upper panel of Figure~\ref{fig:Mpeak_Mbh}).  }
    \label{fig:1d_posteriors}
\end{figure}

To elucidate the mass-dependent transition behavior, we project the two-dimensional $M_{\rm BH}$--$t_{\rm tr}$ distribution (Figure~\ref{fig:fitted_simulation}) onto one-dimensional posteriors. In Figure~\ref{fig:1d_posteriors}, the upper panel shows the distribution of transition times for systems within three narrow (within $\pm 0.1$ dex) mass bins ($M_{\rm BH} = 10^6$, $10^7$, and $10^8 M_\odot$); the lower panel shows the black hole mass distributions for systems transitioning within specific time windows (early: $t_{\rm tr} < 100$\,d; intermediate: 300--500\,d; late: $> 1000$\,d), which closely match the observed mass segregation in X-ray spectral states (Figure~\ref{fig:Mpeak_Mbh}, upper panel). These distributions demonstrate that simple time-dependent disk theory quantitatively reproduces the data.

We note that at fixed Eddington ratio, higher-mass black holes produce cooler accretion disks whose peak thermal emission lies below the soft X-ray band. \citet{MumBalb21a} showed that above a critical mass $M_{\rm crit} \sim 2 \times 10^7 M_\odot$, TDE disks become increasingly unlikely to produce detectable thermal X-ray emission, as the disk emission peaks in the EUV band. Consequently, even if a $M_{\rm BH} \sim 10^8 M_\odot$ TDE remains above the $\dot{m} \sim 0.03$ threshold and has not yet formed a corona, it would likely evade detection in soft X-ray surveys. However, once these high-mass systems transition into the hard (coronal) state, they will become X-ray detectable again. \citet{Mum21} argued that X-ray luminosities in both soft (thermal) and hard (coronal) states should be capped at $L_{\rm X, max}\approx 10^{44}\,{\rm erg\,s^{-1}}$, consistent with the observed maximum in our sample (Figure~\ref{fig:lc_Gamma}).


In this respect, our results suggest that TDEs undergo soft-to-hard spectral transitions at $\dot{m}\sim 0.03$, phenomenologically resembling scaled-up XRBs in terms of the critical Eddington ratio. However, the characteristic timescales of these transitions differ markedly: XRB state changes typically occur over hours to a few days, whereas the analogous transitions inferred in TDEs occur over a few to tens of days \citep{Wevers2021, Yao2022}, corresponding to much shorter durations when expressed in units of the black hole dynamical or viscous timescales. The physical origin of this discrepancy remains unclear (but see possible explanations by \citealt{Noda2018}) and warrants further investigation in future theoretical work.

\subsubsection{A Note on Radio Observations}

In the context of XRBs, the soft-to-hard state transition is also associated with mass ejection and radio emission from a compact jet \citep{Fender2004}, and the 5\,GHz radio luminosity $L_{\rm R}$, the 2--10\,keV X-ray luminosity, and black hole mass are observed to follow the ``Fundamental Plane (FP) of black hole activity'' \citep{Gallo2003, Merloni2003, Falcke2004}. It has been shown that some TDEs may also follow the FP \citep{miller2011,Pasham2018,Alexander2026}. We use this empirical relation to predict the expected radio emission from our TDEs and compare it to our observations.

For AT2024kmq at $t_{\rm obs} = 232$\,d, 
the 2--10\,keV X-ray luminosity is ${\rm log}L_{\rm X} = 43.90\pm0.17$. Using the FP relation from \citet{Gultekin2019} and taking the $\sigma_\ast$-inferred black hole mass (Table~\ref{tab:basic_info}), the radio luminosity should be ${\rm log}L_{\rm R} = 39.88_{-0.69}^{+0.72}$, which exceeds the observed 
radio upper limit (\S\ref{subsec:radio}) by $1.3\sigma$. 
Similarly, for AT2024lhc at $t_{\rm obs} = 77$\,d, 
we have ${\rm log}L_{\rm X} = 43.98^{+0.24}_{-0.25}$ and expect ${\rm log}L_{\rm R} = 40.34_{-0.70}^{+0.75}$, which is more luminous than the transient radio upper limit by $3.1\sigma$. 
The tension with the FP predictions suggests either intrinsically weak/absent outflows or an off-axis jet geometry. Further radio monitoring is warranted to confirm or rule out delayed jet brightening expected from off-axis configurations.

We note that the launch of a radio-emitting jet or outflow at $\dot{m} \sim 0.03$ is one of the scenarios proposed by \citet{Giannios2011} for TDEs in the soft-to-hard transition regime. \citet{Alexander2026} recently attempted to test this scenario and found suggestive evidence; however, their estimates of the mass accretion rate, based on the \texttt{MOSFiT} framework, are subject to significant systematic uncertainties (especially at late times). Interestingly, the only X-ray-detected TDE in their sample that is radio underluminous relative to the FP at $t<1$ yr is AT2018fyk, which also has the highest black hole mass of any TDE in their sample [$\log (M_{\rm BH}/M_\odot)=8.0\pm0.3$; \citealt{Wevers2020,Alexander2026}]. AT2018fyk underwent a clear soft-to-hard transition at $\sim100-200$ d \citep{Wevers2021}, but was not detected in the radio until 1991 d post-discovery \citep{Cendes2024}. After its radio detection, AT2018fyk was found to roughly follow the FP \citep{Alexander2026}. This could suggest either that AT2018fyk's radio emission was indeed powered by a jet launched at the time of the X-ray state transition but was suppressed at earlier times due to relativistic beaming and an off-axis viewing angle, or that its radio-emitting jet or outflow was not launched until well after this transition occurred. This example also motivates long-term radio monitoring of AT2024kmq and AT2024lhc to search for analogous behavior.

\subsection{Optical precursor from self-intersecting shocks} \label{subsec:precursor}
AT2024kmq exhibited a red ($T_{\rm bb}\sim 6\times 10^{3}$\,K) and rapidly fading (duration $\sim 10$\,d, Figure~\ref{fig:precursor}) precursor outburst that preceded the main optical peak by $\sim 30$\,d (rest-frame). \citet{Ho2025} proposed two scenarios for this early emission: (1) an optically-thin synchrotron component powered by a forward shock driven by a jet, analogous to GRB afterglows \citep{Matsumoto2023_22cmc}, or (2) expanding debris generated by self-intersection of the stellar streams, where a fraction of the stream kinetic energy is converted to thermal energy. Scenario (1) predicts a rising late-time radio light curve; our non-detection of bright radio emission out to $t_{\rm rest} \sim 1$ yr (Figure~\ref{fig:radio_lc}) disfavors this interpretation for AT2024kmq.

The self-intersection scenario provides a natural explanation for the precursor emission. For the massive black hole powering AT2024kmq  ($M_{\rm BH}\sim 10^{8}\,M_{\odot}$), the collision between incoming and outgoing debris streams (expected to facilitate circularization and disk formation) occurs at $\lesssim 10 r_{\rm g}$ for canonical impact parameters, even for partial disruptions. The relative velocity between the streams at the moment of self-intersection therefore amounts to tens of percent of the speed of light, leading to substantial kinetic energy dissipation and heating of the gas (i.e., the strong nature of the shock implies the downstream gas is radiation-dominated, reaching radiation and gas temperatures on the order of $\sim 10^{6}$\,K). In such scenarios, radiation-hydrodynamical simulations \citep{Jiang2016_self_crossing_shock, Huang2023_stream_collision, Huang2024_stream_collision}, as well as analytical estimates \citep{dai15_intersection_radius, Lu2020, BonnerotLu2021}, suggest that some of the debris can be ejected from the system at substantial fractions of the self-intersection speed. If this self-intersection creates an expanding cloud of debris (also seen in radiation-hydrodynamical simulations; \citealt{Jiang2016_self_crossing_shock}), then \citet{Ho2025} found that many properties of the precursor exhibited by AT2024kmq, including its temperature, duration, and overall energetics, could be explained by the emission from this byproduct of the self-intersection.

AT2024lhc exhibited qualitatively similar behavior, with a luminous, rapidly fading precursor and comparable black hole mass (Figures~\ref{fig:precursor}, \ref{fig:bb_pars}), supporting a similar physical origin. The precursor in AT2024lhc has a higher effective temperature ($T_{\rm bb} \sim 2 \times 10^4$ K, a factor of $\sim3$ larger than AT2024kmq), which remains consistent with the simulations of \citet{Jiang2016_self_crossing_shock} and likely reflects a smaller photospheric radius (and correspondingly higher temperature) when the expanding debris was first detected.

The self-intersection paradigm applies to TDEs across the black hole mass spectrum, raising the question: why are such precursors rare? As suggested by \citet{Ho2025}, lower-mass black holes ($M_{\rm BH} \lesssim 10^7 M_\odot$) have larger self-intersection radii where dissipation is less efficient, producing weaker or absent optical signatures. At these radii, more debris remains bound, with radiation advected inward rather than emitted by outward-propagating material. Since the volumetric rate of TDEs peaks at lower black hole masses \citep{Yao2023}, the same phenomenon may be ubiquitous but simply more difficult to detect in the dominant TDE population.

\subsection{Post-precursor UV/optical Emission} \label{subsec:opt_peak}

\subsubsection{Power Sources of the Main Peak}
The origin of the peak UV/optical emission in TDEs remains actively debated. Available TDE models that produce UV/optical light curves, such as the TDE module in \texttt{MOSFiT} \citep{Mockler2019}, the reprocessing-outflow model of \citet{Matsumoto2021_tde_optical}, and the quasi-spherical cooling envelope model in \texttt{REDBACK} \citep{Sarin2024_TDE}, when applied to TDEs with $M_{\rm BH} > 10^7 M_\odot$ (inferred from host galaxies), systematically underestimate black hole masses \citep{Ramsden2022, Hammerstein2023, Wise2026, Guolo2026}, suggesting missing physics in the models or alternative emission mechanisms. 

The large black hole mass inferred for AT2024kmq and AT2024lhc facilitates prompt self-intersection and ensuing accretion, and the fact that standard accretion disk theory successfully explains the observed X-ray state transitions (\S\ref{subsec:origin_Xrays}) strongly supports accretion as the dominant power source. Indeed, during the entire X-ray monitoring phase of AT2024lhc ($t_{\rm rest} \geq 48$\,d) and at later times in AT2024kmq ($t_{\rm rest} \gtrsim 100$\,d), the X-ray light curves roughly track the UV/optical decline: both decline by about one order of magnitude over $\sim 1$\,yr (Figures~\ref{fig:lc_overview_24lhc}, \ref{fig:lc_overview_24kmq}, \ref{fig:bb_pars}). Since the coronal X-ray emission is powered by Comptonization of disk photons, the observed correlation between X-ray and UV/optical luminosities suggests that both trace the disk accretion rate.

We can test whether accretion power alone accounts for the UV/optical luminosity by comparing the peak bolometric luminosity to the expected accretion luminosity at the state transition threshold. Under the assumptions of \S\ref{subsec:origin_Xrays}, if accretion is the sole power source, we expect $\lambda_{\rm Edd, peak} \lesssim 0.03$ at the optical peak. This prediction is consistent with observations of AT2024lhc (Table~\ref{tab:Lbol}). However, AT2024kmq reaches $\lambda_{\rm Edd, peak} \gtrsim 0.1$, suggesting an additional power source (at least at its UV/optical peak), most likely shocks from stream-disk collisions \citep{Steinberg2024} and circularization shocks within the accretion flow \citep{Huang2026_disk_formation}. 

\subsubsection{AT2024kmq's Late-time Rebrightening}

Late-time rebrightening has been observed in numerous TDEs. They exhibit diverse characteristics in timing, luminosity, and light curve morphology.
Some events show significant, multiple undulations during the decline phase \citep{Wevers2021, Yao2025_radio}, while others exhibit discrete flares with shapes and amplitudes either similar \citep{Payne2021, Lin2024, Somalwar2025_pTDE, Angus2026} or markedly different \citep{Kankare2017, Malyali2021, Yao2023} from the initial outburst.
When well-separated flares display similar characteristics, they are commonly interpreted as repeated partial TDEs  from a single star on an eccentric orbit \citep{Liu2025_pTDE, Zhong2025} or as distinct TDEs from multiple stars \citep{Mandel2015}.

For AT2024kmq, the modest amplitude of the rebrightening favors a single TDE with multi-phase evolution rather than a repeated disruption scenario.
Several physical mechanisms have been proposed to explain such rebrightenings. Some models attribute the initial flare to shock-powered emission and the subsequent rebrightening to delayed accretion disk formation \citep{Chen2022_19avd, Guo2025_tde_reverberation}, while others invoke interactions between disk winds or debris streams and a dusty torus \citep{Jiang2019_10adi, Zhuang2021} --- a scenario particularly relevant for TDEs in radiative-mode AGN.

In the case of AT2024kmq, multiple lines of evidence point to early disk formation (\S\ref{subsec:origin_Xrays}), ruling out delayed accretion as the origin of the rebrightening. 
Furthermore, the WISE color (Table~\ref{tab:basic_info}) argues against the presence of a dusty torus, disfavoring wind-torus interaction scenarios.
Given that the host galaxy harbors a jet-mode AGN, which likely possesses a truncated thin disk (\S\ref{subsubsec:jet_agn}), we speculate that the rebrightening arises from interactions between TDE's newly formed accretion flow and the pre-existing truncated accretion flow.
Future theoretical work exploring TDE-truncated disk interactions may naturally explain the late-time rebrightenings observed in TDEs hosted by radio-loud AGN.

\section{Conclusion} \label{sec:conclusion}

We have presented multi-wavelength observations of two luminous featureless TDEs, AT2024kmq and AT2024lhc. Both events are hosted by massive black holes ($M_{\rm BH} \sim 10^8 M_\odot$) and share distinctive observational signatures. 

They exhibit luminous ($L_{\rm X, peak} \sim 10^{44}$ erg s$^{-1}$), rapidly variable (minimum observed timescale $\sim 1.2$--4.8\,hr) X-ray emission appearing at early times ($t_{\rm rest} < 50$ d, contemporaneous with or preceding the optical peak). Their X-ray spectra are characterized by power-law photon indices $1.7 < \Gamma < 3$, with NuSTAR detecting photons extending to $\sim$20 keV. 
Rapid variability restricts the X-ray emitting region to $R_{\rm X} \lesssim 10 r_{\rm g}$, while the absence of radio emission disfavors relativistic jets. We interpret this emission as inverse Compton scattering in a compact corona, formed within $\sim50$\,d of disruption, far more rapidly than the $\sim$200--1000 d timescales observed in lower-mass ($M_{\rm BH} \sim 10^6$--$10^7 M_\odot$) TDE systems \citep{Yao2022, Hajela2025}. 

For TDEs as a population, we find that the various X-ray spectral states observed can be understood with simple disk theory, assuming that soft-to-hard state transition happens at a time when the Eddington-normalized mass accretion rate declines to the critical value of $\dot m \sim 0.03$ (\S\ref{subsubsec:soft-to-hard}). This observational result confirms the theoretical prediction of \citet{MumBalb21a}. From the perspective of X-ray state transitions, the accretion disks of TDEs behave phenomenologically as scaled-up versions of XRBs, although the fast duration of the transition and the radio non-detections warrant further investigation.

Both AT2024kmq and AT2024lhc also exhibited optical precursor outbursts that preceded the main peak by $\sim$30 d (rest-frame), and are consistent with being powered by shocks generated by debris stream self-intersection. If this interpretation is correct, optical precursors should be more prominent in TDEs around more massive black holes --- a prediction that can be tested with future systematic searches and analyses of TDE precursor emission. 

Through systematic comparison of 49 ZTF TDEs spanning four spectral subtypes (TDE-H+He, TDE-H, TDE-He, and TDE-featureless), we find that featureless TDEs are hosted by high-mass black holes, consistent with the findings of \citet{Hammerstein2023}. Featureless TDEs exhibit robust bimodality in both peak blackbody luminosity and radius. AT2024kmq and AT2024lhc belong to the high-luminosity, large-radius subclass. These two subclasses likely represent two ways (i.e., high source luminosity and small photosphere; Figure~\ref{fig:uvo_sptype_summary1}) to achieve the high ionization required to suppress spectral lines as demonstrated in \cite{Aspegren2026}. Current models to fit the TDE UV/optical light curves are not reliable in the high-mass regime.

\section*{Acknowledgements}

Yuhan Yao, Ryan Chornock, and Erica Hammerstein acknowledge support from the NASA/NuSTAR GO program, grant 80NSSC25K7488. 
Andrew Mummery acknowledges support from the Ambrose Monell Foundation, the W.M. Keck Foundation and the John N. Bahcall Fellowship Fund at the Institute for Advanced Study.
Marat Gilfanov acknowledges support by the Ministry of Science and Higher Education grant 075-15-2024-647.
Eric R. Coughlin acknowledges support from the National Aeronautics and Space Administration through the Astrophysics Theory Program, grant 80NSSC24K0897.
Dheeraj R. Pasham acknowledges the support of the NASA/NuSTAR GO program, grant 80NSSC25K7949.
Kate D. Alexander gratefully acknowledges support provided by the NSF through award AST-2307668 and from the Alfred P. Sloan Foundation. 
Charlotte R. Angus, Matt Nicholl, and Xinyue Sheng are supported by the European Research Council (ERC) under the European Union's Horizon 2020 research and innovation program (grant agreement No. 948381).
Michael W. Coughlin acknowledges support from the National Science Foundation with grant numbers PHY-2117997, PHY-2308862 and PHY-2409481. 

This work is partly based on observations with the NASA/ESA Hubble Space Telescope obtained at the Space Telescope Science Institute, which is operated by the Association of Universities for Research in Astronomy, Incorporated, under NASA contract NAS5-26555. Support for Program number GO-17667 was provided through a grant from the STScI under NASA contract NAS5-26555.

Based on observations obtained with the Samuel Oschin Telescope 48-inch and the 60-inch Telescope at the Palomar Observatory as part of the Zwicky Transient Facility project. ZTF is supported by the National Science Foundation under Grants No. AST-1440341, AST-2034437, and currently Award \#2407588. ZTF receives additional funding from the ZTF partnership. Current members include Caltech, USA; Caltech/IPAC, USA; University of Maryland, USA; University of California, Berkeley, USA; University of Wisconsin at Milwaukee, USA; Cornell University, USA; Drexel University, USA; University of North Carolina at Chapel Hill, USA; Institute of Science and Technology, Austria; National Central University, Taiwan, and OKC, University of Stockholm, Sweden. Operations are conducted by Caltech's Optical Observatory (COO), Caltech/IPAC, and the University of Washington at Seattle, USA.

The Gordon and Betty Moore Foundation, through both the Data-Driven Investigator Program and a dedicated grant, provided critical funding for SkyPortal.
The ZTF forced-photometry service was funded under the Heising-Simons Foundation grant No. 12540303 (PI: Graham).

The National Radio Astronomy Observatory is a facility of the National Science Foundation operated under cooperative agreement by Associated Universities, Inc.

This work has made use of data from the Asteroid Terrestrial-impact Last Alert System (ATLAS) project. The ATLAS project is primarily funded to search for near earth asteroids through NASA grants NN12AR55G, 80NSSC18K0284, and 80NSSC18K1575; byproducts of the NEO search include images and catalogs from the survey area. This work was partially funded by Kepler/K2 grant J1944/80NSSC19K0112 and HST GO-15889, and STFC grants ST/T000198/1 and ST/S006109/1. The ATLAS science products have been made possible through the contributions of the University of Hawaii Institute for Astronomy, the Queen’s University Belfast, the Space Telescope Science Institute, the South African Astronomical Observatory, and The Millennium Institute of Astrophysics (MAS), Chile.

This work uses data obtained with eROSITA telescope onboard SRG observatory. The SRG observatory was built by Roskosmos with the participation of the Deutsches Zentrum für Luft- und Raumfahrt (DLR). The SRG/eROSITA X-ray telescope was built by a consortium of German Institutes led by MPE, and supported by DLR. The SRG spacecraft was designed, built, launched and is operated by the Lavochkin Association and its subcontractors. The science data were downlinked via the Deep Space Network Antennae in Bear Lakes, Ussurijsk, and Baykonur, funded by Roskosmos. The eROSITA data used in this work were processed using the eSASS software system developed by the German eROSITA consortium and proprietary data reduction and analysis software developed by the Russian eROSITA Consortium.

Some of the data presented herein were obtained at Keck Observatory, which is a private 501(c)3 non-profit organization operated as a scientific partnership among the California Institute of Technology, the University of California, and the National Aeronautics and Space Administration. The Observatory was made possible by the generous financial support of the W. M. Keck Foundation. The authors wish to recognize and acknowledge the very significant cultural role and reverence that the summit of Maunakea has always had within the Native Hawaiian community. We are most fortunate to have the opportunity to conduct observations from this mountain. 

A major upgrade of the Kast spectrograph on the Shane 3 m telescope at Lick Observatory, led by Brad Holden, was made possible through gifts from the Heising-Simons Foundation, William and Marina Kast, and the University of California Observatories. Research at Lick Observatory is partially supported by a generous gift from Google.

Based on observations made with the Liverpool Telescope operated on the island of La Palma by Liverpool John Moores University in the Spanish Observatorio del Roque de los Muchachos of the Instituto de Astrofisica de Canarias with financial support from the UK Science and Technology Facilities Council.

Partly based on observations made with the Nordic Optical Telescope, owned in collaboration by the University of Turku and Aarhus University at the Observatorio del Roque de los Muchachos, La Palma, Spain, of the Instituto de Astrofisica de Canarias. The NOT data were obtained under program ID P68-501.

This research has made use of data from the NuSTAR mission, a project led by the California Institute of Technology, managed by the Jet Propulsion Laboratory, and funded by the National Aeronautics and Space Administration. Data analysis was performed using the NuSTAR Data Analysis Software (NuSTARDAS), jointly developed by the ASI Science Data Center (SSDC, Italy) and the California Institute of Technology (USA).

\section*{Data Availability}

Spectra will be uploaded to WiseRep following acceptance. Photometry is listed in online supplementary data tables. Other data presented in this paper that are not in the Appendix will be provided upon request to the author.



\bibliographystyle{mnras}
\bibliography{main}

@ARTICLE{Cendes2024,
       author = {{Cendes}, Yvette and {Berger}, Edo and {Alexander}, Kate and {Laskar}, Tanmoy and {Goodwin}, Adelle},
        title = "{Late-Time Radio Detection of the TDE AT2018fyk}",
      journal = {The Astronomer's Telegram},
         year = 2024,
        month = jun,
       volume = {16650},
        pages = {1},
       adsurl = {https://ui.adsabs.harvard.edu/abs/2024ATel16650....1C}
}

@ARTICLE{Miller2011,
       author = {{M{\"\i}ller}, J.~M. and {G{\"u}ltekin}, K.},
        title = "{X-Ray and Radio Constraints on the Mass of the Black Hole in Swift J164449.3+573451}",
      journal = {\apjl},
         year = 2011,
        month = sep,
       volume = {738},
       number = {1},
          eid = {L13},
        pages = {L13},
          doi = {10.1088/2041-8205/738/1/L13},
archivePrefix = {arXiv},
       eprint = {1106.2502},
 primaryClass = {astro-ph.HE},
       adsurl = {https://ui.adsabs.harvard.edu/abs/2011ApJ...738L..13M}
}

@ARTICLE{Pasham2018,
       author = {{Pasham}, Dheeraj R. and {van Velzen}, Sjoert},
        title = "{Discovery of a Time Lag between the Soft X-Ray and Radio Emission of the Tidal Disruption Flare ASASSN-14li: Evidence for Linear Disk-Jet Coupling}",
      journal = {\apj},
         year = 2018,
        month = mar,
       volume = {856},
       number = {1},
          eid = {1},
        pages = {1},
          doi = {10.3847/1538-4357/aab361},
archivePrefix = {arXiv},
       eprint = {1709.02882},
 primaryClass = {astro-ph.HE},
       adsurl = {https://ui.adsabs.harvard.edu/abs/2018ApJ...856....1P}
}

@ARTICLE{beloborodov92,
       author = {{Beloborodov}, A.~M. and {Illarionov}, A.~F. and {Ivanov}, P.~B. and {Polnarev}, A.~G.},
        title = "{Angular momentum of a supermassive black hole in a dense star cluster}",
      journal = {\mnras},
         year = 1992,
        month = nov,
       volume = {259},
       number = {2},
        pages = {209-217},
          doi = {10.1093/mnras/259.2.209},
       adsurl = {https://ui.adsabs.harvard.edu/abs/1992MNRAS.259..209B}
}

@ARTICLE{Wise2026,
       author = {{Wise}, Jacob L. and {Perley}, Daniel A. and {Sarin}, Nikhil and {Matsumoto}, Tatsuya and {Hinds}, K.-Ryan and {Yao}, Yuhan and {Sollerman}, Jesper and {Schulze}, Steve and {Bochenek}, Aleksandra and {Coughlin}, Michael W. and {De}, Kishalay and {Dekany}, Richard and {Frederick}, Sara and {Fremling}, Christoffer and {Gezari}, Suvi and {Graham}, Matthew J. and {Ho}, Anna Y.~Q. and {Kulkarni}, Shrinivas and {Laher}, Russ R. and {Omand}, Conor and {Johnson}, Natalya and {Sharma}, Yashvi and {Taggart}, Kirsty and {Ward}, Charlotte and {Wold}, Avery and {Yan}, Lin},
        title = "{AT2019cmw: a highly luminous, cooling featureless TDE candidate from the disruption of a high mass star in an early-type galaxy}",
      journal = {\mnras},
         year = 2026,
        month = mar,
       volume = {546},
       number = {3},
          eid = {stag130},
        pages = {stag130},
          doi = {10.1093/mnras/stag130},
archivePrefix = {arXiv},
       eprint = {2507.07380},
 primaryClass = {astro-ph.HE},
       adsurl = {https://ui.adsabs.harvard.edu/abs/2026MNRAS.546ag130W}
}

@ARTICLE{st80,
       author = {{Sunyaev}, R.~A. and {Titarchuk}, L.~G.},
        title = "{Comptonization of X-Rays in Plasma Clouds - Typical Radiation Spectra}",
      journal = {\aap},
         year = 1980,
        month = jun,
       volume = {86},
        pages = {121},
       adsurl = {https://ui.adsabs.harvard.edu/abs/1980A&A....86..121S}
}

@ARTICLE{gm14,
       author = {{Gilfanov}, Marat and {Merloni}, Andrea},
        title = "{Observational Appearance of Black Holes in X-Ray Binaries and AGN}",
      journal = {\ssr},
         year = 2014,
        month = sep,
       volume = {183},
       number = {1-4},
        pages = {121-148},
          doi = {10.1007/s11214-014-0071-5},
       adsurl = {https://ui.adsabs.harvard.edu/abs/2014SSRv..183..121G}
}

@ARTICLE{sazonov21,
       author = {{Sazonov}, S. and {Gilfanov}, M. and {Medvedev}, P. and {Yao}, Y. and {Khorunzhev}, G. and {Semena}, A. and {Sunyaev}, R. and {Burenin}, R. and {Lyapin}, A. and {Meshcheryakov}, A. and {Uskov}, G. and {Zaznobin}, I. and {Postnov}, K.~A. and {Dodin}, A.~V. and {Belinski}, A.~A. and {Cherepashchuk}, A.~M. and {Eselevich}, M. and {Dodonov}, S.~N. and {Grokhovskaya}, A.~A. and {Kotov}, S.~S. and {Bikmaev}, I.~F. and {Zhuchkov}, R. Ya and {Gumerov}, R.~I. and {van Velzen}, S. and {Kulkarni}, S.},
        title = "{First tidal disruption events discovered by SRG/eROSITA: X-ray/optical properties and X-ray luminosity function at z < 0.6}",
      journal = {\mnras},
         year = 2021,
        month = dec,
       volume = {508},
       number = {3},
        pages = {3820-3847},
          doi = {10.1093/mnras/stab2843},
archivePrefix = {arXiv},
       eprint = {2108.02449},
 primaryClass = {astro-ph.HE},
       adsurl = {https://ui.adsabs.harvard.edu/abs/2021MNRAS.508.3820S}
}

@INCOLLECTION{gilfanov10,
       author = {{Gilfanov}, M.},
        title = "{X-Ray Emission from Black-Hole Binaries}",
    booktitle = {The Jet Paradigm},
    publisher = {Springer},
    address   = {Berlin},
         year = 2010,
       editor = {{Belloni}, Tomaso},
       volume = {794},
        pages = {17},
          doi = {10.1007/978-3-540-76937-8_2},
       adsurl = {https://ui.adsabs.harvard.edu/abs/2010LNP...794...17G}
}

@INCOLLECTION{Belloni2010,
       author = {{Belloni}, T.~M.},
        title = "{States and Transitions in Black Hole Binaries}",
    booktitle = {The Jet Paradigm},
    publisher = {Springer},
    address   = {Berlin},
         year = 2010,
       editor = {{Belloni}, Tomaso},
       volume = {794},
        pages = {53},
          doi = {10.1007/978-3-540-76937-8_3},
       adsurl = {https://ui.adsabs.harvard.edu/abs/2010LNP...794...53B}
}

@ARTICLE{dorazio19,
       author = {{D'Orazio}, Daniel J. and {Loeb}, Abraham and {Guillochon}, James},
        title = "{Constraining the stellar mass function from the deficiency of tidal disruption flares in the nuclei of massive galaxies}",
      journal = {\mnras},
         year = 2019,
        month = may,
       volume = {485},
       number = {3},
        pages = {4413-4422},
          doi = {10.1093/mnras/stz652},
archivePrefix = {arXiv},
       eprint = {1807.00029},
 primaryClass = {astro-ph.HE},
       adsurl = {https://ui.adsabs.harvard.edu/abs/2019MNRAS.485.4413D}
}

@ARTICLE{Mum21,
       author = {{Mummery}, Andrew},
        title = "{A maximum X-ray luminosity scale of disc-dominated tidal destruction events}",
      journal = {\mnras},
         year = 2021,
        month = jul,
       volume = {504},
       number = {4},
        pages = {5144-5154},
          doi = {10.1093/mnras/stab1187},
archivePrefix = {arXiv},
       eprint = {2104.06203},
 primaryClass = {astro-ph.HE},
       adsurl = {https://ui.adsabs.harvard.edu/abs/2021MNRAS.504.5144M}
}

@ARTICLE{MumBalb21a,
       author = {{Mummery}, Andrew and {Balbus}, Steven A.},
        title = "{An upper observable black hole mass scale for tidal destruction events with thermal X-ray spectra}",
      journal = {\mnras},
         year = 2021,
        month = aug,
       volume = {505},
       number = {2},
        pages = {1629-1644},
          doi = {10.1093/mnras/stab1141},
archivePrefix = {arXiv},
       eprint = {2104.06177},
 primaryClass = {astro-ph.HE},
       adsurl = {https://ui.adsabs.harvard.edu/abs/2021MNRAS.505.1629M}
}

@ARTICLE{MumBalb21b,
       author = {{Mummery}, Andrew and {Balbus}, Steven A.},
        title = "{Hard X-ray emission from a Compton scattering corona in large black hole mass tidal disruption events}",
      journal = {\mnras},
         year = 2021,
        month = jul,
       volume = {504},
       number = {4},
        pages = {4730-4742},
          doi = {10.1093/mnras/stab1184},
archivePrefix = {arXiv},
       eprint = {2104.06195},
 primaryClass = {astro-ph.HE},
       adsurl = {https://ui.adsabs.harvard.edu/abs/2021MNRAS.504.4730M}
}

@ARTICLE{coughlin22,
       author = {{Coughlin}, Eric R. and {Nixon}, C.~J.},
        title = "{On the Impact of Relativistic Gravity on the Rate of Tidal Disruption Events}",
      journal = {\apj},
         year = 2022,
        month = sep,
       volume = {936},
       number = {1},
          eid = {70},
        pages = {70},
          doi = {10.3847/1538-4357/ac85b3},
archivePrefix = {arXiv},
       eprint = {2207.14301},
 primaryClass = {astro-ph.HE},
       adsurl = {https://ui.adsabs.harvard.edu/abs/2022ApJ...936...70C}
}

@ARTICLE{lacy82,
       author = {{Lacy}, J.~H. and {Townes}, C.~H. and {Hollenbach}, D.~J.},
        title = "{The nature of the central parsec of the Galaxy}",
      journal = {\apj},
         year = 1982,
        month = nov,
       volume = {262},
        pages = {120-134},
          doi = {10.1086/160402},
       adsurl = {https://ui.adsabs.harvard.edu/abs/1982ApJ...262..120L}
}

@INPROCEEDINGS{Gabriel_04,
       author = {{Gabriel}, C. and {Denby}, M. and {Fyfe}, D.~J. and {Hoar}, J. and {Ibarra}, A. and {Ojero}, E. and {Osborne}, J. and {Saxton}, R.~D. and {Lammers}, U. and {Vacanti}, G.},
        title = "{The XMM-Newton SAS - Distributed Development and Maintenance of a Large Science Analysis System: A Critical Analysis}",
    booktitle = {Astronomical Data Analysis Software and Systems (ADASS) XIII},
         year = 2004,
       editor = {{Ochsenbein}, Francois and {Allen}, Mark G. and {Egret}, Daniel},
       series = {Astronomical Society of the Pacific Conference Series},
       volume = {314},
        month = jul,
        pages = {759},
       adsurl = {https://ui.adsabs.harvard.edu/abs/2004ASPC..314..759G}
}

@ARTICLE{Cenko2016,
       author = {{Cenko}, S. Bradley and {Cucchiara}, Antonino and {Roth}, Nathaniel and {Veilleux}, Sylvain and {Prochaska}, J. Xavier and {Yan}, Lin and {Guillochon}, James and {Maksym}, W. Peter and {Arcavi}, Iair and {Butler}, Nathaniel R. and {Filippenko}, Alexei V. and {Fruchter}, Andrew S. and {Gezari}, Suvi and {Kasen}, Daniel and {Levan}, Andrew J. and {Miller}, Jon M. and {Pasham}, Dheeraj R. and {Ramirez-Ruiz}, Enrico and {Strubbe}, Linda E. and {Tanvir}, Nial R. and {Tombesi}, Francesco},
        title = "{An Ultraviolet Spectrum of the Tidal Disruption Flare ASASSN-14li}",
      journal = {\apjl},
         year = 2016,
        month = feb,
       volume = {818},
       number = {2},
          eid = {L32},
        pages = {L32},
          doi = {10.3847/2041-8205/818/2/L32},
archivePrefix = {arXiv},
       eprint = {1601.03331},
 primaryClass = {astro-ph.HE},
       adsurl = {https://ui.adsabs.harvard.edu/abs/2016ApJ...818L..32C}
}

@ARTICLE{Komossa1999,
       author = {{Komossa}, Stefanie and {Bade}, Norbert},
        title = "{The giant X-ray outbursts in NGC 5905 and IC 3599:() hfill Follow-up observations and outburst scenarios}",
      journal = {\aap},
         year = 1999,
        month = mar,
       volume = {343},
        pages = {775-787},
          doi = {10.48550/arXiv.astro-ph/9901141},
archivePrefix = {arXiv},
       eprint = {astro-ph/9901141},
 primaryClass = {astro-ph},
       adsurl = {https://ui.adsabs.harvard.edu/abs/1999A&A...343..775K}
}

@ARTICLE{Kormendy2013,
       author = {{Kormendy}, John and {Ho}, Luis C.},
        title = "{Coevolution (Or Not) of Supermassive Black Holes and Host Galaxies}",
      journal = {\araa},
         year = 2013,
        month = aug,
       volume = {51},
       number = {1},
        pages = {511-653},
          doi = {10.1146/annurev-astro-082708-101811},
archivePrefix = {arXiv},
       eprint = {1304.7762},
 primaryClass = {astro-ph.CO},
       adsurl = {https://ui.adsabs.harvard.edu/abs/2013ARA&A..51..511K}
}

@INPROCEEDINGS{Steele2004,
       author = {{Steele}, Iain A. and {Smith}, Robert J. and {Rees}, Paul C. and
         {Baker}, Ian P. and {Bates}, S.~D. and {Bode}, Michael F. and
         {Bowman}, Mark K. and {Carter}, Dave and {Etherton}, Jason and
         {Ford}, Martyn J.},
        title = "{The Liverpool Telescope: performance and first results}",
    booktitle = {Ground-based Telescopes},
         year = "2004",
       editor = {{Oschmann}, Jacobus M., Jr.},
       series = {Society of Photo-Optical Instrumentation Engineers (SPIE) Conference Series},
       volume = {5489},
        month = "Oct",
        pages = {679-692},
          doi = {10.1117/12.551456},
       adsurl = {https://ui.adsabs.harvard.edu/abs/2004SPIE.5489..679S}
}

@ARTICLE{Eftekhari2024,
       author = {{Eftekhari}, T. and {Tchekhovskoy}, A. and {Alexander}, K.~D. and {Berger}, E. and {Chornock}, R. and {Laskar}, T. and {Margutti}, R. and {Yao}, Y. and {Cendes}, Y. and {Gomez}, S. and {Hajela}, A. and {Pasham}, D.~R.},
        title = "{Late-time X-Ray Observations of the Jetted Tidal Disruption Event AT2022cmc: The Relativistic Jet Shuts Off}",
      journal = {\apj},
         year = 2024,
        month = oct,
       volume = {974},
       number = {2},
          eid = {149},
        pages = {149},
          doi = {10.3847/1538-4357/ad72ea},
archivePrefix = {arXiv},
       eprint = {2404.10036},
 primaryClass = {astro-ph.HE},
       adsurl = {https://ui.adsabs.harvard.edu/abs/2024ApJ...974..149E}
}

@ARTICLE{Shakura1973,
       author = {{Shakura}, N.~I. and {Sunyaev}, R.~A.},
        title = "{Black holes in binary systems. Observational appearance.}",
      journal = {\aap},
         year = 1973,
        month = jan,
       volume = {24},
        pages = {337-355},
       adsurl = {https://ui.adsabs.harvard.edu/abs/1973A&A....24..337S}
}

@ARTICLE{Kara2025,
       author = {{Kara}, Erin and {Garc{\'\i}a}, Javier},
        title = "{Supermassive Black Holes in X-Rays: From Standard Accretion to Extreme Transients}",
      journal = {\araa},
         year = 2025,
        month = aug,
       volume = {63},
       number = {1},
        pages = {379-430},
          doi = {10.1146/annurev-astro-071221-052844},
archivePrefix = {arXiv},
       eprint = {2503.22791},
 primaryClass = {astro-ph.HE},
       adsurl = {https://ui.adsabs.harvard.edu/abs/2025ARA&A..63..379K}
}

@ARTICLE{Dunn2010,
       author = {{Dunn}, R.~J.~H. and {Fender}, R.~P. and {K{\"o}rding}, E.~G. and {Belloni}, T. and {Cabanac}, C.},
        title = "{A global spectral study of black hole X-ray binaries}",
      journal = {\mnras},
         year = 2010,
        month = mar,
       volume = {403},
       number = {1},
        pages = {61-82},
          doi = {10.1111/j.1365-2966.2010.16114.x},
archivePrefix = {arXiv},
       eprint = {0912.0142},
 primaryClass = {astro-ph.HE},
       adsurl = {https://ui.adsabs.harvard.edu/abs/2010MNRAS.403...61D}
}

@ARTICLE{Maccarone2003,
       author = {{Maccarone}, T.~J.},
        title = "{Do X-ray binary spectral state transition luminosities vary?}",
      journal = {\aap},
         year = 2003,
        month = oct,
       volume = {409},
        pages = {697-706},
          doi = {10.1051/0004-6361:20031146},
archivePrefix = {arXiv},
       eprint = {astro-ph/0308036},
 primaryClass = {astro-ph},
       adsurl = {https://ui.adsabs.harvard.edu/abs/2003A&A...409..697M}
}

@ARTICLE{Wang2011,
       author = {{Wang}, Ting-Gui and {Zhou}, Hong-Yan and {Wang}, Li-Fan and {Lu}, Hong-Lin and {Xu}, Dawei},
        title = "{Transient Superstrong Coronal Lines and Broad Bumps in the Galaxy SDSS J074820.67+471214.3}",
      journal = {\apj},
         year = 2011,
        month = oct,
       volume = {740},
       number = {2},
          eid = {85},
        pages = {85},
          doi = {10.1088/0004-637X/740/2/85},
archivePrefix = {arXiv},
       eprint = {1108.2790},
 primaryClass = {astro-ph.CO},
       adsurl = {https://ui.adsabs.harvard.edu/abs/2011ApJ...740...85W}
}

@ARTICLE{Wang2012,
       author = {{Wang}, Ting-Gui and {Zhou}, Hong-Yan and {Komossa}, S. and {Wang}, Hui-Yuan and {Yuan}, Weimin and {Yang}, Chenwei},
        title = "{Extreme Coronal Line Emitters: Tidal Disruption of Stars by Massive Black Holes in Galactic Nuclei?}",
      journal = {\apj},
         year = 2012,
        month = apr,
       volume = {749},
       number = {2},
          eid = {115},
        pages = {115},
          doi = {10.1088/0004-637X/749/2/115},
archivePrefix = {arXiv},
       eprint = {1202.1064},
 primaryClass = {astro-ph.HE},
       adsurl = {https://ui.adsabs.harvard.edu/abs/2012ApJ...749..115W}
}

@ARTICLE{Komossa2008,
       author = {{Komossa}, S. and {Zhou}, H. and {Wang}, T. and {Ajello}, M. and {Ge}, J. and {Greiner}, J. and {Lu}, H. and {Salvato}, M. and {Saxton}, R. and {Shan}, H. and {Xu}, D. and {Yuan}, W.},
        title = "{Discovery of Superstrong, Fading, Iron Line Emission and Double-peaked Balmer Lines of the Galaxy SDSS J095209.56+214313.3: The Light Echo of a Huge Flare}",
      journal = {\apjl},
         year = 2008,
        month = may,
       volume = {678},
       number = {1},
        pages = {L13},
          doi = {10.1086/588281},
archivePrefix = {arXiv},
       eprint = {0804.2670},
 primaryClass = {astro-ph},
       adsurl = {https://ui.adsabs.harvard.edu/abs/2008ApJ...678L..13K}
}

@INPROCEEDINGS{Phinney1989,
       author = {{Phinney}, E.~S.},
        title = "{Manifestations of a Massive Black Hole in the Galactic Center}",
    booktitle = {The Center of the Galaxy},
         year = 1989,
       editor = {{Morris}, Mark},
       series = {IAU Symposium},
       volume = {136},
        month = jan,
        pages = {543},
       adsurl = {https://ui.adsabs.harvard.edu/abs/1989IAUS..136..543P}
}

@ARTICLE{Rees1988,
       author = {{Rees}, Martin J.},
        title = "{Tidal disruption of stars by black holes of {}10$^{6}$-{}10$^{8}$ solar masses in nearby galaxies}",
      journal = {\nat},
         year = 1988,
        month = jun,
       volume = {333},
       number = {6173},
        pages = {523-528},
          doi = {10.1038/333523a0},
       adsurl = {https://ui.adsabs.harvard.edu/abs/1988Natur.333..523R}
}

@ARTICLE{Auchettl2017,
       author = {{Auchettl}, Katie and {Guillochon}, James and {Ramirez-Ruiz}, Enrico},
        title = "{New Physical Insights about Tidal Disruption Events from a Comprehensive Observational Inventory at X-Ray Wavelengths}",
      journal = {\apj},
         year = 2017,
        month = apr,
       volume = {838},
       number = {2},
          eid = {149},
        pages = {149},
          doi = {10.3847/1538-4357/aa633b},
archivePrefix = {arXiv},
       eprint = {1611.02291},
 primaryClass = {astro-ph.HE},
       adsurl = {https://ui.adsabs.harvard.edu/abs/2017ApJ...838..149A}
}

@ARTICLE{vanVelzen2021,
       author = {{van Velzen}, Sjoert and {Gezari}, Suvi and {Hammerstein}, Erica and {Roth}, Nathaniel and {Frederick}, Sara and {Ward}, Charlotte and {Hung}, Tiara and {Cenko}, S. Bradley and {Stein}, Robert and {Perley}, Daniel A. and {Taggart}, Kirsty and {Foley}, Ryan J. and {Sollerman}, Jesper and {Blagorodnova}, Nadejda and {Andreoni}, Igor and {Bellm}, Eric C. and {Brinnel}, Valery and {De}, Kishalay and {Dekany}, Richard and {Feeney}, Michael and {Fremling}, Christoffer and {Giomi}, Matteo and {Golkhou}, V. Zach and {Graham}, Matthew J. and {Ho}, Anna. Y.~Q. and {Kasliwal}, Mansi M. and {Kilpatrick}, Charles D. and {Kulkarni}, Shrinivas R. and {Kupfer}, Thomas and {Laher}, Russ R. and {Mahabal}, Ashish and {Masci}, Frank J. and {Miller}, Adam A. and {Nordin}, Jakob and {Riddle}, Reed and {Rusholme}, Ben and {van Santen}, Jakob and {Sharma}, Yashvi and {Shupe}, David L. and {Soumagnac}, Maayane T.},
        title = "{Seventeen Tidal Disruption Events from the First Half of ZTF Survey Observations: Entering a New Era of Population Studies}",
      journal = {\apj},
         year = 2021,
        month = feb,
       volume = {908},
       number = {1},
          eid = {4},
        pages = {4},
          doi = {10.3847/1538-4357/abc258},
archivePrefix = {arXiv},
       eprint = {2001.01409},
 primaryClass = {astro-ph.HE},
       adsurl = {https://ui.adsabs.harvard.edu/abs/2021ApJ...908....4V}
}

@ARTICLE{Remillard2006,
       author = {{Remillard}, Ronald A. and {McClintock}, Jeffrey E.},
        title = "{X-Ray Properties of Black-Hole Binaries}",
      journal = {\araa},
         year = 2006,
        month = sep,
       volume = {44},
       number = {1},
        pages = {49-92},
          doi = {10.1146/annurev.astro.44.051905.092532},
archivePrefix = {arXiv},
       eprint = {astro-ph/0606352},
 primaryClass = {astro-ph},
       adsurl = {https://ui.adsabs.harvard.edu/abs/2006ARA&A..44...49R}
}

@INPROCEEDINGS{Gendreau2016,
       author = {{Gendreau}, Keith C. and {Arzoumanian}, Zaven and {Adkins}, Phillip W. and
         {Albert}, Cheryl L. and {Anders}, John F. and {Aylward}, Andrew T. and
         {Baker}, Charles L. and {Balsamo}, Erin R. and {Bamford}, William A. and
         {Benegalrao}, Suyog S. and {Berry}, Daniel L. and {Bhalwani}, Shiraz and
         {Black}, J. Kevin and {Blaurock}, Carl and {Bronke}, Ginger M. and
         {Brown}, Gary L. and {Budinoff}, Jason G. and {Cantwell}, Jeffrey D. and
         {Cazeau}, Thoniel and {Chen}, Philip T. and {Clement}, Thomas G. and
         {Colangelo}, Andrew T. and {Coleman}, Jerry S. and
         {Coopersmith}, Jonathan D. and {Dehaven}, William E. and
         {Doty}, John P. and {Egan}, Mark D. and {Enoto}, Teruaki and
         {Fan}, Terry W. and {Ferro}, Deneen M. and {Foster}, Richard and
         {Galassi}, Nicholas M. and {Gallo}, Luis D. and {Green}, Chris M. and
         {Grosh}, Dave and {Ha}, Kong Q. and {Hasouneh}, Monther A. and
         {Heefner}, Kristofer B. and {Hestnes}, Phyllis and {Hoge}, Lisa J. and
         {Jacobs}, Tawanda M. and {J{\o}rgensen}, John L. and
         {Kaiser}, Michael A. and {Kellogg}, James W. and {Kenyon}, Steven J. and
         {Koenecke}, Richard G. and {Kozon}, Robert P. and {LaMarr}, Beverly and
         {Lambertson}, Mike D. and {Larson}, Anne M. and {Lentine}, Steven and
         {Lewis}, Jesse H. and {Lilly}, Michael G. and {Liu}, Kuochia Alice and
         {Malonis}, Andrew and {Manthripragada}, Sridhar S. and
         {Markwardt}, Craig B. and {Matonak}, Bryan D. and {Mcginnis}, Isaac E. and
         {Miller}, Roger L. and {Mitchell}, Alissa L. and {Mitchell}, Jason W. and
         {Mohammed}, Jelila S. and {Monroe}, Charles A. and
         {Montt de Garcia}, Kristina M. and {Mul{\'e}}, Peter D. and
         {Nagao}, Louis T. and {Ngo}, Son N. and {Norris}, Eric D. and
         {Norwood}, Dwight A. and {Novotka}, Joseph and {Okajima}, Takashi and
         {Olsen}, Lawrence G. and {Onyeachu}, Chimaobi O. and
         {Orosco}, Henry Y. and {Peterson}, Jacqualine R. and
         {Pevear}, Kristina N. and {Pham}, Karen K. and {Pollard}, Sue E. and
         {Pope}, John S. and {Powers}, Daniel F. and {Powers}, Charles E. and
         {Price}, Samuel R. and {Prigozhin}, Gregory Y. and
         {Ramirez}, Julian B. and {Reid}, Winston J. and {Remillard}, Ronald A. and
         {Rogstad}, Eric M. and {Rosecrans}, Glenn P. and {Rowe}, John N. and
         {Sager}, Jennifer A. and {Sanders}, Claude A. and {Savadkin}, Bruce and
         {Saylor}, Maxine R. and {Schaeffer}, Alexander F. and
         {Schweiss}, Nancy S. and {Semper}, Sean R. and {Serlemitsos}, Peter J. and
         {Shackelford}, Larry V. and {Soong}, Yang and {Struebel}, Jonathan and
         {Vezie}, Michael L. and {Villasenor}, Joel S. and
         {Winternitz}, Luke B. and {Wofford}, George I. and
         {Wright}, Michael R. and {Yang}, Mike Y. and {Yu}, Wayne H.},
        title = "{The Neutron star Interior Composition Explorer (NICER): design and development}",
    booktitle = {Space Telescopes and Instrumentation 2016: Ultraviolet to Gamma Ray},
         year = 2016,
       series = {Society of Photo-Optical Instrumentation Engineers (SPIE) Conference Series},
       volume = {9905},
        month = jul,
          eid = {99051H},
        pages = {99051H},
          doi = {10.1117/12.2231304},
       adsurl = {https://ui.adsabs.harvard.edu/abs/2016SPIE.9905E..1HG}
}

@article{pypeit:joss_pub,
    doi = {10.21105/joss.02308},
    url = {https://doi.org/10.21105/joss.02308},
    year = {2020},
    publisher = {The Open Journal},
    volume = {5},
    number = {56},
    pages = {2308},
    author = {J. Xavier Prochaska and Joseph F. Hennawi and Kyle B. Westfall and Ryan J. Cooke and Feige Wang and Tiffany Hsyu and Frederick B. Davies and Emanuele Paolo Farina and Debora Pelliccia},
    title = {PypeIt: The Python Spectroscopic Data Reduction Pipeline},
    journal = {Journal of Open Source Software}
}

@ARTICLE{Cash1979,
       author = {{Cash}, W.},
        title = "{Parameter estimation in astronomy through application of the likelihood ratio.}",
      journal = {\apj},
         year = 1979,
        month = mar,
       volume = {228},
        pages = {939-947},
          doi = {10.1086/156922},
       adsurl = {https://ui.adsabs.harvard.edu/abs/1979ApJ...228..939C}
}

@ARTICLE{Foreman-Mackey2013,
       author = {{Foreman-Mackey}, Daniel and {Hogg}, David W. and {Lang}, Dustin and
         {Goodman}, Jonathan},
        title = "{emcee: The MCMC Hammer}",
      journal = {Publications of the Astronomical Society of the Pacific},
         year = "2013",
        month = "Mar",
       volume = {125},
       number = {925},
        pages = {306},
          doi = {10.1086/670067},
archivePrefix = {arXiv},
       eprint = {1202.3665},
 primaryClass = {astro-ph.IM},
       adsurl = {https://ui.adsabs.harvard.edu/abs/2013PASP..125..306F}
}

@ARTICLE{Bardeen1972,
       author = {{Bardeen}, James M. and {Press}, William H. and {Teukolsky}, Saul A.},
        title = "{Rotating Black Holes: Locally Nonrotating Frames, Energy Extraction, and Scalar Synchrotron Radiation}",
      journal = {\apj},
         year = 1972,
        month = dec,
       volume = {178},
        pages = {347-370},
          doi = {10.1086/151796},
       adsurl = {https://ui.adsabs.harvard.edu/abs/1972ApJ...178..347B}
}

@ARTICLE{Mummery2025_FitTeD,
       author = {{Mummery}, Andrew and {Nathan}, Edward and {Ingram}, Adam and {Gardner}, M.},
        title = "{Fitting transients with discs (FitTeD): a public light curve and spectral fitting package based on evolving relativistic discs}",
      journal = {\mnras},
         year = 2025,
        month = sep,
          doi = {10.1093/mnras/staf1565},
archivePrefix = {arXiv},
       eprint = {2408.15048},
 primaryClass = {astro-ph.HE},
       adsurl = {https://ui.adsabs.harvard.edu/abs/2025MNRAS.tmp.1519M}
}

@ARTICLE{Huang2023_stream_collision,
       author = {{Huang}, Xiaoshan and {Davis}, Shane W. and {Jiang}, Yan-fei},
        title = "{A Bright First Day for Tidal Disruption Events}",
      journal = {\apj},
         year = 2023,
        month = aug,
       volume = {953},
       number = {1},
          eid = {117},
        pages = {117},
          doi = {10.3847/1538-4357/ace0be},
archivePrefix = {arXiv},
       eprint = {2303.17443},
 primaryClass = {astro-ph.HE},
       adsurl = {https://ui.adsabs.harvard.edu/abs/2023ApJ...953..117H}
}

@ARTICLE{Huang2024_stream_collision,
       author = {{Huang}, Xiaoshan and {Davis}, Shane W. and {Jiang}, Yan-fei},
        title = "{Pre-peak Emission in Tidal Disruption Events}",
      journal = {\apj},
         year = 2024,
        month = oct,
       volume = {974},
       number = {2},
          eid = {165},
        pages = {165},
          doi = {10.3847/1538-4357/ad6c39},
archivePrefix = {arXiv},
       eprint = {2404.18446},
 primaryClass = {astro-ph.HE},
       adsurl = {https://ui.adsabs.harvard.edu/abs/2024ApJ...974..165H}
}

@ARTICLE{Guo2025_tde_reverberation,
       author = {{Guo}, Hengxiao and {Sun}, Jingbo and {Li}, Shuangliang and {Jiang}, Yan-Fei and {Wang}, Tinggui and {Bu}, Defu and {Jiang}, Ning and {Wang}, Yanan and {Yao}, Yuhan and {Shen}, Rongfeng and {Gu}, Minfeng and {Sun}, Mouyuan},
        title = "{Reverberation Evidence for Stream Collision and Delayed Disk Formation in Tidal Disruption Events}",
      journal = {\apj},
         year = 2025,
        month = feb,
       volume = {979},
       number = {2},
          eid = {235},
        pages = {235},
          doi = {10.3847/1538-4357/ada274},
archivePrefix = {arXiv},
       eprint = {2312.06771},
 primaryClass = {astro-ph.HE},
       adsurl = {https://ui.adsabs.harvard.edu/abs/2025ApJ...979..235G}
}

@ARTICLE{Mummery2024_fundamental_scaling,
       author = {{Mummery}, Andrew and {van Velzen}, Sjoert and {Nathan}, Edward and {Ingram}, Adam and {Hammerstein}, Erica and {Fraser-Taliente}, Ludovic and {Balbus}, Steven},
        title = "{Fundamental scaling relationships revealed in the optical light curves of tidal disruption events}",
      journal = {\mnras},
         year = 2024,
        month = jan,
       volume = {527},
       number = {2},
        pages = {2452-2489},
          doi = {10.1093/mnras/stad3001},
archivePrefix = {arXiv},
       eprint = {2308.08255},
 primaryClass = {astro-ph.HE},
       adsurl = {https://ui.adsabs.harvard.edu/abs/2024MNRAS.527.2452M}
}

@ARTICLE{Alush2025_plateau,
       author = {{Alush}, Yael and {Stone}, Nicholas C. and {van Velzen}, Sjoert},
        title = "{How Flat is a Plateau? Evolution of Late-Time TDE Disks}",
      journal = {arXiv e-prints},
         year = 2025,
        month = oct,
          eid = {arXiv:2510.24696},
        pages = {arXiv:2510.24696},
          doi = {10.48550/arXiv.2510.24696},
archivePrefix = {arXiv},
       eprint = {2510.24696},
 primaryClass = {astro-ph.HE},
       adsurl = {https://ui.adsabs.harvard.edu/abs/2025arXiv251024696A}
}

@ARTICLE{Sarin2024_TDE,
       author = {{Sarin}, Nikhil and {Metzger}, Brian D.},
        title = "{Tidal Disruption Events through the Lens of the Cooling Envelope Model}",
      journal = {\apjl},
         year = 2024,
        month = jan,
       volume = {961},
       number = {1},
          eid = {L19},
        pages = {L19},
          doi = {10.3847/2041-8213/ad16d8},
archivePrefix = {arXiv},
       eprint = {2307.15121},
 primaryClass = {astro-ph.HE},
       adsurl = {https://ui.adsabs.harvard.edu/abs/2024ApJ...961L..19S}
}

@ARTICLE{Mummery2020_14li,
       author = {{Mummery}, Andrew and {Balbus}, Steven A.},
        title = "{The spectral evolution of disc dominated tidal disruption events}",
      journal = {\mnras},
         year = 2020,
        month = mar,
       volume = {492},
       number = {4},
        pages = {5655-5674},
          doi = {10.1093/mnras/staa192},
archivePrefix = {arXiv},
       eprint = {1912.06577},
 primaryClass = {astro-ph.HE},
       adsurl = {https://ui.adsabs.harvard.edu/abs/2020MNRAS.492.5655M}
}

@ARTICLE{Laskar2014,
       author = {{Laskar}, Tanmoy and {Berger}, Edo and {Tanvir}, Nial and {Zauderer}, B. Ashley and {Margutti}, Raffaella and {Levan}, Andrew and {Perley}, Daniel and {Fong}, Wen-fai and {Wiersema}, Klaas and {Menten}, Karl and {Hrudkova}, Marie},
        title = "{GRB 120521C at z \raisebox{-0.5ex}\textasciitilde 6 and the Properties of High-redshift {\ensuremath{\gamma}}-Ray Bursts}",
      journal = {\apj},
         year = 2014,
        month = jan,
       volume = {781},
       number = {1},
          eid = {1},
        pages = {1},
          doi = {10.1088/0004-637X/781/1/1},
archivePrefix = {arXiv},
       eprint = {1307.6586},
 primaryClass = {astro-ph.HE},
       adsurl = {https://ui.adsabs.harvard.edu/abs/2014ApJ...781....1L}
}

@ARTICLE{Mummery2024_spin,
       author = {{Mummery}, Andrew},
        title = "{The maximum mass of a black hole which can tidally disrupt a star: measuring black hole spins with tidal disruption events}",
      journal = {\mnras},
         year = 2024,
        month = jan,
       volume = {527},
       number = {3},
        pages = {6233-6252},
          doi = {10.1093/mnras/stad3636},
archivePrefix = {arXiv},
       eprint = {2312.00557},
 primaryClass = {gr-qc},
       adsurl = {https://ui.adsabs.harvard.edu/abs/2024MNRAS.527.6233M}
}

@ARTICLE{DESI2022_instrument,
       author = {{DESI Collaboration} and {Abareshi}, B. and {Aguilar}, J. and {Ahlen}, S. and {Alam}, Shadab and {Alexander}, David M. and {Alfarsy}, R. and {Allen}, L. and {Allende Prieto}, C. and {Alves}, O. and {Ameel}, J. and {Armengaud}, E. and {Asorey}, J. and {Aviles}, Alejandro and {Bailey}, S. and {Balaguera-Antol{\'\i}nez}, A. and {Ballester}, O. and {Baltay}, C. and {Bault}, A. and {Beltran}, S.~F. and {Benavides}, B. and {BenZvi}, S. and {Berti}, A. and {Besuner}, R. and {Beutler}, Florian and {Bianchi}, D. and {Blake}, C. and {Blanc}, P. and {Blum}, R. and {Bolton}, A. and {Bose}, S. and {Bramall}, D. and {Brieden}, S. and {Brodzeller}, A. and {Brooks}, D. and {Brownewell}, C. and {Buckley-Geer}, E. and {Cahn}, R.~N. and {Cai}, Z. and {Canning}, R. and {Capasso}, R. and {Carnero Rosell}, A. and {Carton}, P. and {Casas}, R. and {Castander}, F.~J. and {Cervantes-Cota}, J.~L. and {Chabanier}, S. and {Chaussidon}, E. and {Chuang}, C. and {Circosta}, C. and {Cole}, S. and {Cooper}, A.~P. and {da Costa}, L. and {Cousinou}, M. -C. and {Cuceu}, A. and {Davis}, T.~M. and {Dawson}, K. and {de la Cruz-Noriega}, R. and {de la Macorra}, A. and {de Mattia}, A. and {Della Costa}, J. and {Demmer}, P. and {Derwent}, M. and {Dey}, A. and {Dey}, B. and {Dhungana}, G. and {Ding}, Z. and {Dobson}, C. and {Doel}, P. and {Donald-McCann}, J. and {Donaldson}, J. and {Douglass}, K. and {Duan}, Y. and {Dunlop}, P. and {Edelstein}, J. and {Eftekharzadeh}, S. and {Eisenstein}, D.~J. and {Enriquez-Vargas}, M. and {Escoffier}, S. and {Evatt}, M. and {Fagrelius}, P. and {Fan}, X. and {Fanning}, K. and {Fawcett}, V.~A. and {Ferraro}, S. and {Ereza}, J. and {Flaugher}, B. and {Font-Ribera}, A. and {Forero-Romero}, J.~E. and {Frenk}, C.~S. and {Fromenteau}, S. and {G{\"a}nsicke}, B.~T. and {Garcia-Quintero}, C. and {Garrison}, L. and {Gazta{\~n}aga}, E. and {Gerardi}, F. and {Gil-Mar{\'\i}n}, H. and {Gontcho A Gontcho}, S. and {Gonzalez-Morales}, Alma X. and {Gonzalez-de-Rivera}, G. and {Gonzalez-Perez}, V. and {Gordon}, C. and {Graur}, O. and {Green}, D. and {Grove}, C. and {Gruen}, D. and {Gutierrez}, G. and {Guy}, J. and {Hahn}, C. and {Harris}, S. and {Herrera}, D. and {Herrera-Alcantar}, Hiram K. and {Honscheid}, K. and {Howlett}, C. and {Huterer}, D. and {Ir{\v{s}}i{\v{c}}}, V. and {Ishak}, M. and {Jelinsky}, P. and {Jiang}, L. and {Jimenez}, J. and {Jing}, Y.~P. and {Joyce}, R. and {Jullo}, E. and {Juneau}, S. and {Kara{\c{c}}ayl{\i}}, N.~G. and {Karamanis}, M. and {Karcher}, A. and {Karim}, T. and {Kehoe}, R. and {Kent}, S. and {Kirkby}, D. and {Kisner}, T. and {Kitaura}, F. and {Koposov}, S.~E. and {Kov{\'a}cs}, A. and {Kremin}, A. and {Krolewski}, Alex and {L'Huillier}, B. and {Lahav}, O. and {Lambert}, A. and {Lamman}, C. and {Lan}, Ting-Wen and {Landriau}, M. and {Lane}, S. and {Lang}, D. and {Lange}, J.~U. and {Lasker}, J. and {Le Guillou}, L. and {Leauthaud}, A. and {Le Van Suu}, A. and {Levi}, Michael E. and {Li}, T.~S. and {Magneville}, C. and {Manera}, M. and {Manser}, Christopher J. and {Marshall}, B. and {Martini}, Paul and {McCollam}, W. and {McDonald}, P. and {Meisner}, Aaron M. and {Mena-Fern{\'a}ndez}, J. and {Meneses-Rizo}, J. and {Mezcua}, M. and {Miller}, T. and {Miquel}, R. and {Montero-Camacho}, P. and {Moon}, J. and {Moustakas}, J. and {Mueller}, E. and {Mu{\~n}oz-Guti{\'e}rrez}, Andrea and {Myers}, Adam D. and {Nadathur}, S. and {Najita}, J. and {Napolitano}, L. and {Neilsen}, E. and {Newman}, Jeffrey A. and {Nie}, J.~D. and {Ning}, Y. and {Niz}, G. and {Norberg}, P. and {Noriega}, Hern{\'a}n E. and {O'Brien}, T. and {Obuljen}, A. and {Palanque-Delabrouille}, N. and {Palmese}, A. and {Zhiwei}, P. and {Pappalardo}, D. and {PENG}, X. and {Percival}, W.~J. and {Perruchot}, S. and {Pogge}, R. and {Poppett}, C. and {Porredon}, A. and {Prada}, F. and {Prochaska}, J. and {Pucha}, R. and {P{\'e}rez-Fern{\'a}ndez}, A. and {P{\'e}rez-R{\`a}fols}, I. and {Rabinowitz}, D. and {Raichoor}, A.},
        title = "{Overview of the Instrumentation for the Dark Energy Spectroscopic Instrument}",
      journal = {\aj},
         year = 2022,
        month = nov,
       volume = {164},
       number = {5},
          eid = {207},
        pages = {207},
          doi = {10.3847/1538-3881/ac882b},
archivePrefix = {arXiv},
       eprint = {2205.10939},
 primaryClass = {astro-ph.IM},
       adsurl = {https://ui.adsabs.harvard.edu/abs/2022AJ....164..207D}
}

@MISC{Moustakas2023,
  author = {{Moustakas}, John and {Scholte}, Dirk and {Dey}, Biprateep and {Khederlarian}, Ashod},
  title = "{FastSpecFit: Fast spectral synthesis and emission-line fitting of DESI spectra}",
   howpublished = {Astrophysics Source Code Library, record ascl:2308.005},
  year = 2023,
  month = aug,
  eid = {ascl:2308.005},
  pages = {ascl:2308.005},
  archivePrefix = {ascl},
  eprint = {2308.005},
  adsurl = {https://ui.adsabs.harvard.edu/abs/2023ascl.soft08005M}
}

@ARTICLE{Hajela2025,
       author = {{Hajela}, A. and {Alexander}, K.~D. and {Margutti}, R. and {Chornock}, R. and {Bietenholz}, M. and {Christy}, C.~T. and {Stroh}, M. and {Terreran}, G. and {Saxton}, R. and {Komossa}, S. and {Bright}, J.~S. and {Ramirez-Ruiz}, E. and {Coppejans}, D.~L. and {Leung}, J.~K. and {Cendes}, Y. and {Wiston}, E. and {Laskar}, T. and {Horesh}, A. and {Schroeder}, G. and {A.~J.}, Nayana and {Wieringa}, M.~H. and {Velez}, N. and {Berger}, E. and {Blanchard}, P.~K. and {Eftekhari}, T. and {Gomez}, S. and {Nicholl}, M. and {Sears}, H. and {Zauderer}, B.~A.},
        title = "{Eight Years of Light from ASASSN-15oi: Toward Understanding the Late-time Evolution of TDEs}",
      journal = {\apj},
         year = 2025,
        month = apr,
       volume = {983},
       number = {1},
          eid = {29},
        pages = {29},
          doi = {10.3847/1538-4357/adb620},
archivePrefix = {arXiv},
       eprint = {2407.19019},
 primaryClass = {astro-ph.HE},
       adsurl = {https://ui.adsabs.harvard.edu/abs/2025ApJ...983...29H}
}

@ARTICLE{Johnson2021,
       author = {{Johnson}, Benjamin D. and {Leja}, Joel and {Conroy}, Charlie and {Speagle}, Joshua S.},
        title = "{Stellar Population Inference with Prospector}",
      journal = {\apjs},
         year = 2021,
        month = jun,
       volume = {254},
       number = {2},
          eid = {22},
        pages = {22},
          doi = {10.3847/1538-4365/abef67},
archivePrefix = {arXiv},
       eprint = {2012.01426},
 primaryClass = {astro-ph.GA},
       adsurl = {https://ui.adsabs.harvard.edu/abs/2021ApJS..254...22J}
}

@ARTICLE{Berger2026,
       author = {{Berger}, Vera and {Kara}, Erin and {Chakraborty}, Joheen and {Masterson}, Megan and {Burdge}, Kevin},
        title = "{Disk-to-Corona State Transition and Extreme X-ray Variability in the Tidal Disruption Event AT2019teq}",
      journal = {arXiv e-prints},
         year = 2026,
        month = jan,
          eid = {arXiv:2601.04311},
        pages = {arXiv:2601.04311},
          doi = {10.48550/arXiv.2601.04311},
archivePrefix = {arXiv},
       eprint = {2601.04311},
 primaryClass = {astro-ph.HE},
       adsurl = {https://ui.adsabs.harvard.edu/abs/2026arXiv260104311B}
}

@ARTICLE{VahdatMotlagh2019,
       author = {{Vahdat Motlagh}, A. and {Kalemci}, E. and {Maccarone}, T.~J.},
        title = "{Investigating state transition luminosities of Galactic black hole transients in the outburst decay}",
      journal = {\mnras},
         year = 2019,
        month = may,
       volume = {485},
       number = {2},
        pages = {2744-2758},
          doi = {10.1093/mnras/stz569},
archivePrefix = {arXiv},
       eprint = {1903.00837},
 primaryClass = {astro-ph.HE},
       adsurl = {https://ui.adsabs.harvard.edu/abs/2019MNRAS.485.2744V}
}

@ARTICLE{Fender2004,
       author = {{Fender}, R.~P. and {Belloni}, T.~M. and {Gallo}, E.},
        title = "{Towards a unified model for black hole X-ray binary jets}",
      journal = {\mnras},
         year = 2004,
        month = dec,
       volume = {355},
       number = {4},
        pages = {1105-1118},
          doi = {10.1111/j.1365-2966.2004.08384.x},
archivePrefix = {arXiv},
       eprint = {astro-ph/0409360},
 primaryClass = {astro-ph},
       adsurl = {https://ui.adsabs.harvard.edu/abs/2004MNRAS.355.1105F}
}

@MISC{Debes2024,
       author = {{Debes}, John and {Sankrit}, Ravi and {Fischer}, Travis and {Frazer}, Elaine and {Hirschauer}, Alec and {Rowlands}, Kate and {Burger}, Matthew and {Swaters}, Robert and {Jedrzejewski}, Robert and {Gomez}, Sierrra and {Dos Santos}, Leonardo and {Hernandez}, Svea and {Miller}, Lauren and {Payne}, Anna and {Rafelski}, Marc and {Wevers}, Thomas and {Anderson}, Sara and {Bair}, Tom and {Bello}, Kathryn and {Carlberg}, Joleen and {Charlow}, Brian and {Cortese}, Andrew and {Dencheva}, Nadia and {Ellis}, Tracy and {Falk}, Ben and {Fleming}, Scott and {Forshay}, Peter and {Gilani}, Syed and {Hall}, Patty and {Kimball}, Tim and {Kelley}, Talya and {Kidwell}, Richard and {Kotler}, Jenn and {Kovacs}, Aiden and {James}, Bethan and {Rahmani}, Christopher and {Rodriguez}, David and {Roman-Duval}, Julia and {Soderblom}, David and {Sherbert}, Lisa and {Welty}, Dan and {Wolfe}, David},
        title = "{The Hubble Advanced Spectral Product (HASP) Program}",
 howpublished = {Instrument Science Report COS 2024-01, 31 pages},
         year = 2024,
        month = jan,
        pages = {1},
       adsurl = {https://ui.adsabs.harvard.edu/abs/2024cos..rept....1D}
}

@ARTICLE{Woodgate1998,
       author = {{Woodgate}, B.~E. and {Kimble}, R.~A. and {Bowers}, C.~W. and {Kraemer}, S. and {Kaiser}, M.~E. and {Danks}, A.~C. and {Grady}, J.~F. and {Loiacono}, J.~J. and {Brumfield}, M. and {Feinberg}, L. and {Gull}, T.~R. and {Heap}, S.~R. and {Maran}, S.~P. and {Lindler}, D. and {Hood}, D. and {Meyer}, W. and {Vanhouten}, C. and {Argabright}, V. and {Franka}, S. and {Bybee}, R. and {Dorn}, D. and {Bottema}, M. and {Woodruff}, R. and {Michika}, D. and {Sullivan}, J. and {Hetlinger}, J. and {Ludtke}, C. and {Stocker}, R. and {Delamere}, A. and {Rose}, D. and {Becker}, I. and {Garner}, H. and {Timothy}, J.~G. and {Blouke}, M. and {Joseph}, C.~L. and {Hartig}, G. and {Green}, R.~F. and {Jenkins}, E.~B. and {Linsky}, J.~L. and {Hutchings}, J.~B. and {Moos}, H.~W. and {Boggess}, A. and {Roesler}, F. and {Weistrop}, D.},
        title = "{The Space Telescope Imaging Spectrograph Design}",
      journal = {\pasp},
         year = 1998,
        month = oct,
       volume = {110},
       number = {752},
        pages = {1183-1204},
          doi = {10.1086/316243},
       adsurl = {https://ui.adsabs.harvard.edu/abs/1998PASP..110.1183W}
}

@ARTICLE{Matsumoto2021_tde_optical,
       author = {{Matsumoto}, Tatsuya and {Piran}, Tsvi},
        title = "{Limits on mass outflow from optical tidal disruption events}",
      journal = {\mnras},
         year = 2021,
        month = apr,
       volume = {502},
       number = {3},
        pages = {3385-3393},
          doi = {10.1093/mnras/stab240},
archivePrefix = {arXiv},
       eprint = {2009.01240},
 primaryClass = {astro-ph.HE},
       adsurl = {https://ui.adsabs.harvard.edu/abs/2021MNRAS.502.3385M}
}

@ARTICLE{Noda2018,
       author = {{Noda}, Hirofumi and {Done}, Chris},
        title = "{Explaining changing-look AGN with state transition triggered by rapid mass accretion rate drop}",
      journal = {\mnras},
         year = 2018,
        month = nov,
       volume = {480},
       number = {3},
        pages = {3898-3906},
          doi = {10.1093/mnras/sty2032},
archivePrefix = {arXiv},
       eprint = {1805.07873},
 primaryClass = {astro-ph.GA},
       adsurl = {https://ui.adsabs.harvard.edu/abs/2018MNRAS.480.3898N}
}

@ARTICLE{Simard2011,
       author = {{Simard}, Luc and {Mendel}, J. Trevor and {Patton}, David R. and {Ellison}, Sara L. and {McConnachie}, Alan W.},
        title = "{A Catalog of Bulge+disk Decompositions and Updated Photometry for 1.12 Million Galaxies in the Sloan Digital Sky Survey}",
      journal = {\apjs},
         year = 2011,
        month = sep,
       volume = {196},
       number = {1},
          eid = {11},
        pages = {11},
          doi = {10.1088/0067-0049/196/1/11},
archivePrefix = {arXiv},
       eprint = {1107.1518},
 primaryClass = {astro-ph.CO},
       adsurl = {https://ui.adsabs.harvard.edu/abs/2011ApJS..196...11S}
}

@ARTICLE{Du2022,
       author = {{Du}, Peizhi and {Ega{\~n}a-Ugrinovic}, Daniel and {Essig}, Rouven and {Fragione}, Giacomo and {Perna}, Rosalba},
        title = "{Searching for ultra-light bosons and constraining black hole spin distributions with stellar tidal disruption events}",
      journal = {Nature Communications},
         year = 2022,
        month = aug,
       volume = {13},
          eid = {4626},
        pages = {4626},
          doi = {10.1038/s41467-022-32301-4},
archivePrefix = {arXiv},
       eprint = {2202.01215},
 primaryClass = {hep-ph},
       adsurl = {https://ui.adsabs.harvard.edu/abs/2022NatCo..13.4626D}
}

@ARTICLE{silverman2012,
	author = {{Silverman}, Jeffrey M. and {Foley}, Ryan J. and {Filippenko}, Alexei V. and {Ganeshalingam}, Mohan and {Barth}, Aaron J. and {Chornock}, Ryan and {Griffith}, Christopher V. and {Kong}, Jason J. and {Lee}, Nicholas and {Leonard}, Douglas C. and {Matheson}, Thomas and {Miller}, Emily G. and {Steele}, Thea N. and {Barris}, Brian J. and {Bloom}, Joshua S. and {Cobb}, Bethany E. and {Coil}, Alison L. and {Desroches}, Louis-Benoit and {Gates}, Elinor L. and {Ho}, Luis C. and {Jha}, Saurabh W. and {Kandrashoff}, Michael T. and {Li}, Weidong and {Mandel}, Kaisey S. and {Modjaz}, Maryam and {Moore}, Matthew R. and {Mostardi}, Robin E. and {Papenkova}, Marina S. and {Park}, Sung and {Perley}, Daniel A. and {Poznanski}, Dovi and {Reuter}, Cassie A. and {Scala}, James and {Serduke}, Franklin J.~D. and {Shields}, Joseph C. and {Swift}, Brandon J. and {Tonry}, John L. and {Van Dyk}, Schuyler D. and {Wang}, Xiaofeng and {Wong}, Diane S.},
	title = "{Berkeley Supernova Ia Program - I. Observations, data reduction and spectroscopic sample of 582 low-redshift Type Ia supernovae}",
	journal = {\mnras},
	year = 2012,
	month = sep,
	volume = {425},
	number = {3},
	pages = {1789-1818},
	doi = {10.1111/j.1365-2966.2012.21270.x},
	archivePrefix = {arXiv},
	eprint = {1202.2128},
	primaryClass = {astro-ph.CO},
	adsurl = {https://ui.adsabs.harvard.edu/abs/2012MNRAS.425.1789S}
}

@ARTICLE{Yao2025_radio,
       author = {{Yao}, Yuhan and {Alexander}, Kate D. and {Lu}, Wenbin and {Somalwar}, Jean J. and {Ravi}, Vikram and {Chornock}, Ryan and {Margutti}, Raffaella and {Perley}, Daniel A. and {Miller-Jones}, James C.~A. and {Beniamini}, Paz and {Nayana}, A.~J. and {Bloom}, Joshua S. and {Christy}, Collin T. and {Graham}, Matthew J. and {Groom}, Steven L. and {Hammerstein}, Erica and {Helou}, George and {Kasliwal}, Mansi M. and {Kulkarni}, S.~R. and {Laher}, Russ R. and {Mahabal}, Ashish A. and {Neveu}, J{\'e}r{\'e}my and {Riddle}, Reed and {Smith}, Roger and {van Velzen}, Sjoert},
        title = "{Optically Overluminous Tidal Disruption Events: Outflow Properties and Implications for Extremely Relativistic Disruptions}",
      journal = {\apj},
         year = 2025,
        month = nov,
       volume = {993},
       number = {2},
          eid = {198},
        pages = {198},
          doi = {10.3847/1538-4357/ae08b3},
archivePrefix = {arXiv},
       eprint = {2507.06453},
 primaryClass = {astro-ph.HE},
       adsurl = {https://ui.adsabs.harvard.edu/abs/2025ApJ...993..198Y}
}

@ARTICLE{Conroy2009,
       author = {{Conroy}, Charlie and {Gunn}, James E. and {White}, Martin},
        title = "{The Propagation of Uncertainties in Stellar Population Synthesis Modeling. I. The Relevance of Uncertain Aspects of Stellar Evolution and the Initial Mass Function to the Derived Physical Properties of Galaxies}",
      journal = {\apj},
         year = 2009,
        month = jul,
       volume = {699},
       number = {1},
        pages = {486-506},
          doi = {10.1088/0004-637X/699/1/486},
archivePrefix = {arXiv},
       eprint = {0809.4261},
 primaryClass = {astro-ph},
       adsurl = {https://ui.adsabs.harvard.edu/abs/2009ApJ...699..486C}
}

@ARTICLE{Hills1975,
       author = {{Hills}, J.~G.},
        title = "{Possible power source of Seyfert galaxies and QSOs}",
      journal = {\nat},
         year = 1975,
        month = mar,
       volume = {254},
       number = {5498},
        pages = {295-298},
          doi = {10.1038/254295a0},
       adsurl = {https://ui.adsabs.harvard.edu/abs/1975Natur.254..295H}
}

@ARTICLE{Fremling2016,
       author = {{Fremling}, C. and {Sollerman}, J. and {Taddia}, F. and {Ergon}, M. and {Fraser}, M. and {Karamehmetoglu}, E. and {Valenti}, S. and {Jerkstrand}, A. and {Arcavi}, I. and {Bufano}, F. and {Elias Rosa}, N. and {Filippenko}, A.~V. and {Fox}, D. and {Gal-Yam}, A. and {Howell}, D.~A. and {Kotak}, R. and {Mazzali}, P. and {Milisavljevic}, D. and {Nugent}, P.~E. and {Nyholm}, A. and {Pian}, E. and {Smartt}, S.},
        title = "{PTF12os and iPTF13bvn. Two stripped-envelope supernovae from low-mass progenitors in NGC 5806}",
      journal = {\aap},
         year = 2016,
        month = sep,
       volume = {593},
          eid = {A68},
        pages = {A68},
          doi = {10.1051/0004-6361/201628275},
archivePrefix = {arXiv},
       eprint = {1606.03074},
 primaryClass = {astro-ph.HE},
       adsurl = {https://ui.adsabs.harvard.edu/abs/2016A&A...593A..68F}
}

@ARTICLE{Harrison2013,
       author = {{Harrison}, Fiona A. and {Craig}, William W. and {Christensen}, Finn E. and
         {Hailey}, Charles J. and {Zhang}, William W. and {Boggs}, Steven E. and
         {Stern}, Daniel and {Cook}, W. Rick and {Forster}, Karl and
         {Giommi}, Paolo and {Grefenstette}, Brian W. and {Kim}, Yunjin and
         {Kitaguchi}, Takao and {Koglin}, Jason E. and {Madsen}, Kristin K. and
         {Mao}, Peter H. and {Miyasaka}, Hiromasa and {Mori}, Kaya and
         {Perri}, Matteo and {Pivovaroff}, Michael J. and {Puccetti}, Simonetta and
         {Rana}, Vikram R. and {Westergaard}, Niels J. and {Willis}, Jason and
         {Zoglauer}, Andreas and {An}, Hongjun and {Bachetti}, Matteo and
         {Barri{\`e}re}, Nicolas M. and {Bellm}, Eric C. and {Bhalerao}, Varun and
         {Brejnholt}, Nicolai F. and {Fuerst}, Felix and {Liebe}, Carl C. and
         {Markwardt}, Craig B. and {Nynka}, Melania and {Vogel}, Julia K. and
         {Walton}, Dominic J. and {Wik}, Daniel R. and {Alexander}, David M. and
         {Cominsky}, Lynn R. and {Hornschemeier}, Ann E. and {Hornstrup}, Allan and
         {Kaspi}, Victoria M. and {Madejski}, Greg M. and {Matt}, Giorgio and
         {Molendi}, Silvano and {Smith}, David M. and {Tomsick}, John A. and
         {Ajello}, Marco and {Ballantyne}, David R. and {Balokovi{\'c}}, Mislav and
         {Barret}, Didier and {Bauer}, Franz E. and {Blandford}, Roger D. and
         {Brandt}, W. Niel and {Brenneman}, Laura W. and {Chiang}, James and
         {Chakrabarty}, Deepto and {Chenevez}, Jerome and {Comastri}, Andrea and
         {Dufour}, Francois and {Elvis}, Martin and {Fabian}, Andrew C. and
         {Farrah}, Duncan and {Fryer}, Chris L. and {Gotthelf}, Eric V. and
         {Grindlay}, Jonathan E. and {Helfand}, David J. and {Krivonos}, Roman and
         {Meier}, David L. and {Miller}, Jon M. and {Natalucci}, Lorenzo and
         {Ogle}, Patrick and {Ofek}, Eran O. and {Ptak}, Andrew and
         {Reynolds}, Stephen P. and {Rigby}, Jane R. and
         {Tagliaferri}, Gianpiero and {Thorsett}, Stephen E. and
         {Treister}, Ezequiel and {Urry}, C. Megan},
        title = "{The Nuclear Spectroscopic Telescope Array (NuSTAR) High-energy X-Ray Mission}",
      journal = {\apj},
         year = 2013,
        month = jun,
       volume = {770},
       number = {2},
          eid = {103},
        pages = {103},
          doi = {10.1088/0004-637X/770/2/103},
archivePrefix = {arXiv},
       eprint = {1301.7307},
 primaryClass = {astro-ph.IM},
       adsurl = {https://ui.adsabs.harvard.edu/abs/2013ApJ...770..103H}
}

@ARTICLE{Yao2022,
       author = {{Yao}, Yuhan and {Lu}, Wenbin and {Guolo}, Muryel and {Pasham}, Dheeraj R. and {Gezari}, Suvi and {Gilfanov}, Marat and {Gendreau}, Keith C. and {Harrison}, Fiona and {Cenko}, S. Bradley and {Kulkarni}, S.~R. and {Miller}, Jon M. and {Walton}, Dominic J. and {Garc{\'\i}a}, Javier A. and {van Velzen}, Sjoert and {Alexander}, Kate D. and {Miller-Jones}, James C.~A. and {Nicholl}, Matt and {Hammerstein}, Erica and {Medvedev}, Pavel and {Stern}, Daniel and {Ravi}, Vikram and {Sunyaev}, R. and {Bloom}, Joshua S. and {Graham}, Matthew J. and {Kool}, Erik C. and {Mahabal}, Ashish A. and {Masci}, Frank J. and {Purdum}, Josiah and {Rusholme}, Ben and {Sharma}, Yashvi and {Smith}, Roger and {Sollerman}, Jesper},
        title = "{The Tidal Disruption Event AT2021ehb: Evidence of Relativistic Disk Reflection, and Rapid Evolution of the Disk-Corona System}",
      journal = {\apj},
         year = 2022,
        month = sep,
       volume = {937},
       number = {1},
          eid = {8},
        pages = {8},
          doi = {10.3847/1538-4357/ac898a},
archivePrefix = {arXiv},
       eprint = {2206.12713},
 primaryClass = {astro-ph.HE},
       adsurl = {https://ui.adsabs.harvard.edu/abs/2022ApJ...937....8Y}
}

@ARTICLE{Yao2023,
       author = {{Yao}, Yuhan and {Ravi}, Vikram and {Gezari}, Suvi and {van Velzen}, Sjoert and {Lu}, Wenbin and {Schulze}, Steve and {Somalwar}, Jean J. and {Kulkarni}, S.~R. and {Hammerstein}, Erica and {Nicholl}, Matt and {Graham}, Matthew J. and {Perley}, Daniel A. and {Cenko}, S. Bradley and {Stein}, Robert and {Ricarte}, Angelo and {Chadayammuri}, Urmila and {Quataert}, Eliot and {Bellm}, Eric C. and {Bloom}, Joshua S. and {Dekany}, Richard and {Drake}, Andrew J. and {Groom}, Steven L. and {Mahabal}, Ashish A. and {Prince}, Thomas A. and {Riddle}, Reed and {Rusholme}, Ben and {Sharma}, Yashvi and {Sollerman}, Jesper and {Yan}, Lin},
        title = "{Tidal Disruption Event Demographics with the Zwicky Transient Facility: Volumetric Rates, Luminosity Function, and Implications for the Local Black Hole Mass Function}",
      journal = {\apjl},
         year = 2023,
        month = sep,
       volume = {955},
       number = {1},
          eid = {L6},
        pages = {L6},
          doi = {10.3847/2041-8213/acf216},
archivePrefix = {arXiv},
       eprint = {2303.06523},
 primaryClass = {astro-ph.HE},
       adsurl = {https://ui.adsabs.harvard.edu/abs/2023ApJ...955L...6Y}
}

@ARTICLE{Hammerstein2023,
       author = {{Hammerstein}, Erica and {van Velzen}, Sjoert and {Gezari}, Suvi and {Cenko}, S. Bradley and {Yao}, Yuhan and {Ward}, Charlotte and {Frederick}, Sara and {Villanueva}, Natalia and {Somalwar}, Jean J. and {Graham}, Matthew J. and {Kulkarni}, Shrinivas R. and {Stern}, Daniel and {Andreoni}, Igor and {Bellm}, Eric C. and {Dekany}, Richard and {Dhawan}, Suhail and {Drake}, Andrew J. and {Fremling}, Christoffer and {Gatkine}, Pradip and {Groom}, Steven L. and {Ho}, Anna Y.~Q. and {Kasliwal}, Mansi M. and {Karambelkar}, Viraj and {Kool}, Erik C. and {Masci}, Frank J. and {Medford}, Michael S. and {Perley}, Daniel A. and {Purdum}, Josiah and {Roestel}, Jan van and {Sharma}, Yashvi and {Sollerman}, Jesper and {Taggart}, Kirsty and {Yan}, Lin},
        title = "{The Final Season Reimagined: 30 Tidal Disruption Events from the ZTF-I Survey}",
      journal = {\apj},
         year = 2023,
        month = jan,
       volume = {942},
       number = {1},
          eid = {9},
        pages = {9},
          doi = {10.3847/1538-4357/aca283},
archivePrefix = {arXiv},
       eprint = {2203.01461},
 primaryClass = {astro-ph.HE},
       adsurl = {https://ui.adsabs.harvard.edu/abs/2023ApJ...942....9H}
}

@ARTICLE{Oke1995,
       author = {{Oke}, J.~B. and {Cohen}, J.~G. and {Carr}, M. and {Cromer}, J. and
         {Dingizian}, A. and {Harris}, F.~H. and {Labrecque}, S. and
         {Lucinio}, R. and {Schaal}, W. and {Epps}, H. and {Miller}, J.},
        title = "{The Keck Low-Resolution Imaging Spectrometer}",
      journal = {\pasp},
         year = "1995",
        month = "Apr",
       volume = {107},
        pages = {375},
          doi = {10.1086/133562},
       adsurl = {https://ui.adsabs.harvard.edu/abs/1995PASP..107..375O}
}

@ARTICLE{Shingles2021,
       author = {{Shingles}, L. and {Smith}, K.~W. and {Young}, D.~R. and {Smartt}, S.~J. and {Tonry}, J. and {Denneau}, L. and {Heinze}, A. and {Weiland}, H. and {Flewelling}, H. and {Stalder}, B. and {Clocchiatti}, A. and {F{\"o}rster}, F. and {Pignata}, G. and {Rest}, A. and {Anderson}, J. and {Stubbs}, C. and {Erasmus}, N.},
        title = "{Release of the ATLAS Forced Photometry server for public use}",
      journal = {Transient Name Server AstroNote},
         year = 2021,
        month = jan,
       volume = {7},
        pages = {1-7},
       adsurl = {https://ui.adsabs.harvard.edu/abs/2021TNSAN...7....1S}
}

@ARTICLE{Smith2020,
       author = {{Smith}, K.~W. and {Smartt}, S.~J. and {Young}, D.~R. and {Tonry}, J.~L. and {Denneau}, L. and {Flewelling}, H. and {Heinze}, A.~N. and {Weiland}, H.~J. and {Stalder}, B. and {Rest}, A. and {Stubbs}, C.~W. and {Anderson}, J.~P. and {Chen}, T. -W. and {Clark}, P. and {Do}, A. and {F{\"o}rster}, F. and {Fulton}, M. and {Gillanders}, J. and {McBrien}, O.~R. and {O'Neill}, D. and {Srivastav}, S. and {Wright}, D.~E.},
        title = "{Design and Operation of the ATLAS Transient Science Server}",
      journal = {\pasp},
         year = 2020,
        month = aug,
       volume = {132},
       number = {1014},
          eid = {085002},
        pages = {085002},
          doi = {10.1088/1538-3873/ab936e},
archivePrefix = {arXiv},
       eprint = {2003.09052},
 primaryClass = {astro-ph.IM},
       adsurl = {https://ui.adsabs.harvard.edu/abs/2020PASP..132h5002S}
}

@ARTICLE{Ho2025,
       author = {{Ho}, Anna Y.~Q. and {Yao}, Yuhan and {Matsumoto}, Tatsuya and {Schroeder}, Genevieve and {Coughlin}, Eric and {Perley}, Daniel A. and {Andreoni}, Igor and {Bellm}, Eric C. and {Chen}, Tracy X. and {Chornock}, Ryan and {Covarrubias}, Sofia and {Das}, Kaustav and {Fremling}, Christoffer and {Gilfanov}, Marat and {Hinds}, K.~R. and {Jarvis}, Dan and {Kasliwal}, Mansi M. and {Liu}, Chang and {Lyman}, Joseph D. and {Masci}, Frank J. and {Prince}, Thomas A. and {Ravi}, Vikram and {Rich}, R. Michael and {Riddle}, Reed and {Sevilla}, Jason and {Smith}, Roger and {Sollerman}, Jesper and {Somalwar}, Jean J. and {Srinivasaragavan}, Gokul P. and {Sunyaev}, Rashid and {Vail}, Jada L. and {Wise}, Jacob L. and {Yun}, Sol Bin},
        title = "{A Luminous Red Optical Flare and Hard X-ray Emission in the Tidal Disruption Event AT2024kmq}",
      journal = {arXiv e-prints},
         year = 2025,
        month = feb,
          eid = {arXiv:2502.07885},
        pages = {arXiv:2502.07885},
          doi = {10.48550/arXiv.2502.07885},
archivePrefix = {arXiv},
       eprint = {2502.07885},
 primaryClass = {astro-ph.HE},
       adsurl = {https://ui.adsabs.harvard.edu/abs/2025arXiv250207885H}
}

@ARTICLE{Guolo2024,
       author = {{Guolo}, Muryel and {Gezari}, Suvi and {Yao}, Yuhan and {van Velzen}, Sjoert and {Hammerstein}, Erica and {Cenko}, S. Bradley and {Tokayer}, Yarone M.},
        title = "{A Systematic Analysis of the X-Ray Emission in Optically Selected Tidal Disruption Events: Observational Evidence for the Unification of the Optically and X-Ray-selected Populations}",
      journal = {\apj},
         year = 2024,
        month = may,
       volume = {966},
       number = {2},
          eid = {160},
        pages = {160},
          doi = {10.3847/1538-4357/ad2f9f},
archivePrefix = {arXiv},
       eprint = {2308.13019},
 primaryClass = {astro-ph.HE},
       adsurl = {https://ui.adsabs.harvard.edu/abs/2024ApJ...966..160G}
}

@ARTICLE{Willingale2013,
   author = {{Willingale}, R. and {Starling}, R.~L.~C. and {Beardmore}, A.~P. and
        {Tanvir}, N.~R. and {O'Brien}, P.~T.},
    title = "{Calibration of X-ray absorption in our Galaxy}",
  journal = {\mnras},
archivePrefix = "arXiv",
   eprint = {1303.0843},
 primaryClass = "astro-ph.HE",
     year = 2013,
    month = may,
   volume = 431,
    pages = {394-404},
      doi = {10.1093/mnras/stt175},
   adsurl = {http://adsabs.harvard.edu/abs/2013MNRAS.431..394W}
}

@ARTICLE{Kaastra2016,
       author = {{Kaastra}, J.~S. and {Bleeker}, J.~A.~M.},
        title = "{Optimal binning of X-ray spectra and response matrix design}",
      journal = {\aap},
         year = 2016,
        month = mar,
       volume = {587},
          eid = {A151},
        pages = {A151},
          doi = {10.1051/0004-6361/201527395},
archivePrefix = {arXiv},
       eprint = {1601.05309},
 primaryClass = {astro-ph.IM},
       adsurl = {https://ui.adsabs.harvard.edu/abs/2016A&A...587A.151K}
}

@ARTICLE{Sfaradi2024,
       author = {{Sfaradi}, Itai and {Beniamini}, Paz and {Horesh}, Assaf and {Piran}, Tsvi and {Bright}, Joe and {Rhodes}, Lauren and {Williams}, David R.~A. and {Fender}, Rob and {Leung}, James K. and {Murphy}, Tara and {Green}, Dave A.},
        title = "{An off-axis relativistic jet seen in the long lasting delayed radio flare of the TDE AT 2018hyz}",
      journal = {\mnras},
         year = 2024,
        month = jan,
       volume = {527},
       number = {3},
        pages = {7672-7680},
          doi = {10.1093/mnras/stad3717},
archivePrefix = {arXiv},
       eprint = {2308.01965},
 primaryClass = {astro-ph.HE},
       adsurl = {https://ui.adsabs.harvard.edu/abs/2024MNRAS.527.7672S}
}

@ARTICLE{Aspegren2026,
       author = {{Aspegren}, Olivia and {Kasen}, Daniel},
        title = "{The Emission and Suppression of Line Features in Luminous Transients}",
      journal = {arXiv e-prints},
         year = 2026,
        month = jan,
          eid = {arXiv:2601.00947},
        pages = {arXiv:2601.00947},
          doi = {10.48550/arXiv.2601.00947},
archivePrefix = {arXiv},
       eprint = {2601.00947},
 primaryClass = {astro-ph.HE},
       adsurl = {https://ui.adsabs.harvard.edu/abs/2026arXiv260100947A}
}

@ARTICLE{Ivezic2019,
       author = {{Ivezi{\'c}}, {\v{Z}}eljko and {Kahn}, Steven M. and {Tyson}, J. Anthony and {Abel}, Bob and {Acosta}, Emily and {Allsman}, Robyn and {Alonso}, David and {AlSayyad}, Yusra and {Anderson}, Scott F. and {Andrew}, John and {Angel}, James Roger P. and {Angeli}, George Z. and {Ansari}, Reza and {Antilogus}, Pierre and {Araujo}, Constanza and {Armstrong}, Robert and {Arndt}, Kirk T. and {Astier}, Pierre and {Aubourg}, {\'E}ric and {Auza}, Nicole and {Axelrod}, Tim S. and {Bard}, Deborah J. and {Barr}, Jeff D. and {Barrau}, Aurelian and {Bartlett}, James G. and {Bauer}, Amanda E. and {Bauman}, Brian J. and {Baumont}, Sylvain and {Bechtol}, Ellen and {Bechtol}, Keith and {Becker}, Andrew C. and {Becla}, Jacek and {Beldica}, Cristina and {Bellavia}, Steve and {Bianco}, Federica B. and {Biswas}, Rahul and {Blanc}, Guillaume and {Blazek}, Jonathan and {Blandford}, Roger D. and {Bloom}, Josh S. and {Bogart}, Joanne and {Bond}, Tim W. and {Booth}, Michael T. and {Borgland}, Anders W. and {Borne}, Kirk and {Bosch}, James F. and {Boutigny}, Dominique and {Brackett}, Craig A. and {Bradshaw}, Andrew and {Brandt}, William Nielsen and {Brown}, Michael E. and {Bullock}, James S. and {Burchat}, Patricia and {Burke}, David L. and {Cagnoli}, Gianpietro and {Calabrese}, Daniel and {Callahan}, Shawn and {Callen}, Alice L. and {Carlin}, Jeffrey L. and {Carlson}, Erin L. and {Chandrasekharan}, Srinivasan and {Charles-Emerson}, Glenaver and {Chesley}, Steve and {Cheu}, Elliott C. and {Chiang}, Hsin-Fang and {Chiang}, James and {Chirino}, Carol and {Chow}, Derek and {Ciardi}, David R. and {Claver}, Charles F. and {Cohen-Tanugi}, Johann and {Cockrum}, Joseph J. and {Coles}, Rebecca and {Connolly}, Andrew J. and {Cook}, Kem H. and {Cooray}, Asantha and {Covey}, Kevin R. and {Cribbs}, Chris and {Cui}, Wei and {Cutri}, Roc and {Daly}, Philip N. and {Daniel}, Scott F. and {Daruich}, Felipe and {Daubard}, Guillaume and {Daues}, Greg and {Dawson}, William and {Delgado}, Francisco and {Dellapenna}, Alfred and {de Peyster}, Robert and {de Val-Borro}, Miguel and {Digel}, Seth W. and {Doherty}, Peter and {Dubois}, Richard and {Dubois-Felsmann}, Gregory P. and {Durech}, Josef and {Economou}, Frossie and {Eifler}, Tim and {Eracleous}, Michael and {Emmons}, Benjamin L. and {Fausti Neto}, Angelo and {Ferguson}, Henry and {Figueroa}, Enrique and {Fisher-Levine}, Merlin and {Focke}, Warren and {Foss}, Michael D. and {Frank}, James and {Freemon}, Michael D. and {Gangler}, Emmanuel and {Gawiser}, Eric and {Geary}, John C. and {Gee}, Perry and {Geha}, Marla and {Gessner}, Charles J.~B. and {Gibson}, Robert R. and {Gilmore}, D. Kirk and {Glanzman}, Thomas and {Glick}, William and {Goldina}, Tatiana and {Goldstein}, Daniel A. and {Goodenow}, Iain and {Graham}, Melissa L. and {Gressler}, William J. and {Gris}, Philippe and {Guy}, Leanne P. and {Guyonnet}, Augustin and {Haller}, Gunther and {Harris}, Ron and {Hascall}, Patrick A. and {Haupt}, Justine and {Hernandez}, Fabio and {Herrmann}, Sven and {Hileman}, Edward and {Hoblitt}, Joshua and {Hodgson}, John A. and {Hogan}, Craig and {Howard}, James D. and {Huang}, Dajun and {Huffer}, Michael E. and {Ingraham}, Patrick and {Innes}, Walter R. and {Jacoby}, Suzanne H. and {Jain}, Bhuvnesh and {Jammes}, Fabrice and {Jee}, M. James and {Jenness}, Tim and {Jernigan}, Garrett and {Jevremovi{\'c}}, Darko and {Johns}, Kenneth and {Johnson}, Anthony S. and {Johnson}, Margaret W.~G. and {Jones}, R. Lynne and {Juramy-Gilles}, Claire and {Juri{\'c}}, Mario and {Kalirai}, Jason S. and {Kallivayalil}, Nitya J. and {Kalmbach}, Bryce and {Kantor}, Jeffrey P. and {Karst}, Pierre and {Kasliwal}, Mansi M. and {Kelly}, Heather and {Kessler}, Richard and {Kinnison}, Veronica and {Kirkby}, David and {Knox}, Lloyd and {Kotov}, Ivan V. and {Krabbendam}, Victor L. and {Krughoff}, K. Simon and {Kub{\'a}nek}, Petr and {Kuczewski}, John and {Kulkarni}, Shri and {Ku}, John and {Kurita}, Nadine R. and {Lage}, Craig S. and {Lambert}, Ron and {Lange}, Travis and {Langton}, J. Brian and {Le Guillou}, Laurent and {Levine}, Deborah and {Liang}, Ming and {Lim}, Kian-Tat and {Lintott}, Chris J. and {Long}, Kevin E. and {Lopez}, Margaux and {Lotz}, Paul J. and {Lupton}, Robert H. and {Lust}, Nate B. and {MacArthur}, Lauren A. and {Mahabal}, Ashish and {Mandelbaum}, Rachel and {Markiewicz}, Thomas W. and {Marsh}, Darren S. and {Marshall}, Philip J. and {Marshall}, Stuart and {May}, Morgan and {McKercher}, Robert and {McQueen}, Michelle and {Meyers}, Joshua and {Migliore}, Myriam and {Miller}, Michelle and {Mills}, David J.},
        title = "{LSST: From Science Drivers to Reference Design and Anticipated Data Products}",
      journal = {\apj},
         year = 2019,
        month = mar,
       volume = {873},
       number = {2},
          eid = {111},
        pages = {111},
          doi = {10.3847/1538-4357/ab042c},
archivePrefix = {arXiv},
       eprint = {0805.2366},
 primaryClass = {astro-ph},
       adsurl = {https://ui.adsabs.harvard.edu/abs/2019ApJ...873..111I}
}

@ARTICLE{Huang2024_tde_spin,
       author = {{Huang}, Hao-Tse and {Lu}, Wenbin},
        title = "{Tidal disruption rate suppression by the event horizon of spinning black holes}",
      journal = {\mnras},
         year = 2024,
        month = jan,
       volume = {527},
       number = {2},
        pages = {1865-1883},
          doi = {10.1093/mnras/stad3269},
archivePrefix = {arXiv},
       eprint = {2301.00259},
 primaryClass = {astro-ph.GA},
       adsurl = {https://ui.adsabs.harvard.edu/abs/2024MNRAS.527.1865H}
}

@ARTICLE{Charalampopoulos2022,
       author = {{Charalampopoulos}, P. and {Leloudas}, G. and {Malesani}, D.~B. and {Wevers}, T. and {Arcavi}, I. and {Nicholl}, M. and {Pursiainen}, M. and {Lawrence}, A. and {Anderson}, J.~P. and {Benetti}, S. and {Cannizzaro}, G. and {Chen}, T.-W. and {Galbany}, L. and {Gromadzki}, M. and {Guti{\'e}rrez}, C.~P. and {Inserra}, C. and {Jonker}, P.~G. and {M{\"u}ller-Bravo}, T.~E. and {Onori}, F. and {Short}, P. and {Sollerman}, J. and {Young}, D.~R.},
        title = "{A detailed spectroscopic study of tidal disruption events}",
      journal = {\aap},
         year = 2022,
        month = mar,
       volume = {659},
          eid = {A34},
        pages = {A34},
          doi = {10.1051/0004-6361/202142122},
archivePrefix = {arXiv},
       eprint = {2109.00016},
 primaryClass = {astro-ph.HE},
       adsurl = {https://ui.adsabs.harvard.edu/abs/2022A&A...659A..34C}
}

@ARTICLE{Guolo2026,
       author = {{Guolo}, M. and {Mummery}, A. and {van Velzen}, S. and {Gezari}, S. and {Nicholl}, M. and {Yao}, Y. and {Karmen}, M. and {Ajay}, Y. and {Wevers}, T. and {LeBaron}, N. and {Chornock}, R.},
        title = "{Compact Accretion Disks in the Aftermath of Tidal Disruption Events: Parameter Inference from Joint X-ray Spectra and UV/Optical Photometry Fitting}",
      journal = {arXiv e-prints},
         year = 2025,
        month = oct,
          eid = {arXiv:2510.26774},
        pages = {arXiv:2510.26774},
          doi = {10.48550/arXiv.2510.26774},
archivePrefix = {arXiv},
       eprint = {2510.26774},
 primaryClass = {astro-ph.HE},
       adsurl = {https://ui.adsabs.harvard.edu/abs/2025arXiv251026774G}
}

@ARTICLE{Miller1993,
       author = {{Miller}, J. S. and {Stone}, R. P. S. },
        title = "{Lick Obs. Tech. Rep. 66.}",
      journal = {Lick Observatory Techical Reports},
      volume = 66,
         year = 1993
}

@ARTICLE{vanderWalt2019,
       author = {{van der Walt}, St{\'e}fan and {Crellin-Quick}, Arien and {Bloom}, Joshua},
        title = "{SkyPortal: An Astronomical Data Platform}",
      journal = {The Journal of Open Source Software},
         year = 2019,
        month = may,
       volume = {4},
       number = {37},
          eid = {1247},
        pages = {1247},
          doi = {10.21105/joss.01247},
       adsurl = {https://ui.adsabs.harvard.edu/abs/2019JOSS....4.1247V}
}

@ARTICLE{Yao2024_24lhc_astronote,
       author = {{Yao}, Y. and {Chornock}, R. and {Guo}, X. and {LeBaron}, N. and {Margutti}, R. and {Ravi}, V. and {Somalwar}, J.},
        title = "{Classification of AT 2024lhc as a Tidal Disruption Event with Luminous and Variable X-ray Emission}",
      journal = {Transient Name Server AstroNote},
         year = 2024,
        month = jul,
       volume = {177},
        pages = {1},
       adsurl = {https://ui.adsabs.harvard.edu/abs/2024TNSAN.177....1Y}
}

@INPROCEEDINGS{Fruscione2006,
       author = {{Fruscione}, Antonella and {McDowell}, Jonathan C. and {Allen}, Glenn E. and {Brickhouse}, Nancy S. and {Burke}, Douglas J. and {Davis}, John E. and {Durham}, Nick and {Elvis}, Martin and {Galle}, Elizabeth C. and {Harris}, Daniel E. and {Huenemoerder}, David P. and {Houck}, John C. and {Ishibashi}, Bish and {Karovska}, Margarita and {Nicastro}, Fabrizio and {Noble}, Michael S. and {Nowak}, Michael A. and {Primini}, Frank A. and {Siemiginowska}, Aneta and {Smith}, Randall K. and {Wise}, Michael},
        title = "{CIAO: Chandra's data analysis system}",
    booktitle = {Observatory Operations: Strategies, Processes, and Systems},
         year = 2006,
       editor = {{Silva}, David R. and {Doxsey}, Rodger E.},
       series = {Society of Photo-Optical Instrumentation Engineers (SPIE) Conference Series},
       volume = {6270},
        month = jun,
          eid = {62701V},
        pages = {62701V},
          doi = {10.1117/12.671760},
       adsurl = {https://ui.adsabs.harvard.edu/abs/2006SPIE.6270E..1VF}
}

@ARTICLE{Yao2024_22cmc,
       author = {{Yao}, Yuhan and {Lu}, Wenbin and {Harrison}, Fiona and {Kulkarni}, S.~R. and {Gezari}, Suvi and {Guolo}, Muryel and {Cenko}, S. Bradley and {Ho}, Anna Y.~Q.},
        title = "{The On-axis Jetted Tidal Disruption Event AT2022cmc: X-Ray Observations and Broadband Spectral Modeling}",
      journal = {\apj},
         year = 2024,
        month = apr,
       volume = {965},
       number = {1},
          eid = {39},
        pages = {39},
          doi = {10.3847/1538-4357/ad2b6b},
archivePrefix = {arXiv},
       eprint = {2308.09834},
 primaryClass = {astro-ph.HE},
       adsurl = {https://ui.adsabs.harvard.edu/abs/2024ApJ...965...39Y}
}

@INPROCEEDINGS{Garmire2003,
       author = {{Garmire}, Gordon P. and {Bautz}, Mark W. and {Ford}, Peter G. and {Nousek}, John A. and {Ricker}, George R., Jr.},
        title = "{Advanced CCD imaging spectrometer (ACIS) instrument on the Chandra X-ray Observatory}",
    booktitle = {X-Ray and Gamma-Ray Telescopes and Instruments for Astronomy.},
         year = 2003,
       editor = {{Truemper}, Joachim E. and {Tananbaum}, Harvey D.},
       series = {Society of Photo-Optical Instrumentation Engineers (SPIE) Conference Series},
       volume = {4851},
        month = mar,
        pages = {28-44},
          doi = {10.1117/12.461599},
       adsurl = {https://ui.adsabs.harvard.edu/abs/2003SPIE.4851...28G}
}

@ARTICLE{Gunn2006,
       author = {{Gunn}, James E. and {Siegmund}, Walter A. and {Mannery}, Edward J. and {Owen}, Russell E. and {Hull}, Charles L. and {Leger}, R. French and {Carey}, Larry N. and {Knapp}, Gillian R. and {York}, Donald G. and {Boroski}, William N. and {Kent}, Stephen M. and {Lupton}, Robert H. and {Rockosi}, Constance M. and {Evans}, Michael L. and {Waddell}, Patrick and {Anderson}, John E. and {Annis}, James and {Barentine}, John C. and {Bartoszek}, Larry M. and {Bastian}, Steven and {Bracker}, Stephen B. and {Brewington}, Howard J. and {Briegel}, Charles I. and {Brinkmann}, Jon and {Brown}, Yorke J. and {Carr}, Michael A. and {Czarapata}, Paul C. and {Drennan}, Craig C. and {Dombeck}, Thomas and {Federwitz}, Glenn R. and {Gillespie}, Bruce A. and {Gonzales}, Carlos and {Hansen}, Sten U. and {Harvanek}, Michael and {Hayes}, Jeffrey and {Jordan}, Wendell and {Kinney}, Ellyne and {Klaene}, Mark and {Kleinman}, S.~J. and {Kron}, Richard G. and {Kresinski}, Jurek and {Lee}, Glenn and {Limmongkol}, Siriluk and {Lindenmeyer}, Carl W. and {Long}, Daniel C. and {Loomis}, Craig L. and {McGehee}, Peregrine M. and {Mantsch}, Paul M. and {Neilsen}, Jr., Eric H. and {Neswold}, Richard M. and {Newman}, Peter R. and {Nitta}, Atsuko and {Peoples}, Jr., John and {Pier}, Jeffrey R. and {Prieto}, Peter S. and {Prosapio}, Angela and {Rivetta}, Claudio and {Schneider}, Donald P. and {Snedden}, Stephanie and {Wang}, Shu-i.},
        title = "{The 2.5 m Telescope of the Sloan Digital Sky Survey}",
      journal = {\aj},
         year = 2006,
        month = apr,
       volume = {131},
       number = {4},
        pages = {2332-2359},
          doi = {10.1086/500975},
archivePrefix = {arXiv},
       eprint = {astro-ph/0602326},
 primaryClass = {astro-ph},
       adsurl = {https://ui.adsabs.harvard.edu/abs/2006AJ....131.2332G}
}

@ARTICLE{Rest2025,
       author = {{Rest}, S. and {Rest}, A. and {Kilpatrick}, C.~D. and {Jencson}, J.~E. and {von Coelln}, S. and {Strolger}, L. and {Smartt}, S. and {Anderson}, J.~P. and {Clocchiatti}, A. and {Coulter}, D.~A. and {Denneau}, L. and {Gomez}, S. and {Heinze}, A. and {Ridden-Harper}, R. and {Smith}, K.~W. and {Stalder}, B. and {Tonry}, J.~L. and {Wang}, Q. and {Zenati}, Y.},
        title = "{ATClean: A Novel Method for Detecting Low-luminosity Transients and Application to Pre-explosion Counterparts from SN 2023ixf}",
      journal = {\apj},
         year = 2025,
        month = feb,
       volume = {979},
       number = {2},
          eid = {114},
        pages = {114},
          doi = {10.3847/1538-4357/ad973d},
archivePrefix = {arXiv},
       eprint = {2405.03747},
 primaryClass = {astro-ph.HE},
       adsurl = {https://ui.adsabs.harvard.edu/abs/2025ApJ...979..114R}
}

@ARTICLE{Chornock2024_24lhc_CR,
       author = {{Chornock}, R.},
        title = "{TRex Transient Classification Report for 2024-07-03}",
      journal = {Transient Name Server Classification Report},
         year = 2024,
        month = jul,
       volume = {2024-2237},
        pages = {1},
       adsurl = {https://ui.adsabs.harvard.edu/abs/2024TNSCR2237....1C}
}

@ARTICLE{Coughlin2023,
       author = {{Coughlin}, Michael W. and {Bloom}, Joshua S. and {Nir}, Guy and {Antier}, Sarah and {du Laz}, Theophile Jegou and {van der Walt}, St{\'e}fan and {Crellin-Quick}, Arien and {Culino}, Thomas and {Duev}, Dmitry A. and {Goldstein}, Daniel A. and {Healy}, Brian F. and {Karambelkar}, Viraj and {Lilleboe}, Jada and {Shin}, Kyung Min and {Singer}, Leo P. and {Ahumada}, Tom{\'a}s and {Anand}, Shreya and {Bellm}, Eric C. and {Dekany}, Richard and {Graham}, Matthew J. and {Kasliwal}, Mansi M. and {Kostadinova}, Ivona and {Kiendrebeogo}, R. Weizmann and {Kulkarni}, Shrinivas R. and {Jenkins}, Sydney and {LeBaron}, Natalie and {Mahabal}, Ashish A. and {Neill}, James D. and {Parazin}, B. and {Peloton}, Julien and {Perley}, Daniel A. and {Riddle}, Reed and {Rusholme}, Ben and {van Santen}, Jakob and {Sollerman}, Jesper and {Stein}, Robert and {Turpin}, D. and {Wold}, Avery and {Amat}, Carla and {Bonnefon}, Adrien and {Bonnefoy}, Adrien and {Flament}, Manon and {Kerkow}, Frank and {Kishore}, Sulekha and {Jani}, Shloke and {Mahanty}, Stephen K. and {Liu}, C{\'e}line and {Llinares}, Laura and {Makarison}, Jolyane and {Olli{\'e}ric}, Alix and {Perez}, In{\`e}s and {Pont}, Lydie and {Sharma}, Vyom},
        title = "{A Data Science Platform to Enable Time-domain Astronomy}",
      journal = {\apjs},
         year = 2023,
        month = aug,
       volume = {267},
       number = {2},
          eid = {31},
        pages = {31},
          doi = {10.3847/1538-4365/acdee1},
archivePrefix = {arXiv},
       eprint = {2305.00108},
 primaryClass = {astro-ph.IM},
       adsurl = {https://ui.adsabs.harvard.edu/abs/2023ApJS..267...31C}
}

@ARTICLE{Brightman2021,
       author = {{Brightman}, Murray and {Ward}, Charlotte and {Stern}, Daniel and {Mooley}, Kunal and {De}, Kishalay and {Gezari}, Suvi and {Van Velzen}, Sjoert and {Andreoni}, Igor and {Graham}, Matthew and {Masci}, Frank J. and {Riddle}, Reed and {Zolkower}, Jeffry},
        title = "{A Luminous X-Ray Transient in SDSS J143359.16+400636.0: A Likely Tidal Disruption Event}",
      journal = {\apj},
         year = 2021,
        month = mar,
       volume = {909},
       number = {2},
          eid = {102},
        pages = {102},
          doi = {10.3847/1538-4357/abde34},
archivePrefix = {arXiv},
       eprint = {2010.12587},
 primaryClass = {astro-ph.HE},
       adsurl = {https://ui.adsabs.harvard.edu/abs/2021ApJ...909..102B}
}

@ARTICLE{Ahumada2020,
       author = {{Ahumada}, Romina and {Allende Prieto}, Carlos and {Almeida}, Andr{\'e}s and {Anders}, Friedrich and {Anderson}, Scott F. and {Andrews}, Brett H. and {Anguiano}, Borja and {Arcodia}, Riccardo and {Armengaud}, Eric and {Aubert}, Marie and {Avila}, Santiago and {Avila-Reese}, Vladimir and {Badenes}, Carles and {Balland}, Christophe and {Barger}, Kat and {Barrera-Ballesteros}, Jorge K. and {Basu}, Sarbani and {Bautista}, Julian and {Beaton}, Rachael L. and {Beers}, Timothy C. and {Benavides}, B. Izamar T. and {Bender}, Chad F. and {Bernardi}, Mariangela and {Bershady}, Matthew and {Beutler}, Florian and {Bidin}, Christian Moni and {Bird}, Jonathan and {Bizyaev}, Dmitry and {Blanc}, Guillermo A. and {Blanton}, Michael R. and {Boquien}, M{\'e}d{\'e}ric and {Borissova}, Jura and {Bovy}, Jo and {Brandt}, W.~N. and {Brinkmann}, Jonathan and {Brownstein}, Joel R. and {Bundy}, Kevin and {Bureau}, Martin and {Burgasser}, Adam and {Burtin}, Etienne and {Cano-D{\'\i}az}, Mariana and {Capasso}, Raffaella and {Cappellari}, Michele and {Carrera}, Ricardo and {Chabanier}, Sol{\`e}ne and {Chaplin}, William and {Chapman}, Michael and {Cherinka}, Brian and {Chiappini}, Cristina and {Doohyun Choi}, Peter and {Chojnowski}, S. Drew and {Chung}, Haeun and {Clerc}, Nicolas and {Coffey}, Damien and {Comerford}, Julia M. and {Comparat}, Johan and {da Costa}, Luiz and {Cousinou}, Marie-Claude and {Covey}, Kevin and {Crane}, Jeffrey D. and {Cunha}, Katia and {Ilha}, Gabriele da Silva and {Dai}, Yu Sophia and {Damsted}, Sanna B. and {Darling}, Jeremy and {Davidson}, Jr., James W. and {Davies}, Roger and {Dawson}, Kyle and {De}, Nikhil and {de la Macorra}, Axel and {De Lee}, Nathan and {Queiroz}, Anna B{\'a}rbara de Andrade and {Deconto Machado}, Alice and {de la Torre}, Sylvain and {Dell'Agli}, Flavia and {du Mas des Bourboux}, H{\'e}lion and {Diamond-Stanic}, Aleksandar M. and {Dillon}, Sean and {Donor}, John and {Drory}, Niv and {Duckworth}, Chris and {Dwelly}, Tom and {Ebelke}, Garrett and {Eftekharzadeh}, Sarah and {Davis Eigenbrot}, Arthur and {Elsworth}, Yvonne P. and {Eracleous}, Mike and {Erfanianfar}, Ghazaleh and {Escoffier}, Stephanie and {Fan}, Xiaohui and {Farr}, Emily and {Fern{\'a}ndez-Trincado}, Jos{\'e} G. and {Feuillet}, Diane and {Finoguenov}, Alexis and {Fofie}, Patricia and {Fraser-McKelvie}, Amelia and {Frinchaboy}, Peter M. and {Fromenteau}, Sebastien and {Fu}, Hai and {Galbany}, Llu{\'\i}s and {Garcia}, Rafael A. and {Garc{\'\i}a-Hern{\'a}ndez}, D.~A. and {Garma Oehmichen}, Luis Alberto and {Ge}, Junqiang and {Geimba Maia}, Marcio Antonio and {Geisler}, Doug and {Gelfand}, Joseph and {Goddy}, Julian and {Gonzalez-Perez}, Violeta and {Grabowski}, Kathleen and {Green}, Paul and {Grier}, Catherine J. and {Guo}, Hong and {Guy}, Julien and {Harding}, Paul and {Hasselquist}, Sten and {Hawken}, Adam James and {Hayes}, Christian R. and {Hearty}, Fred and {Hekker}, S. and {Hogg}, David W. and {Holtzman}, Jon A. and {Horta}, Danny and {Hou}, Jiamin and {Hsieh}, Bau-Ching and {Huber}, Daniel and {Hunt}, Jason A.~S. and {Ider Chitham}, J. and {Imig}, Julie and {Jaber}, Mariana and {Jimenez Angel}, Camilo Eduardo and {Johnson}, Jennifer A. and {Jones}, Amy M. and {J{\"o}nsson}, Henrik and {Jullo}, Eric and {Kim}, Yerim and {Kinemuchi}, Karen and {Kirkpatrick}, IV, Charles C. and {Kite}, George W. and {Klaene}, Mark and {Kneib}, Jean-Paul and {Kollmeier}, Juna A. and {Kong}, Hui and {Kounkel}, Marina and {Krishnarao}, Dhanesh and {Lacerna}, Ivan and {Lan}, Ting-Wen and {Lane}, Richard R. and {Law}, David R. and {Le Goff}, Jean-Marc and {Leung}, Henry W. and {Lewis}, Hannah and {Li}, Cheng and {Lian}, Jianhui and {Lin}, Lihwai and {Long}, Dan and {Longa-Pe{\~n}a}, Pen{\'e}lope and {Lundgren}, Britt and {Lyke}, Brad W. and {Mackereth}, J. Ted and {MacLeod}, Chelsea L. and {Majewski}, Steven R. and {Manchado}, Arturo and {Maraston}, Claudia and {Martini}, Paul and {Masseron}, Thomas and {Masters}, Karen L. and {Mathur}, Savita and {McDermid}, Richard M. and {Merloni}, Andrea and {Merrifield}, Michael and {M{\'e}sz{\'a}ros}, Szabolcs and {Miglio}, Andrea and {Minniti}, Dante and {Minsley}, Rebecca and {Miyaji}, Takamitsu and {Mohammad}, Faizan Gohar and {Mosser}, Benoit and {Mueller}, Eva-Maria and {Muna}, Demitri and {Mu{\~n}oz-Guti{\'e}rrez}, Andrea and {Myers}, Adam D. and {Nadathur}, Seshadri and {Nair}, Preethi and {Nandra}, Kirpal and {Correa do Nascimento}, Janaina and {Nevin}, Rebecca Jean and {Newman}, Jeffrey A. and {Nidever}, David L. and {Nitschelm}, Christian and {Noterdaeme}, Pasquier and {O'Connell}, Julia E. and {Olmstead}, Matthew D. and {Oravetz}, Daniel and {Oravetz}, Audrey and {Osorio}, Yeisson and {Pace}, Zachary J. and {Padilla}, Nelson and {Palanque-Delabrouille}, Nathalie and {Palicio}, Pedro A.},
        title = "{The 16th Data Release of the Sloan Digital Sky Surveys: First Release from the APOGEE-2 Southern Survey and Full Release of eBOSS Spectra}",
      journal = {\apjs},
         year = 2020,
        month = jul,
       volume = {249},
       number = {1},
          eid = {3},
        pages = {3},
          doi = {10.3847/1538-4365/ab929e},
archivePrefix = {arXiv},
       eprint = {1912.02905},
 primaryClass = {astro-ph.GA},
       adsurl = {https://ui.adsabs.harvard.edu/abs/2020ApJS..249....3A}
}

@ARTICLE{Hammerstein2023_KCWI,
       author = {{Hammerstein}, Erica and {Cenko}, S. Bradley and {Gezari}, Suvi and {Veilleux}, Sylvain and {O'Connor}, Brendan and {van Velzen}, Sjoert and {Ward}, Charlotte and {Yao}, Yuhan and {Graham}, Matthew},
        title = "{Integral Field Spectroscopy of 13 Tidal Disruption Event Hosts from the Zwicky Transient Facility Survey}",
      journal = {\apj},
         year = 2023,
        month = nov,
       volume = {957},
       number = {2},
          eid = {86},
        pages = {86},
          doi = {10.3847/1538-4357/acfb84},
archivePrefix = {arXiv},
       eprint = {2307.15705},
 primaryClass = {astro-ph.CO},
       adsurl = {https://ui.adsabs.harvard.edu/abs/2023ApJ...957...86H}
}

@ARTICLE{Frederick2021,
       author = {{Frederick}, Sara and {Gezari}, Suvi and {Graham}, Matthew J. and {Sollerman}, Jesper and {van Velzen}, Sjoert and {Perley}, Daniel A. and {Stern}, Daniel and {Ward}, Charlotte and {Hammerstein}, Erica and {Hung}, Tiara and {Yan}, Lin and {Andreoni}, Igor and {Bellm}, Eric C. and {Duev}, Dmitry A. and {Kowalski}, Marek and {Mahabal}, Ashish A. and {Masci}, Frank J. and {Medford}, Michael and {Rusholme}, Ben and {Smith}, Roger and {Walters}, Richard},
        title = "{A Family Tree of Optical Transients from Narrow-line Seyfert 1 Galaxies}",
      journal = {\apj},
         year = 2021,
        month = oct,
       volume = {920},
       number = {1},
          eid = {56},
        pages = {56},
          doi = {10.3847/1538-4357/ac110f},
archivePrefix = {arXiv},
       eprint = {2010.08554},
 primaryClass = {astro-ph.HE},
       adsurl = {https://ui.adsabs.harvard.edu/abs/2021ApJ...920...56F}
}

@ARTICLE{Wiseman2025,
       author = {{Wiseman}, P. and {Williams}, R.~D. and {Arcavi}, I. and {Galbany}, L. and {Graham}, M.~J. and {H{\"o}nig}, S. and {Newsome}, M. and {Subrayan}, B. and {Sullivan}, M. and {Wang}, Y. and {Ili{\'c}}, D. and {Nicholl}, M. and {Oates}, S. and {Petrushevska}, T. and {Smith}, K.~W.},
        title = "{A systematically selected sample of luminous, long-duration, ambiguous nuclear transients}",
      journal = {\mnras},
         year = 2025,
        month = feb,
       volume = {537},
       number = {2},
        pages = {2024-2045},
          doi = {10.1093/mnras/staf116},
archivePrefix = {arXiv},
       eprint = {2406.11552},
 primaryClass = {astro-ph.HE},
       adsurl = {https://ui.adsabs.harvard.edu/abs/2025MNRAS.537.2024W}
}

@ARTICLE{Hinkle2022,
       author = {{Hinkle}, Jason T. and {Holoien}, Thomas W. -S. and {Shappee}, Benjamin. J. and {Neustadt}, Jack M.~M. and {Auchettl}, Katie and {Vallely}, Patrick J. and {Shahbandeh}, Melissa and {Kluge}, Matthias and {Kochanek}, Christopher S. and {Stanek}, K.~Z. and {Huber}, Mark E. and {Post}, Richard S. and {Bersier}, David and {Ashall}, Christopher and {Tucker}, Michael A. and {Williams}, Jonathan P. and {de Jaeger}, Thomas and {Do}, Aaron and {Fausnaugh}, Michael and {Gruen}, Daniel and {Hopp}, Ulrich and {Myles}, Justin and {Obermeier}, Christian and {Payne}, Anna V. and {Thompson}, Todd A.},
        title = "{The Curious Case of ASASSN-20hx: A Slowly Evolving, UV- and X-Ray-Luminous, Ambiguous Nuclear Transient}",
      journal = {\apj},
         year = 2022,
        month = may,
       volume = {930},
       number = {1},
          eid = {12},
        pages = {12},
          doi = {10.3847/1538-4357/ac5f54},
archivePrefix = {arXiv},
       eprint = {2108.03245},
 primaryClass = {astro-ph.HE},
       adsurl = {https://ui.adsabs.harvard.edu/abs/2022ApJ...930...12H}
}

@ARTICLE{Zhu2025,
       author = {{Zhu}, Jiazheng and {Jiang}, Ning and {Wang}, Yibo and {Wang}, Tinggui and {Sun}, Luming and {Zhong}, Shiyan and {Yao}, Yuhan and {Chornock}, Ryan and {Dai}, Lixin and {Lyu}, Jianwei and {Shu}, Xinwen and {Fremling}, Christoffer and {Hammerstein}, Erica and {Huang}, Shifeng and {Li}, Wenkai and {You}, Bei},
        title = "{Ultraviolet Spectral Evidence for Ansky as a Slowly Evolving Featureless Tidal Disruption Event with Quasiperiodic Eruptions}",
      journal = {\apjl},
         year = 2025,
        month = nov,
       volume = {994},
       number = {1},
          eid = {L16},
        pages = {L16},
          doi = {10.3847/2041-8213/ae19ea},
archivePrefix = {arXiv},
       eprint = {2510.22211},
 primaryClass = {astro-ph.HE},
       adsurl = {https://ui.adsabs.harvard.edu/abs/2025ApJ...994L..16Z}
}

@ARTICLE{Ramsden2022,
       author = {{Ramsden}, Paige and {Lanning}, Daniel and {Nicholl}, Matt and {McGee}, Sean L.},
        title = "{The bulge masses of TDE host galaxies and their scaling with black hole mass}",
      journal = {\mnras},
         year = 2022,
        month = sep,
       volume = {515},
       number = {1},
        pages = {1146-1157},
          doi = {10.1093/mnras/stac1810},
archivePrefix = {arXiv},
       eprint = {2201.02650},
 primaryClass = {astro-ph.HE},
       adsurl = {https://ui.adsabs.harvard.edu/abs/2022MNRAS.515.1146R}
}

@ARTICLE{Mockler2019,
       author = {{Mockler}, Brenna and {Guillochon}, James and {Ramirez-Ruiz}, Enrico},
        title = "{Weighing Black Holes Using Tidal Disruption Events}",
      journal = {\apj},
         year = 2019,
        month = feb,
       volume = {872},
       number = {2},
          eid = {151},
        pages = {151},
          doi = {10.3847/1538-4357/ab010f},
archivePrefix = {arXiv},
       eprint = {1801.08221},
 primaryClass = {astro-ph.HE},
       adsurl = {https://ui.adsabs.harvard.edu/abs/2019ApJ...872..151M}
}

@ARTICLE{Steinberg2024,
       author = {{Steinberg}, Elad and {Stone}, Nicholas C.},
        title = "{Stream-disk shocks as the origins of peak light in tidal disruption events}",
      journal = {\nat},
         year = 2024,
        month = jan,
       volume = {625},
       number = {7995},
        pages = {463-467},
          doi = {10.1038/s41586-023-06875-y},
archivePrefix = {arXiv},
       eprint = {2206.10641},
 primaryClass = {astro-ph.HE},
       adsurl = {https://ui.adsabs.harvard.edu/abs/2024Natur.625..463S}
}

@ARTICLE{Jiang2025,
       author = {{Jiang}, Yan-Fei and {Blaes}, Omer and {Kaul}, Ish and {Zhang}, Lizhong},
        title = "{Radiation and Magnetic Pressure Support in Accretion Disks Around Supermassive Black Holes and the Physical Origin of the Extreme-ultraviolet to Soft X-Ray Spectrum}",
      journal = {\apj},
         year = 2025,
        month = jul,
       volume = {988},
       number = {1},
          eid = {43},
        pages = {43},
          doi = {10.3847/1538-4357/addecb},
archivePrefix = {arXiv},
       eprint = {2505.09671},
 primaryClass = {astro-ph.HE},
       adsurl = {https://ui.adsabs.harvard.edu/abs/2025ApJ...988...43J}
}

@ARTICLE{Blandford1981,
       author = {{Blandford}, R.~D. and {Payne}, D.~G.},
        title = "{Compton scattering in a converging fluid flow. I - The transfer equation. II - Radiation-dominated shock}",
      journal = {\mnras},
         year = 1981,
        month = mar,
       volume = {194},
        pages = {1033-1039},
          doi = {10.1093/mnras/194.4.1033},
       adsurl = {https://ui.adsabs.harvard.edu/abs/1981MNRAS.194.1033B}
}

@ARTICLE{Fabian2016,
       author = {{Fabian}, A.~C.},
        title = "{The innermost extremes of black hole accretion}",
      journal = {Astronomische Nachrichten},
         year = 2016,
        month = may,
       volume = {337},
       number = {4-5},
        pages = {375},
          doi = {10.1002/asna.201612316},
archivePrefix = {arXiv},
       eprint = {1511.08596},
 primaryClass = {astro-ph.HE},
       adsurl = {https://ui.adsabs.harvard.edu/abs/2016AN....337..375F}
}

@ARTICLE{Yuan2014,
       author = {{Yuan}, Feng and {Narayan}, Ramesh},
        title = "{Hot Accretion Flows Around Black Holes}",
      journal = {\araa},
         year = 2014,
        month = aug,
       volume = {52},
        pages = {529-588},
          doi = {10.1146/annurev-astro-082812-141003},
archivePrefix = {arXiv},
       eprint = {1401.0586},
 primaryClass = {astro-ph.HE},
       adsurl = {https://ui.adsabs.harvard.edu/abs/2014ARA&A..52..529Y}
}

@ARTICLE{Chakraborty2024,
       author = {{Chakraborty}, Joheen and {Arcodia}, Riccardo and {Kara}, Erin and {Miniutti}, Giovanni and {Giustini}, Margherita and {Tetarenko}, Alexandra J. and {Rhodes}, Lauren and {Franchini}, Alessia and {Bonetti}, Matteo and {Burdge}, Kevin B. and {Goodwin}, Adelle J. and {Maccarone}, Thomas J. and {Merloni}, Andrea and {Ponti}, Gabriele and {Remillard}, Ronald A. and {Saxton}, Richard D.},
        title = "{Testing EMRI Models for Quasi-periodic Eruptions with 3.5 yr of Monitoring eRO-QPE1}",
      journal = {\apj},
         year = 2024,
        month = apr,
       volume = {965},
       number = {1},
          eid = {12},
        pages = {12},
          doi = {10.3847/1538-4357/ad2941},
archivePrefix = {arXiv},
       eprint = {2402.08722},
 primaryClass = {astro-ph.HE},
       adsurl = {https://ui.adsabs.harvard.edu/abs/2024ApJ...965...12C}
}

@ARTICLE{Gultekin2019,
       author = {{G{\"u}ltekin}, Kayhan and {King}, Ashley L. and {Cackett}, Edward M. and {Nyland}, Kristina and {Miller}, Jon M. and {Di Matteo}, Tiziana and {Markoff}, Sera and {Rupen}, Michael P.},
        title = "{The Fundamental Plane of Black Hole Accretion and Its Use as a Black Hole-Mass Estimator}",
      journal = {\apj},
         year = 2019,
        month = jan,
       volume = {871},
       number = {1},
          eid = {80},
        pages = {80},
          doi = {10.3847/1538-4357/aaf6b9},
archivePrefix = {arXiv},
       eprint = {1901.02530},
 primaryClass = {astro-ph.HE},
       adsurl = {https://ui.adsabs.harvard.edu/abs/2019ApJ...871...80G}
}

@ARTICLE{Falcke2004,
       author = {{Falcke}, H. and {K{\"o}rding}, E. and {Markoff}, S.},
        title = "{A scheme to unify low-power accreting black holes. Jet-dominated accretion flows and the radio/X-ray correlation}",
      journal = {\aap},
         year = 2004,
        month = feb,
       volume = {414},
        pages = {895-903},
          doi = {10.1051/0004-6361:20031683},
archivePrefix = {arXiv},
       eprint = {astro-ph/0305335},
 primaryClass = {astro-ph},
       adsurl = {https://ui.adsabs.harvard.edu/abs/2004A&A...414..895F}
}

@ARTICLE{Merloni2003,
       author = {{Merloni}, Andrea and {Heinz}, Sebastian and {di Matteo}, Tiziana},
        title = "{A Fundamental Plane of black hole activity}",
      journal = {\mnras},
         year = 2003,
        month = nov,
       volume = {345},
       number = {4},
        pages = {1057-1076},
          doi = {10.1046/j.1365-2966.2003.07017.x},
archivePrefix = {arXiv},
       eprint = {astro-ph/0305261},
 primaryClass = {astro-ph},
       adsurl = {https://ui.adsabs.harvard.edu/abs/2003MNRAS.345.1057M}
}

@ARTICLE{Graham2019,
       author = {{Graham}, Matthew J. and {Kulkarni}, S.~R. and {Bellm}, Eric C. and {Adams}, Scott M. and {Barbarino}, Cristina and {Blagorodnova}, Nadejda and {Bodewits}, Dennis and {Bolin}, Bryce and {Brady}, Patrick R. and {Cenko}, S. Bradley and {Chang}, Chan-Kao and {Coughlin}, Michael W. and {De}, Kishalay and {Eadie}, Gwendolyn and {Farnham}, Tony L. and {Feindt}, Ulrich and {Franckowiak}, Anna and {Fremling}, Christoffer and {Gezari}, Suvi and {Ghosh}, Shaon and {Goldstein}, Daniel A. and {Golkhou}, V. Zach and {Goobar}, Ariel and {Ho}, Anna Y.~Q. and {Huppenkothen}, Daniela and {Ivezi{\'c}}, {\v{Z}}eljko and {Jones}, R. Lynne and {Juric}, Mario and {Kaplan}, David L. and {Kasliwal}, Mansi M. and {Kelley}, Michael S.~P. and {Kupfer}, Thomas and {Lee}, Chien-De and {Lin}, Hsing Wen and {Lunnan}, Ragnhild and {Mahabal}, Ashish A. and {Miller}, Adam A. and {Ngeow}, Chow-Choong and {Nugent}, Peter and {Ofek}, Eran O. and {Prince}, Thomas A. and {Rauch}, Ludwig and {van Roestel}, Jan and {Schulze}, Steve and {Singer}, Leo P. and {Sollerman}, Jesper and {Taddia}, Francesco and {Yan}, Lin and {Ye}, Quan-Zhi and {Yu}, Po-Chieh and {Barlow}, Tom and {Bauer}, James and {Beck}, Ron and {Belicki}, Justin and {Biswas}, Rahul and {Brinnel}, Valery and {Brooke}, Tim and {Bue}, Brian and {Bulla}, Mattia and {Burruss}, Rick and {Connolly}, Andrew and {Cromer}, John and {Cunningham}, Virginia and {Dekany}, Richard and {Delacroix}, Alex and {Desai}, Vandana and {Duev}, Dmitry A. and {Feeney}, Michael and {Flynn}, David and {Frederick}, Sara and {Gal-Yam}, Avishay and {Giomi}, Matteo and {Groom}, Steven and {Hacopians}, Eugean and {Hale}, David and {Helou}, George and {Henning}, John and {Hover}, David and {Hillenbrand}, Lynne A. and {Howell}, Justin and {Hung}, Tiara and {Imel}, David and {Ip}, Wing-Huen and {Jackson}, Edward and {Kaspi}, Shai and {Kaye}, Stephen and {Kowalski}, Marek and {Kramer}, Emily and {Kuhn}, Michael and {Landry}, Walter and {Laher}, Russ R. and {Mao}, Peter and {Masci}, Frank J. and {Monkewitz}, Serge and {Murphy}, Patrick and {Nordin}, Jakob and {Patterson}, Maria T. and {Penprase}, Bryan and {Porter}, Michael and {Rebbapragada}, Umaa and {Reiley}, Dan and {Riddle}, Reed and {Rigault}, Mickael and {Rodriguez}, Hector and {Rusholme}, Ben and {van Santen}, Jakob and {Shupe}, David L. and {Smith}, Roger M. and {Soumagnac}, Maayane T. and {Stein}, Robert and {Surace}, Jason and {Szkody}, Paula and {Terek}, Scott and {Van Sistine}, Angela and {van Velzen}, Sjoert and {Vestrand}, W. Thomas and {Walters}, Richard and {Ward}, Charlotte and {Zhang}, Chaoran and {Zolkower}, Jeffry},
        title = "{The Zwicky Transient Facility: Science Objectives}",
      journal = {\pasp},
         year = 2019,
        month = jul,
       volume = {131},
       number = {1001},
        pages = {078001},
          doi = {10.1088/1538-3873/ab006c},
archivePrefix = {arXiv},
       eprint = {1902.01945},
 primaryClass = {astro-ph.IM},
       adsurl = {https://ui.adsabs.harvard.edu/abs/2019PASP..131g8001G}
}

@ARTICLE{Masci2023,
       author = {{Masci}, Frank J. and {Laher}, Russ R. and {Rusholme}, Benjamin and {Shupe}, David and {Paladini}, Roberta and {Groom}, Steve and {Wold}, Avery and {Miller}, Adam A. and {Drake}, Andrew},
        title = "{A New Forced Photometry Service for the Zwicky Transient Facility}",
      journal = {arXiv e-prints},
         year = 2023,
        month = may,
          eid = {arXiv:2305.16279},
        pages = {arXiv:2305.16279},
          doi = {10.48550/arXiv.2305.16279},
archivePrefix = {arXiv},
       eprint = {2305.16279},
 primaryClass = {astro-ph.IM},
       adsurl = {https://ui.adsabs.harvard.edu/abs/2023arXiv230516279M}
}

@ARTICLE{Chambers2016,
       author = {{Chambers}, K.~C. and {Magnier}, E.~A. and {Metcalfe}, N. and {Flewelling}, H.~A. and {Huber}, M.~E. and {Waters}, C.~Z. and {Denneau}, L. and {Draper}, P.~W. and {Farrow}, D. and {Finkbeiner}, D.~P. and {Holmberg}, C. and {Koppenhoefer}, J. and {Price}, P.~A. and {Rest}, A. and {Saglia}, R.~P. and {Schlafly}, E.~F. and {Smartt}, S.~J. and {Sweeney}, W. and {Wainscoat}, R.~J. and {Burgett}, W.~S. and {Chastel}, S. and {Grav}, T. and {Heasley}, J.~N. and {Hodapp}, K.~W. and {Jedicke}, R. and {Kaiser}, N. and {Kudritzki}, R. -P. and {Luppino}, G.~A. and {Lupton}, R.~H. and {Monet}, D.~G. and {Morgan}, J.~S. and {Onaka}, P.~M. and {Shiao}, B. and {Stubbs}, C.~W. and {Tonry}, J.~L. and {White}, R. and {Ba{\~n}ados}, E. and {Bell}, E.~F. and {Bender}, R. and {Bernard}, E.~J. and {Boegner}, M. and {Boffi}, F. and {Botticella}, M.~T. and {Calamida}, A. and {Casertano}, S. and {Chen}, W. -P. and {Chen}, X. and {Cole}, S. and {Deacon}, N. and {Frenk}, C. and {Fitzsimmons}, A. and {Gezari}, S. and {Gibbs}, V. and {Goessl}, C. and {Goggia}, T. and {Gourgue}, R. and {Goldman}, B. and {Grant}, P. and {Grebel}, E.~K. and {Hambly}, N.~C. and {Hasinger}, G. and {Heavens}, A.~F. and {Heckman}, T.~M. and {Henderson}, R. and {Henning}, T. and {Holman}, M. and {Hopp}, U. and {Ip}, W. -H. and {Isani}, S. and {Jackson}, M. and {Keyes}, C.~D. and {Koekemoer}, A.~M. and {Kotak}, R. and {Le}, D. and {Liska}, D. and {Long}, K.~S. and {Lucey}, J.~R. and {Liu}, M. and {Martin}, N.~F. and {Masci}, G. and {McLean}, B. and {Mindel}, E. and {Misra}, P. and {Morganson}, E. and {Murphy}, D.~N.~A. and {Obaika}, A. and {Narayan}, G. and {Nieto-Santisteban}, M.~A. and {Norberg}, P. and {Peacock}, J.~A. and {Pier}, E.~A. and {Postman}, M. and {Primak}, N. and {Rae}, C. and {Rai}, A. and {Riess}, A. and {Riffeser}, A. and {Rix}, H.~W. and {R{\"o}ser}, S. and {Russel}, R. and {Rutz}, L. and {Schilbach}, E. and {Schultz}, A.~S.~B. and {Scolnic}, D. and {Strolger}, L. and {Szalay}, A. and {Seitz}, S. and {Small}, E. and {Smith}, K.~W. and {Soderblom}, D.~R. and {Taylor}, P. and {Thomson}, R. and {Taylor}, A.~N. and {Thakar}, A.~R. and {Thiel}, J. and {Thilker}, D. and {Unger}, D. and {Urata}, Y. and {Valenti}, J. and {Wagner}, J. and {Walder}, T. and {Walter}, F. and {Watters}, S.~P. and {Werner}, S. and {Wood-Vasey}, W.~M. and {Wyse}, R.},
        title = "{The Pan-STARRS1 Surveys}",
      journal = {arXiv e-prints},
         year = 2016,
        month = dec,
          eid = {arXiv:1612.05560},
        pages = {arXiv:1612.05560},
          doi = {10.48550/arXiv.1612.05560},
archivePrefix = {arXiv},
       eprint = {1612.05560},
 primaryClass = {astro-ph.IM},
       adsurl = {https://ui.adsabs.harvard.edu/abs/2016arXiv161205560C}
}

@ARTICLE{Beniamini2023,
       author = {{Beniamini}, Paz and {Piran}, Tsvi and {Matsumoto}, Tatsuya},
        title = "{Swift J1644+57 as an off-axis Jet}",
      journal = {\mnras},
         year = 2023,
        month = sep,
       volume = {524},
       number = {1},
        pages = {1386-1395},
          doi = {10.1093/mnras/stad1950},
archivePrefix = {arXiv},
       eprint = {2305.06370},
 primaryClass = {astro-ph.HE},
       adsurl = {https://ui.adsabs.harvard.edu/abs/2023MNRAS.524.1386B}
}

@ARTICLE{Matsumoto2023_22cmc,
       author = {{Matsumoto}, Tatsuya and {Metzger}, Brian D.},
        title = "{Synchrotron afterglow model for AT 2022cmc: jetted tidal disruption event or engine-powered supernova?}",
      journal = {\mnras},
         year = 2023,
        month = jul,
       volume = {522},
       number = {3},
        pages = {4028-4037},
          doi = {10.1093/mnras/stad1182},
archivePrefix = {arXiv},
       eprint = {2301.11939},
 primaryClass = {astro-ph.HE},
       adsurl = {https://ui.adsabs.harvard.edu/abs/2023MNRAS.522.4028M}
}

@ARTICLE{Andreoni2022,
       author = {{Andreoni}, Igor and {Coughlin}, Michael W. and {Perley}, Daniel A. and {Yao}, Yuhan and {Lu}, Wenbin and {Cenko}, S. Bradley and {Kumar}, Harsh and {Anand}, Shreya and {Ho}, Anna Y.~Q. and {Kasliwal}, Mansi M. and {de Ugarte Postigo}, Antonio and {Sagu{\'e}s-Carracedo}, Ana and {Schulze}, Steve and {Kann}, D. Alexander and {Kulkarni}, S.~R. and {Sollerman}, Jesper and {Tanvir}, Nial and {Rest}, Armin and {Izzo}, Luca and {Somalwar}, Jean J. and {Kaplan}, David L. and {Ahumada}, Tom{\'a}s and {Anupama}, G.~C. and {Auchettl}, Katie and {Barway}, Sudhanshu and {Bellm}, Eric C. and {Bhalerao}, Varun and {Bloom}, Joshua S. and {Bremer}, Michael and {Bulla}, Mattia and {Burns}, Eric and {Campana}, Sergio and {Chandra}, Poonam and {Charalampopoulos}, Panos and {Cooke}, Jeff and {D'Elia}, Valerio and {Das}, Kaustav Kashyap and {Dobie}, Dougal and {Ag{\"u}{\'\i} Fern{\'a}ndez}, Jos{\'e} Feliciano and {Freeburn}, James and {Fremling}, Cristoffer and {Gezari}, Suvi and {Goode}, Simon and {Graham}, Matthew J. and {Hammerstein}, Erica and {Karambelkar}, Viraj R. and {Kilpatrick}, Charles D. and {Kool}, Erik C. and {Krips}, Melanie and {Laher}, Russ R. and {Leloudas}, Giorgos and {Levan}, Andrew and {Lundquist}, Michael J. and {Mahabal}, Ashish A. and {Medford}, Michael S. and {Miller}, M. Coleman and {M{\"o}ller}, Anais and {Mooley}, Kunal P. and {Nayana}, A.~J. and {Nir}, Guy and {Pang}, Peter T.~H. and {Paraskeva}, Emmy and {Perley}, Richard A. and {Petitpas}, Glen and {Pursiainen}, Miika and {Ravi}, Vikram and {Ridden-Harper}, Ryan and {Riddle}, Reed and {Rigault}, Mickael and {Rodriguez}, Antonio C. and {Rusholme}, Ben and {Sharma}, Yashvi and {Smith}, I.~A. and {Stein}, Robert D. and {Th{\"o}ne}, Christina and {Tohuvavohu}, Aaron and {Valdes}, Frank and {van Roestel}, Jan and {Vergani}, Susanna D. and {Wang}, Qinan and {Zhang}, Jielai},
        title = "{A very luminous jet from the disruption of a star by a massive black hole}",
      journal = {\nat},
         year = 2022,
        month = dec,
       volume = {612},
       number = {7940},
        pages = {430-434},
          doi = {10.1038/s41586-022-05465-8},
archivePrefix = {arXiv},
       eprint = {2211.16530},
 primaryClass = {astro-ph.HE},
       adsurl = {https://ui.adsabs.harvard.edu/abs/2022Natur.612..430A}
}

@ARTICLE{Nicholl2023,
       author = {{Nicholl}, M. and {Srivastav}, S. and {Fulton}, M.~D. and {Gomez}, S. and {Huber}, M.~E. and {Oates}, S.~R. and {Ramsden}, P. and {Rhodes}, L. and {Smartt}, S.~J. and {Smith}, K.~W. and {Aamer}, A. and {Anderson}, J.~P. and {Bauer}, F.~E. and {Berger}, E. and {de Boer}, T. and {Chambers}, K.~C. and {Charalampopoulos}, P. and {Chen}, T. -W. and {Fender}, R.~P. and {Fraser}, M. and {Gao}, H. and {Green}, D.~A. and {Galbany}, L. and {Gompertz}, B.~P. and {Gromadzki}, M. and {Guti{\'e}rrez}, C.~P. and {Howell}, D.~A. and {Inserra}, C. and {Jonker}, P.~G. and {Kopsacheili}, M. and {Lowe}, T.~B. and {Magnier}, E.~A. and {McCully}, C. and {McGee}, S.~L. and {Moore}, T. and {M{\"u}ller-Bravo}, T.~E. and {Newsome}, M. and {Gonzalez}, E. Padilla and {Pellegrino}, C. and {Pessi}, T. and {Pursiainen}, M. and {Rest}, A. and {Ridley}, E.~J. and {Shappee}, B.~J. and {Sheng}, X. and {Smith}, G.~P. and {Terreran}, G. and {Tucker}, M.~A. and {Vink{\'o}}, J. and {Wainscoat}, R.~J. and {Wiseman}, P. and {Young}, D.~R.},
        title = "{AT 2022aedm and a New Class of Luminous, Fast-cooling Transients in Elliptical Galaxies}",
      journal = {\apjl},
         year = 2023,
        month = sep,
       volume = {954},
       number = {1},
          eid = {L28},
        pages = {L28},
          doi = {10.3847/2041-8213/acf0ba},
archivePrefix = {arXiv},
       eprint = {2307.02556},
 primaryClass = {astro-ph.HE},
       adsurl = {https://ui.adsabs.harvard.edu/abs/2023ApJ...954L..28N}
}

@ARTICLE{Bellm2019b,
       author = {{Bellm}, Eric C. and {Kulkarni}, Shrinivas R. and {Graham}, Matthew J. and
         {Dekany}, Richard and {Smith}, Roger M. and {Riddle}, Reed and
         {Masci}, Frank J. and {Helou}, George and {Prince}, Thomas A. and
         {Adams}, Scott M.},
        title = "{The Zwicky Transient Facility: System Overview, Performance, and First Results}",
      journal = {\pasp},
         year = "2019",
        month = "Jan",
       volume = {131},
       number = {995},
        pages = {018002},
          doi = {10.1088/1538-3873/aaecbe},
archivePrefix = {arXiv},
       eprint = {1902.01932},
 primaryClass = {astro-ph.IM},
       adsurl = {https://ui.adsabs.harvard.edu/abs/2019PASP..131a8002B}
}

@ARTICLE{Roming2005,
       author = {{Roming}, Peter W.~A. and {Kennedy}, Thomas E. and {Mason}, Keith O. and
         {Nousek}, John A. and {Ahr}, Lindy and {Bingham}, Richard E. and
         {Broos}, Patrick S. and {Carter}, Mary J. and {Hancock}, Barry K. and
         {Huckle}, Howard E.},
        title = "{The Swift Ultra-Violet/Optical Telescope}",
      journal = {\ssr},
         year = "2005",
        month = "Oct",
       volume = {120},
       number = {3-4},
        pages = {95-142},
          doi = {10.1007/s11214-005-5095-4},
archivePrefix = {arXiv},
       eprint = {astro-ph/0507413},
 primaryClass = {astro-ph},
       adsurl = {https://ui.adsabs.harvard.edu/abs/2005SSRv..120...95R}
}

@ARTICLE{Burrows2005,
       author = {{Burrows}, David N. and {Hill}, J.~E. and {Nousek}, J.~A. and
         {Kennea}, J.~A. and {Wells}, A. and {Osborne}, J.~P. and
         {Abbey}, A.~F. and {Beardmore}, A. and {Mukerjee}, K. and
         {Short}, A.~D.~T. and {Chincarini}, G. and {Campana}, S. and
         {Citterio}, O. and {Moretti}, A. and {Pagani}, C. and
         {Tagliaferri}, G. and {Giommi}, P. and {Capalbi}, M. and
         {Tamburelli}, F. and {Angelini}, L. and {Cusumano}, G. and
         {Br{\"a}uninger}, H.~W. and {Burkert}, W. and {Hartner}, G.~D.},
        title = "{The Swift X-Ray Telescope}",
      journal = {\ssr},
         year = 2005,
        month = oct,
       volume = {120},
       number = {3-4},
        pages = {165-195},
          doi = {10.1007/s11214-005-5097-2},
archivePrefix = {arXiv},
       eprint = {astro-ph/0508071},
 primaryClass = {astro-ph},
       adsurl = {https://ui.adsabs.harvard.edu/abs/2005SSRv..120..165B}
}

@ARTICLE{Alexander2026,
       author = {{Alexander}, Kate D. and {Margutti}, Raffaella and {Gomez}, Sebastian and {Stroh}, Michael and {Chornock}, Ryan and {Laskar}, Tanmoy and {Cendes}, Y. and {Berger}, Edo and {Eftekhari}, Tarraneh and {Franz}, Noah and {Hajela}, Aprajita and {Metzger}, B.~D. and {Terreran}, Giacomo and {Bietenholz}, Michael and {Christy}, Collin and {de Colle}, Fabio and {Komossa}, S. and {Nicholl}, Matt and {Ramirez-Ruiz}, Enrico and {Saxton}, Richard and {Schroeder}, Genevieve and {Williams}, Peter and {Wu}, William},
        title = "{The Multi-Wavelength Context of Delayed Radio Emission in TDEs: Evidence for Accretion-Driven Outflows}",
      journal = {arXiv e-prints},
         year = 2025,
        month = jun,
          eid = {arXiv:2506.12729},
        pages = {arXiv:2506.12729},
          doi = {10.48550/arXiv.2506.12729},
archivePrefix = {arXiv},
       eprint = {2506.12729},
 primaryClass = {astro-ph.HE},
       adsurl = {https://ui.adsabs.harvard.edu/abs/2025arXiv250612729A}
}

@ARTICLE{Gallo2003,
       author = {{Gallo}, E. and {Fender}, R.~P. and {Pooley}, G.~G.},
        title = "{A universal radio-X-ray correlation in low/hard state black hole binaries}",
      journal = {\mnras},
         year = 2003,
        month = sep,
       volume = {344},
       number = {1},
        pages = {60-72},
          doi = {10.1046/j.1365-8711.2003.06791.x},
archivePrefix = {arXiv},
       eprint = {astro-ph/0305231},
 primaryClass = {astro-ph},
       adsurl = {https://ui.adsabs.harvard.edu/abs/2003MNRAS.344...60G}
}

@ARTICLE{Giannios2011,
       author = {{Giannios}, Dimitrios and {Metzger}, Brian D.},
        title = "{Radio transients from stellar tidal disruption by massive black holes}",
      journal = {\mnras},
         year = 2011,
        month = sep,
       volume = {416},
       number = {3},
        pages = {2102-2107},
          doi = {10.1111/j.1365-2966.2011.19188.x},
archivePrefix = {arXiv},
       eprint = {1102.1429},
 primaryClass = {astro-ph.HE},
       adsurl = {https://ui.adsabs.harvard.edu/abs/2011MNRAS.416.2102G}
}

@ARTICLE{Yuan2004,
       author = {{Yuan}, Feng and {Narayan}, Ramesh},
        title = "{On the Nature of X-Ray-Bright, Optically Normal Galaxies}",
      journal = {\apj},
         year = 2004,
        month = sep,
       volume = {612},
       number = {2},
        pages = {724-728},
          doi = {10.1086/422802},
archivePrefix = {arXiv},
       eprint = {astro-ph/0401117},
 primaryClass = {astro-ph},
       adsurl = {https://ui.adsabs.harvard.edu/abs/2004ApJ...612..724Y}
}

@ARTICLE{Onori2022,
       author = {{Onori}, F. and {Cannizzaro}, G. and {Jonker}, P.~G. and {Kim}, M. and {Nicholl}, M. and {Mattila}, S. and {Reynolds}, T.~M. and {Fraser}, M. and {Wevers}, T. and {Brocato}, E. and {Anderson}, J.~P. and {Carini}, R. and {Charalampopoulos}, P. and {Clark}, P. and {Gromadzki}, M. and {Guti{\'e}rrez}, C.~P. and {Ihanec}, N. and {Inserra}, C. and {Lawrence}, A. and {Leloudas}, G. and {Lundqvist}, P. and {M{\"u}ller-Bravo}, T.~E. and {Piranomonte}, S. and {Pursiainen}, M. and {Rybicki}, K.~A. and {Somero}, A. and {Young}, D.~R. and {Chambers}, K.~C. and {Gao}, H. and {de Boer}, T.~J.~L. and {Magnier}, E.~A.},
        title = "{The nuclear transient AT 2017gge: a tidal disruption event in a dusty and gas-rich environment and the awakening of a dormant SMBH}",
      journal = {\mnras},
         year = 2022,
        month = nov,
       volume = {517},
       number = {1},
        pages = {76-98},
          doi = {10.1093/mnras/stac2673},
archivePrefix = {arXiv},
       eprint = {2206.00049},
 primaryClass = {astro-ph.HE},
       adsurl = {https://ui.adsabs.harvard.edu/abs/2022MNRAS.517...76O}
}

@ARTICLE{Kroupa2001,
       author = {{Kroupa}, Pavel},
        title = "{On the variation of the initial mass function}",
      journal = {\mnras},
         year = 2001,
        month = apr,
       volume = {322},
       number = {2},
        pages = {231-246},
          doi = {10.1046/j.1365-8711.2001.04022.x},
archivePrefix = {arXiv},
       eprint = {astro-ph/0009005},
 primaryClass = {astro-ph},
       adsurl = {https://ui.adsabs.harvard.edu/abs/2001MNRAS.322..231K}
}

@ARTICLE{Zhong2025,
       author = {{Zhong}, Shiyan},
        title = "{Modeling the UV/Optical Light Curve of Re-brightening Tidal Disruption Events}",
      journal = {\apj},
         year = 2025,
        month = apr,
       volume = {983},
       number = {2},
          eid = {131},
        pages = {131},
          doi = {10.3847/1538-4357/adc005},
archivePrefix = {arXiv},
       eprint = {2412.12549},
 primaryClass = {astro-ph.HE},
       adsurl = {https://ui.adsabs.harvard.edu/abs/2025ApJ...983..131Z}
}

@ARTICLE{Zhuang2021,
       author = {{Zhuang}, Jialun and {Shen}, Rong-Feng},
        title = "{The late flare in tidal disruption events due to the interaction of disk wind with dusty torus}",
      journal = {Journal of High Energy Astrophysics},
         year = 2021,
        month = nov,
       volume = {32},
        pages = {11-19},
          doi = {10.1016/j.jheap.2021.06.001},
archivePrefix = {arXiv},
       eprint = {2106.08793},
 primaryClass = {astro-ph.HE},
       adsurl = {https://ui.adsabs.harvard.edu/abs/2021JHEAp..32...11Z}
}

@ARTICLE{Jiang2019_10adi,
       author = {{Jiang}, Ning and {Wang}, Tinggui and {Mou}, Guobin and {Liu}, Hui and {Dou}, Liming and {Sheng}, Zhenfeng and {Wang}, Yibo},
        title = "{Infrared Echo and Late-stage Rebrightening of Nuclear Transient Ps1-10adi: Exploring the Torus with Tidal Disruption Events in Active Galactic Nuclei}",
      journal = {\apj},
         year = 2019,
        month = jan,
       volume = {871},
       number = {1},
          eid = {15},
        pages = {15},
          doi = {10.3847/1538-4357/aaf6b2},
archivePrefix = {arXiv},
       eprint = {1812.01295},
 primaryClass = {astro-ph.GA},
       adsurl = {https://ui.adsabs.harvard.edu/abs/2019ApJ...871...15J}
}

@ARTICLE{Liu2025_pTDE,
       author = {{Liu}, Chang and {Yarza}, Ricardo and {Ramirez-Ruiz}, Enrico},
        title = "{Repeating Partial Tidal Encounters of Sun-like Stars Leading to Their Complete Disruption}",
      journal = {\apj},
         year = 2025,
        month = jan,
       volume = {979},
       number = {1},
          eid = {40},
        pages = {40},
          doi = {10.3847/1538-4357/ad9b0b},
archivePrefix = {arXiv},
       eprint = {2406.01670},
 primaryClass = {astro-ph.HE},
       adsurl = {https://ui.adsabs.harvard.edu/abs/2025ApJ...979...40L}
}

@ARTICLE{Cardelli1989,
       author = {{Cardelli}, Jason A. and {Clayton}, Geoffrey C. and {Mathis}, John S.},
        title = "{The Relationship between Infrared, Optical, and Ultraviolet Extinction}",
      journal = {\apj},
         year = "1989",
        month = "Oct",
       volume = {345},
        pages = {245},
          doi = {10.1086/167900},
       adsurl = {https://ui.adsabs.harvard.edu/abs/1989ApJ...345..245C}
}



\newpage
\noindent $^{1}$ Miller Institute for Basic Research in Science, 206B Stanley Hall, Berkeley, CA 94720, USA\\
$^{2}$ Department of Astronomy, University of California, Berkeley, CA 94720-3411, USA\\
$^{3}$ Berkeley Center for Multi-messenger Research on Astrophysical Transients and Outreach (Multi-RAPTOR), University of California, Berkeley, CA 94720-3411, USA\\
$^{4}$ School of Natural Sciences, Institute for Advanced Study, 1 Einstein Drive, Princeton, NJ 08540, USA\\
$^{5}$ Department of Physics, University of California, 366 Physics North MC 7300, Berkeley, CA 94720, USA\\
$^{6}$ Space Research Institute (IKI), Russian Academy of Sciences, Profsoyuznaya 84/32, Moscow 117997, Russia\\
$^{7}$ Max-Planck-Institut f\"{u}r Astrophysik, Karl-Schwarzschild-Str. 1, D-85741 Garching, Germany\\
$^{8}$ Bloomberg Center for Physics and Astronomy, Johns Hopkins University, 3400 N. Charles St., Baltimore, MD 21218, USA\\
$^{9}$ Department of Physics, Syracuse University, Syracuse, NY 13210,USA\\
$^{10}$ Theoretical Astrophysics Center, University of California, Berkeley, CA 94720, USA\\
$^{11}$ Department of Physics \& Kavli Institute for Astrophysics and Space Research, Massachusetts Institute of Technology, Cambridge, MA 02139, USA\\
$^{12}$ Eureka Scientific, 2452 Delmer Street Suite 100, Oakland, CA 94602, USA\\
$^{13}$ Department of Physics, The George Washington University, Washington, DC 20052, USA\\
$^{14}$ Department of Astronomy/Steward Observatory, 933 North Cherry Avenue, Room N204, Tucson, AZ 85721-0065, USA\\
$^{15}$ Astrophysics Research Centre, School of Mathematics and Physics, Queen’s University Belfast, Belfast BT7 1NN, UK\\
$^{16}$ Center for Astrophysics | Harvard \& Smithsonian, 60 Garden Street, Cambridge, MA 02138-1516, USA\\
$^{17}$ McWilliams Center for Cosmology and Astrophysics, Department of Physics, Carnegie Mellon University, 5000 Forbes Avenue, Pittsburgh, PA 15213, USA\\
$^{18}$ Astrophysics Research Institute, Liverpool John Moores University, 146 Brownlow Hill, Liverpool L3 5RF, UK\\
$^{19}$ Department of Astronomy, Cornell University, Ithaca, NY 14853, USA\\
$^{20}$ California Institute of Technology, TAPIR, Mail Code 350-17, Pasadena, CA 91125, USA\\
$^{21}$ Division of Physics, Mathematics and Astronomy, California Institute of Technology, 1200 E. California Blvd, Pasadena, CA 91125, USA\\
$^{22}$ Anton Pannekoek Institute for Astronomy, University of Amsterdam, Science Park 904, 1098XH Amsterdam, the Netherlands\\
$^{23}$ Department of Physics and Astronomy, University of California, Los Angeles, CA 90095, USA\\
$^{24}$ The Oskar Klein Centre, Department of Astronomy, Stockholm University, AlbaNova, SE-10691, Stockholm, Sweden\\
$^{25}$ Kavli Institute for Particle Astrophysics and Cosmology, Stanford, CA 94305, USA\\
$^{26}$ School of Physics and Astronomy, University of Minnesota, Minneapolis, Minnesota 55455, USA\\
$^{27}$ Cahill Center for Astrophysics, California Institute of Technology, MC 249-17, 1200 E California Boulevard, Pasadena, CA 91125, USA\\
$^{28}$ IPAC, California Institute of Technology, 1200 E. California Blvd, Pasadena, CA 91125, USA\\
$^{30}$ Center for Data Driven Discovery, California Institute of Technology, Pasadena, CA 91125, USA\\
$^{31}$ Caltech Optical Observatories, California Institute of Technology, Pasadena, CA 91125, USA


\appendix

\section{Supplementary Figures}

\begin{figure}
\centering
    \includegraphics[width=0.8\columnwidth]{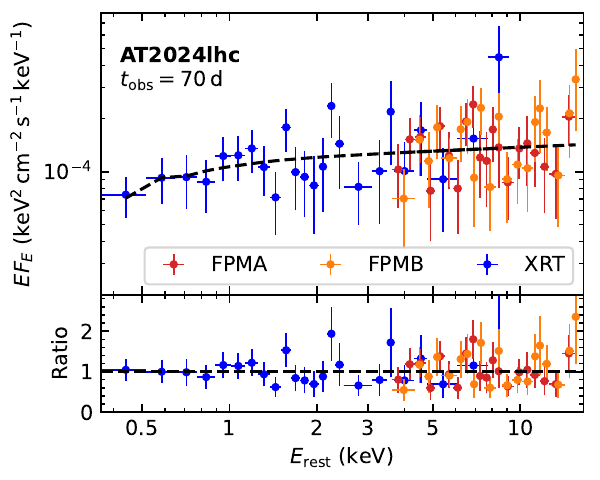}\\
    \includegraphics[width=0.54\columnwidth]{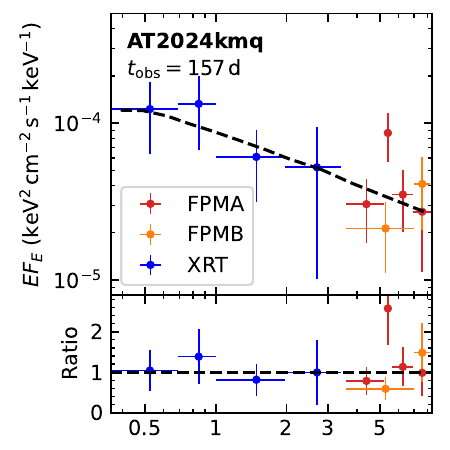}
    \includegraphics[width=0.45\columnwidth]{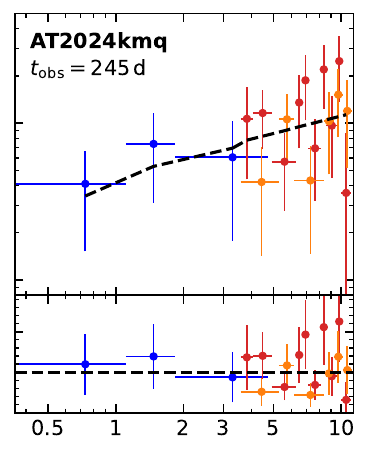}\\
    \includegraphics[width=0.54\columnwidth]{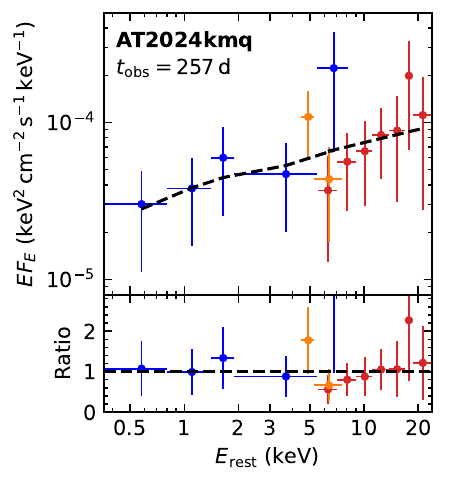}
    \includegraphics[width=0.45\columnwidth]{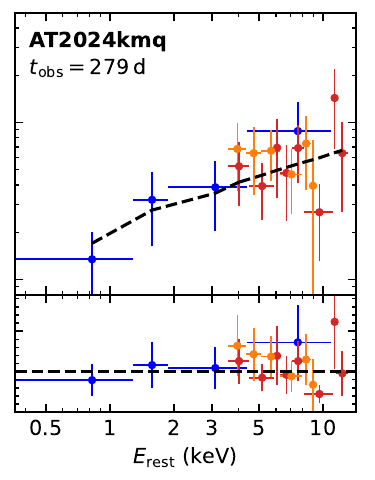}\\
    \caption{Joint spectral fitting between NuSTAR and XRT. The dashed lines show the best-fit absorbed power-law model. \label{fig:nustar}}
\end{figure}

The results of joint NuSTAR+XRT spectral fitting are shown in Figure~\ref{fig:nustar}. 

\section{Supplementary Tables}

A log of optical spectroscopy of AT2024kmq and AT2024lhc is presented in Table~\ref{tab:spec_lowres}.
A log of NuSTAR observations is presented in Table~\ref{tab:nustar}.
A log of XMM-Newton and Chandra observations is presented in Table~\ref{tab:xmm_cxo}.
The VLA observation logs are presented in Tables~\ref{tab:vla_24lhc}--\ref{tab:vla_24kmc}.

\begin{table*}
\caption{Log of optical spectroscopy.
\label{tab:spec_lowres}} 
\begin{threeparttable}
\begin{tabular}{ccrccccc}
\hline
IAU Name
&  Start Date 
& $t_{\rm rest}$
& Telescope
& Instrument
& Wavelength Range
& Slit width
& Exp.\\ 
& (UT)
& (d)
& 
& 
& (\AA)
& (\arcsec)
& (s)\\ 
\hline
AT2024lhc & 2024-06-14.4 & 42 & Shane & Kast & 3525--10500  & 2.0 & 3660/3600$^{a}$ \\
AT2024lhc & 2024-06-28.3 & 54 & Shane & Kast & 3525--10500  & 2.0 & 3660/3600$^{a}$ \\
AT2024lhc &2024-07-02.5 & 57 &Keck I &LRIS  & 3260--10250 & 1.0 & 650/600$^{a}$ \\
AT2024lhc & 2024-07-29.4 & 80 & Keck I & LRIS & 3260--10250 & 0.7 & 600 \\
AT2024lhc & 2024-08-15.4  & 94  &  Shane & Kast &  3525--10500  & 2.0 & 3660/3600$^{a}$ \\
AT2024lhc & 2024-08-28.4  & 104  &  Shane & Kast &  3525--10500  & 2.0 & 3660/3600$^{a}$ \\
AT2024lhc & 2025-03-05.6 & 262 & Keck I & LRIS & 3260--10250 & 1.0 & 1200/1150$^{a}$ \\
AT2024lhc & 2026-06-01.5 & 335 & Keck I & LRIS & 3260--10250 & 1.0 & 2400/2300$^{a}$ \\
\hline
AT2024kmq & 2024-11-28.5$^{b}$ & 153 & Shane & Kast & 4170--6830 & 1.5 & 3566\\
AT2024kmq & 2024-11-29.5$^{b}$ & 154 & Shane & Kast & 4170--6830 & 1.5 & 2145\\
AT2024kmq & 2024-12-08.2 & 161 &  NOT  & ALFOSC & 3500--9700   & 1.0 & 3000\\
AT2024kmq & 2025-06-25.6 &  328  & Keck I & LRIS & 3260--10250 & 1.0 & 2400/2300$^{a}$ \\
\hline
\end{tabular}
\begin{tablenotes}
\item $^{a}$ Exposure times on blue/red sides of the spectrograph.
\item $^{b}$ The red side of Kast was not available on those dates. As such, the 452/3306 grism was used on the blue side to increase the wavelength coverage. These two spectra are combined and displayed as a single spectrum in Figure~\ref{fig:opt_spec_24kmq}.
\end{tablenotes}
\end{threeparttable}
\end{table*}

\begin{table*}
\caption{NuSTAR observation log and joint spectral fitting with XRT. \label{tab:nustar}}
\begin{threeparttable}
\begin{tabular}{cccccccccc}
\hline 
     Name & obsID & Exp. & Start Time & $t_{\rm obs}$& $t_{\rm rest}$ & FPMA Count Rate  & $\Gamma$ & $f_{\rm X}$ & \texttt{cstat/dof}\\
        &    & (ks) & (UT) & (d) & (d) & ($\rm count\,s^{-1}$) &  &($10^{-13}$\,erg\,s$^{-1}$\,cm$^{-2}$) \\
\hline
AT2024kmq & 81001608002$^{a}$ & 67.3 & 2024-11-02.1 & 157 & 132 & $0.0008 \pm 0.0002$ 
&$2.59^{+0.17}_{-0.18}$ & $3.68^{+0.87}_{-0.66}$ & 31.15/33\\
          & 81002638002$^{b}$ & 23.0 & 2025-01-29.5 & 245 & 206 &$0.0026 \pm 0.0004$ 
& $1.63^{+0.20}_{-0.22}$ & $3.66^{+0.61}_{-0.63}$ & 47.68/39 \\
          & 81002638004$^{b}$ & 22.5 & 2025-02-10.9 & 257 & 216 & $0.0015\pm0.0004$ 
& $1.72^{+0.16}_{-0.17}$ & $2.75^{+0.51}_{-0.39}$ & 28.09/47\\
          & 81002638006$^{b}$ & 49.4 & 2025-03-04.6 & 279 & 234 & $0.0012 \pm 0.0002$ 
& $1.59^{+0.16}_{-0.18}$ & $1.91^{+0.20}_{-0.46}$ & 45.25/50\\
\hline
AT2024lhc & 81001628002$^{c}$ & 54.6 & 2024-07-03.1 & 70 & 59 & $0.0037\pm0.0003$ 
&$1.93\pm0.05$ &  $6.46^{+0.42}_{-0.39}$&  81.22/93\\
\hline
\end{tabular}
\begin{tablenotes}
\item The FPMA net count rate is given in the 3--15\,keV energy band. 
\item $^{a}$ PI: Guolo. Jointly modeled with XRT obsID 16717008. 
\item $^{b}$ PI: Pasham. The $t_{\rm obs}=245$\,d observation was jointly modeled with XRT obsID 16717015; 
the $t_{\rm obs}=279$\,d observation was jointly modeled with XRT obsIDs 16717020, 16717016, 16717022;
and the $t_{\rm obs}=257$\,d observation was jointly modeled with XRT obsIDs 16717023, 16717024, 16717025.
\item $^{c}$ PI: Yao. Jointly modeled with XRT obsID 16665006. 
\end{tablenotes}
\end{threeparttable}
\end{table*}

\begin{table*}
    \centering
    \caption{XMM-Newton and Chandra observation log and spectral fitting results. \label{tab:xmm_cxo}}
\begin{threeparttable}
    \begin{tabular}{ccccccccc}
    \hline
       Name & Telescope & obsID  & Exp. & Start Time & $t_{\rm obs}$ & $\Gamma$ & $f_{\rm X}$ & \texttt{cstat/dof}\\
    &  &  & (ks) & (UT) & (d) &  & ($10^{-13}$\,erg\,s$^{-1}$\,cm$^{-2}$) \\
\hline
AT2024kmq & Chandra$^{a}$    & 30886      & 16.7 & 2025-05-05.3 & 340 &  $1.77\pm 0.27$ & $0.99^{+0.11}_{-0.13}$ & 32.43/31 \\
\hline
AT2024lhc & XMM-Newton$^{b}$ & 0942540201 & 14.7 & 2024-09-23.4 & 152 & $2.13\pm 0.06$ & $4.39^{+0.22}_{-0.23}$ & 30.42/26  \\
          & Chandra$^{a}$    &  30885     & 15.9 & 2025-05-06.6 & 377 & $1.74\pm 0.15$ & $3.05^{+0.20}_{-0.17}$ & 43.28/47\\
          & XMM-Newton$^{c}$ & 0953000101 & 7.6 & 2025-08-14.4 & 477 & $2.06^{+0.30}_{-0.29}$ & $0.67_{-0.12}^{+0.26}$  & 14.84/10\\
\hline
\end{tabular}
\end{threeparttable}
\begin{tablenotes}
    \item $^{a}$ PI: Yao. Observed as part of the joint VLA+Chandra program 25A-15.
    \item $^{b}$ PI: Guolo. Observed as part of a dedicated X-
ray follow-up program targeting optically selected TDEs.
    \item $^{c}$ PI: Chornock. Observed as part of a joint HST+XMM-Newton program. 
\end{tablenotes}
\end{table*}

\begin{table}
\caption{Targeted VLA observations of AT2024lhc.\label{tab:vla_24lhc}}
\begin{tabular}{l|c|c|c|c|c|c}
    \hline
	   Date & $t_{\rm obs}$ & $t_{\rm rest}$& Receiver & $\nu_0$ & $f_\nu$ & Array \\ 
		    & (day)  & (day) & (GHz) & ($\mu$Jy) & &Config.\\
    \hline
    2024-07-10.0 & 77 & 64& X & 10.0 & $<11.4$ & B \\
    2024-08-25.8 & 123 & 102 & X & 10.0 & $<10.2$ & B \\
    2024-10-22.7 & 182 & 151 &X & 10.0 & $<10.5$ & A \\
    2025-01-17.4 & 269 & 222 & X & 10.0 & $<11.4$ & A \\
    2025-05-26.2 & 397 & 329 & X & 10.0 & $<7.5$  & C \\
    2025-06-02.1 & 404 & 335 & C & 6.0 &  $<14.4$  & C \\
    \hline
    \end{tabular}
\end{table}

\begin{table*}
\caption{Targeted VLA observations of AT2024kmq.\label{tab:vla_24kmc}}
\begin{tabular}{c|c|c|c|c|c|c|c|c}
    \hline
	    Date & $t_{\rm obs}$ & $t_{\rm rest}$ & Receiver &  $\nu_0$ &Beam Size & Integrated $f_\nu$ & Peak $f_\nu$ & Array Config. \\
	     &  (day)   &  (day)         &              &   (GHz) & (\arcsec)& ($\mu$Jy) &($\mu$Jy) & \\
     \hline
    2025-01-17.6 & 232 & 195 & X & 10.0 & $0.22\times 0.20$ & $117\pm12$ & $66\pm5$  & A\\
    2025-01-17.6 & 232 & 195& C & 6.0 & $0.42\times 0.36$ & $156\pm11$  & $99 \pm 5$ & A \\ 
    2025-02-27.4 & 273 & 229&X & 10.0 & $9.14\times 7.80$ & $312\pm19$ & $169\pm7$ & D\\
    2025-02-27.4 & 273 & 229&C & 6.0 & $15.98\times 12.53$ & $508\pm30$  & $340\pm13$ & D\\
    2025-03-20.4 & 294 & 247&Ku & 15.1 & $5.73\times 5.56$ & $230\pm17$ & $130\pm7$ & D\\
    2025-03-20.4 & 294 & 247&K & 22.1 & $4.53\times 3.43$ & $173\pm17 $ & $94\pm6$ & D\\
    2025-07-02.1 & 398 & 334 &Ku & 15.1 & $1.85\times 1.65$  & $158\pm17$  & $115\pm8$  & C\\
    2025-07-02.2 & 398 & 334 &K  & 22.0 & $1.59\times 1.11$  & $152\pm14$  & $120\pm7$ & C\\
    2025-07-15.1 & 411 & 345 &X  & 10.0 & $2.57\times 2.46$  & $318\pm28$ & $165\pm10$ & C\\
    2025-07-15.1 & 411 & 345 &C  & 6.0 & $4.52\times4.01$  & $561\pm56$ &  $161\pm13$ & C\\
     \hline
\end{tabular}
\end{table*}


\bsp	
\label{lastpage}
\end{document}